\documentclass{aa}

\usepackage{chemfig}
\usepackage[version=4,arrows=pgf]{mhchem}
\mhchemoptions{arrows=pgf{Kite[length=10pt 4,width=0pt 1]}{0.15ex}}
\usepackage{graphicx}
\usepackage{txfonts}
\usepackage{natbib}
\usepackage{siunitx}
\usepackage{makecell}
\usepackage{adjustbox}
\bibpunct[]{(}{)}{;}{a}{}{,}

\begin{document} 

   \title{Exploring molecular complexity with ALMA (EMoCA): Complex isocyanides in Sgr B2(N)}


   \author{E. R. Willis
          \inst{1}
          \and
          R. T. Garrod\inst{1,2}
          \and
          A. Belloche\inst{3}
          \and
          H. S. P. M\"{u}ller\inst{4}
          \and
          C. J. Barger\inst{1}
          \and
          M. Bonfand\inst{3}
          \and
          K. M. Menten\inst{3}
          }

   \institute{Department of Chemistry, University of Virginia, Charlottesville, VA 22904 USA\\
              \email{ew2zb@virginia.edu}
         \and
             Department of Astronomy, University of Virginia, Charlottesville, VA 22904 USA
         \and
             Max-Planck-Institut f\"{u}r Radioastronomie, Auf dem H\"{u}gel 69, 53121 Bonn, Germany
         \and
             I. Physikalisches Institut, Universit\"{a}t zu K\"{o}ln, Z\"{u}lpicher Str. 77, 50937 K\"{o}ln, Germany
             }

   \date{Received XX XXXX 2018; accepted XX XXXX XXXX}

 
  \abstract
   {The Exploring Molecule Complexity with ALMA (EMoCA) survey is an imaging spectral line survey using the Atacama Large Millimeter/submillimeter Array (ALMA) to study the hot-core complex Sagittarius B2(N). Recently, EMoCA revealed the presence of three new hot cores in this complex (N3-N5), in addition to providing detailed spectral data on the previously known hot cores in the complex (N1 and N2). The present study focuses on N2, which is a rich and interesting source for the study of complex molecules whose narrow line widths ameliorate the line confusion problem.}
   {We investigate the column densities and excitation temperatures of cyanide and isocyanide species in Sgr~B2(N2). We then use state-of-the-art chemical models to interpret these observed quantities. We also investigate the effect of varying the cosmic-ray ionization rate ($\zeta$) on the chemistry of these molecules. }
   {We used the EMoCA survey data to search for isocyanides in Sgr B2(N2) and their corresponding cyanide analogs. We then used the coupled three-phase chemical kinetics code {\em MAGICKAL} to simulate their chemistry. Several new species, and over 100 new reactions have been added to the network. In addition, a new single-stage simultaneous collapse/warm-up model has been implemented, thus eliminating the need for the previous two-stage models. A variable, visual extinction-dependent $\zeta$ was also incorporated into the model and tested.}
   {We report the tentative detection of \ce{CH3NC} and \ce{HCCNC} in Sgr B2(N2), which represents the first detection of both species in a hot core of Sgr~B2. In addition, we calculate new upper limits for \ce{C2H5NC}, \ce{C2H3NC}, \ce{HNC3}, and \ce{HC3NH+}. Our updated chemical models can reproduce most observed NC:CN ratios reasonably well depending on the physical parameters chosen. The model that performs best has an extinction-dependent cosmic-ray ionization rate that varies from $\sim$\num{2e-15} s$^{-1}$ at the edge of the cloud to $\sim$\num{1e-16} s$^{-1}$ in the center. Models with higher extinction-dependent $\zeta$ than this model generally do not agree as well, nor do models with a constant $\zeta$ greater than the canonical value of \num{1.3e-17} s$^{-1}$ throughout the source. Radiative transfer models are run using results of the best-fit chemical model. Column densities produced by the radiative transfer models are significantly lower than those determined observationally. Inaccuracy in the observationally determined density and temperature profiles is a possible explanation. Excitation temperatures are well reproduced for the true ``hot core'' molecules, but are more variable for other molecules such as \ce{HC3N}, for which fewer lines exist in ALMA Band 3.}
   {The updated chemical models do a very good job of reproducing the observed abundances ratio of \ce{CH3NC}:\ce{CH3CN} towards Sgr B2(N2), while being consistent with upper limits for other isocyanide/cyanide pairs. \ce{HCCNC}:\ce{HC3N} is poorly reproduced, however. Our results highlight the need for models with $A_V$-depdendent $\zeta$. However, there is still much to be understood about the chemistry of these species, as evidenced by the systematic overproduction of \ce{HCCNC}. Further study is also needed to understand the complex effect of varying $\zeta$ on the chemistry of these species. The new single-stage chemical model should be a powerful tool in analyzing hot-core sources in the future.}


   \maketitle
%

\section{Introduction} 
\label{sec:intro}
In chemistry, a cyanide is any organic molecule that contains the cyano functional group (\ce{-C#N}). Cyanide species have been observed astronomically for some time, beginning with the first detection of the CN radical toward visually bright stars in the optical regime \citep{mckellar40}. Radio telescopes did not detect the CN radical until 30 years later \citep{jefferts70}. More complex cyanides have since been detected in the interstellar medium (ISM), with the identification of hydrogen cyanide \citep[HCN;][]{snyder71} and methyl cyanide (CH$_{3}$CN; \citealt{solomon71}) the following year. Since then, many cyanide species have been found, including vinyl cyanide (\ce{C2H3CN}; \citealt{gardner75}), ethyl cyanide (\ce{C2H5CN}; \citealt{djohnson77}), \emph{n}-propyl cyanide (\emph{n}-\ce{C3H7CN}; \citealt{belloche09}) and \emph{i}-propyl cyanide (\emph{i}-\ce{C3H7CN}, a branched species; \citealt{belloche14}). Most recently, benzonitrile (\emph{c}-C$_{6}$H$_{5}$CN; \citealt{mcguire17}), the first benzene-derived aromatic molecule detected in the ISM through radio astronomy was seen towards the dark cloud TMC-1.

Although these detections are interesting in that they help to reveal the chemical complexity achieved in the ISM, observations of these molecules have also proven useful from a practical standpoint. For example, the HCN($J$=1$-$0) rotational line has been used extensively as a dense gas tracer in external galaxies \citep[e.g.,][]{schirm16,sliwa17,kjohnson18}. HCN is also very abundant in the atmosphere of Titan, and has been used to measure nitrogen fractionation there \citep[e.g.,][]{molter16}. CH$_{3}$CN is regularly used to determine kinetic temperature in star-forming regions \citep{bell14}, and \ce{HC3N} has been shown to be important in observations of ultraluminous infrared galaxies \citep{costagliola15}.

Despite the extent to which cyanides have been studied in the ISM, the isocyanides (molecules that contain the isocyano functional group, \ce{-N#C}) have been comparatively sparsely studied. The first isocyanide detected in the ISM was HNC, along with the related molecule HNCO, toward Sgr B2 \citep{snyder72}. Since these first detections, to the authors' knowledge, only seven other isocyanides  have been unambiguously detected in astronomical sources, most of them (only) in the circumstellar envelope of the carbon-rich asymptotic giant branch star IRC+10216: \ce{CH3NC} \citep{cernicharo88}, \ce{HCCNC} \citep{Kawaguchi92a}, \ce{MgNC} \citep{kawaguchi93}, \ce{AlNC} \citep{ziurys02}, \ce{SiNC} \citep{guelin04}, \ce{HMgNC} \citep{cabezas13}, and \ce{CaNC} \citep{cernicharo19}. \ce{HCCNC} was detected toward TMC-1 \citep{Kawaguchi92a}, and \citet{belloche13} reported a tentative detection of this species towards Sgr~B2(N) with the IRAM 30\,m telescope, with only one uncontaminated line and one blended line. \citet{remijan05} reported the detection of one transition of \ce{CH3NC} with the GBT toward Sgr~B2(N), with two velocity components seen in absorption and emission, respectively. This detection added an additional piece of evidence for the presence of \ce{CH3NC} in Sgr B2, complementing the initial tentative detection of three higher-energy transitions obtained with the IRAM 30\,m toward Sgr B2(OH) by \citet{cernicharo88}. Remijan et al. failed to detect any compact emission of \ce{CH3NC} with BIMA, suggesting that the GBT detection traces a large-scale distribution of \ce{CH3NC} in Sgr~B2.

Even more sparse than the observational efforts for isocyanides have been the modeling efforts, with the exception of HNC. Of the six most complex isocyanides detected, only \ce{HMgNC} has been included in a chemical network, in the aforementioned detection paper \citep{cabezas13}. The interstellar chemistry of \ce{CH3NC} has been investigated before \citep{defrees85}. These latter authors investigated the formation rate of the protonated ions \ce{CH3NCH+} and \ce{CH3CNH+} from the radiative association reaction of \ce{CH3+} and \ce{HCN}. They found, using ab initio quantum theory as well as equilibrium calculations, that the ratio of formation of \ce{CH3NCH+}/\ce{CH3CNH+} after relaxation of the complex should be between 0.1 and 0.4. However, this mechanism, as well as any others involving \ce{CH3NC}, does not appear to have been incorporated into a large chemical network until recent work by some of the present authors \citep{calcutt18}. 

It is important to confront new modeling efforts with state-of-the-art observational data. To this end, we make use of the Exploring Molecular Complexity with ALMA (EMoCA) survey. EMoCA is an imaging spectral line survey conducted towards Sagittarius B2(N). Sgr~B2(N) is a protocluster located in the Galactic center region, at a projected distance of about 100~pc from Sgr~A$^\star$. It contains a number of HII regions, some compact and ultracompact \citep[e.g.,][]{Gaume95}, and Class II methanol masers \citep{Caswell96}, both signposts of ongoing high mass star formation. Sgr~B2(N) also harbors several hot molecular cores at the early stage of high mass star formation which present a high density of spectral lines revealing the presence of numerous complex organic molecules. Many complex organic molecules were first detected toward Sgr B2(N), which motivated its selection as a target for the EMoCA survey. Analysis of the data from EMoCA has led to several important results, including the aforementioned detection of \emph{i}-propyl cyanide \citep{belloche14}, as well as important insights into deuteration levels in Sgr B2(N) \citep{Belloche16}. More recently, three new hot cores have been detected and characterized in Sgr B2(N), signifying further sources to study complex organic molecules (\citealt{bonfand17}, see also \citealt{sanchezmonge17}). This work focuses on using the EMoCA data to search for various nitrogen-containing organic molecules toward Sgr B2(N). We include alkyl cyanides and isocyanides in our search, which are chemical species in which the -CN or -NC group is attached to an alkyl substituent. Alkyl substituents are acyclic saturated hydrocarbons that are missing one hydrogen atom. We also search for simple cyanopolyynes and isocyanopolyynes (e.g., \ce{HC3N and HCCNC}).

This paper aims to be the most comprehensive modeling and observational study of isocyanide chemistry in the ISM to date. We expand on the chemical network for \ce{CH3NC} first introduced in \citet{calcutt18}. Several new molecules have been introduced in the chemical network as well. These include vinyl isocyanide (\ce{C2H3NC}) and ethyl isocyanide (\ce{C2H5NC}), as well as associated radicals (e.g., \ce{CH2NC}). This marks the first time that these molecules have been incorporated into an astrochemical network.

The paper is organized as follows. Section 2 provides information about the spectroscopic predictions used to analyze the observed spectra. Section 3 outlines the observational methods. Section 4 outlines the observational results. Section 5 discusses the additions made to the chemical modeling. Modeling results are presented in Sect. 6. Section 7 contains the discussion, while Sect. 8 is the conclusion.

\section{Observations}
\label{s:observations}
We use data from the the EMoCA spectral line survey performed with the Atacama Large Millimeter/submillimeter Array (ALMA) in its cycles 0 and 1 to search for various nitrogen-containing organic molecules toward the high-mass star-forming region Sgr B2(N). We used the main array with baselines ranging from $\sim$17~m to $\sim$400~m, which imply a maximum recoverable scale of $\sim 20\arcsec$. This scale translates into $\sim$0.8 pc at the adopted distance of 8.3 kpc \citep{Reid14}. The survey covers the frequency range between 84.1 and 
114.4~GHz in five setups. It has a spectral resolution of 488.3 kHz (1.7 to 1.3~km~s$^{-1}$)
and a median angular resolution of 1.6$\arcsec$ ($\sim$0.06 pc or $\sim$13000 au). The achieved rms sensitivity is on the order of 3~mJy~beam$^{-1}$, which translates into an rms sensitivity of 0.1-0.2~K in effective radiation temperature scale depending on the setup. The field was centered at ($\alpha, \delta$)$_{\rm J2000}$= ($17^{\rm h}47^{\rm m}19.87^{\rm s}, -28^\circ22'16''$), half way between the two main hot cores Sgr~B2(N1) and (N2) that are separated by $4.9\arcsec$ ($\sim$0.2~pc) in the north--south direction. Sgr~B2(N1) and (N2) have systemic velocities of $\sim$64 and $\sim$74~km~s$^{-1}$, respectively \citep[e.g.,][]{Belloche08,Belloche16}. The size of the primary beam varies between $69\arcsec$ at 84~GHz and 51$\arcsec$ at 114~GHz \citep{Remijan15}.
A detailed 
description of the observations, the data reduction process, and the method 
used to identify the detected lines and derive column densities was presented 
in \citet{Belloche16}.

\section{Laboratory spectroscopy background}
\label{lab-spec}
Transition frequencies were taken from the catalog of the Cologne Database 
for Molecular Spectroscopy, CDMS, \citep{CDMS_1,CDMS_3} for the most part. 
Other sources of data are mentioned in specific cases.

The CH$_3$CN $\varv _8 = 1$ laboratory data are based on \citet{MeCN_v8_le_2_rot_2015} 
with additional data, especially in the range of our survey, coming from \citet{MeCN-v8=1_1969}. 
The partition function includes energy levels from vibrational states up to $\sim$1700~K 
\citep{MeCN_v8_le_2_rot_2015}, and as such the vibrational contributions are complete at 170~K.
The CH$_3$NC $\varv  = 0$ data are based on \citet{MeNC_v0_etc_IR_Rot_2011}. 
Additional data, also in the range of our survey, were taken from \citet{MeNC-gs_1970}. 
The values of the spectroscopic parameters $A$ and $D_K$, along with vibrational 
information, were taken from \citet{MeNC_IR_1995}. Preliminary CH$_3$NC 
$\varv _8 = 1$ data were calculated from \citet{MeNC_v8eq1-2_IR_Rot_2011}.

The C$_2$H$_5$CN data are based on \citet{EtCN_rot_2009} with additional important 
information especially in the range of our survey from \citet{EtCN_rot_1996} and 
from \citet{EtCN_rot_1994}. Vibrational correction factors to the rotational 
partition function are available via the CDMS documentation. They are based on 
\citet{EtCN_isos_IR_1981}. The C$_2$H$_5$NC data are based on \citet{Margules18} 
with additional low-frequency data \citep{EtNC_rot1968,EtNC_rot_1985,EtNC_rot_1992}.
Vibrational energies of the three lowest vibrational fundamentals were estimated 
from quantum chemical calculations (H.~S.~P. M{\"u}ller, 2017, unpublished) in 
comparison to higher lying fundamentals \citep{EtNC_rot_IR_1969}.

The C$_2$H$_3$CN data are based on \citet{VyCN_div_rot_2008} with particularly 
noteworthy data in the range of our survey from \citet{VyCN_rot_1996}. 
The partition function includes numerous low-lying vibrational states 
(H.~S.~P. M{\"u}ller, 2008, unpublished) and is converged up to $\sim$200~K. 
The vibrational energies were based on \citet{VyCN_IR_1999} and on quantum 
chemical calculations (H.~S.~P. M{\"u}ller, 2008, unpublished). 
These vibrational data are compatible with more recent ones by 
\citet{VyCN_FIR_2015}. This latter study and references therein contain information 
on higher $J$, $K_a$, and frequencies, but more noteworthy on (in part) highly 
excited vibrational states of vinyl cyanide.
The C$_2$H$_3$NC data were taken from the JPL catalog \citep {JPL-catalog_1998} 
but are based on \citet{VyNC_rot_1975}. Additional data were taken from 
\citet{VyNC_rot_1982}. Vibrational corrections to the partition function 
were evaluated from \citet{RNC_IR_2015}.

The HC$_3$N $\varv _7 = 1$ data are based on \citet{HC3N_rot_2000} with 
additional data in the range of our survey from \citet{HC3N_rot_1986}. 
The HCCNC $\varv = 0$ data were taken from the JPL catalog; they are based on \citet{HCCNC_rot_1992} with additional data from \citet{HCCNC_rot_1991}. Vibrational corrections to the partition function were derived from \citet{HCCNC_IR_1992}. The HNC$_3$ data were based on \citet{Vastel18} with additional data from \citet{HNC3_rot_1993}. Vibrational corrections to the 
partition function were derived from \citet{HNC3_IR_2001}. The HC$_3$NH$^+$ data are based on \citet{HC3NH+_etc_rot_2000}. Vibrational corrections to the partition function were derived from \citet{HC3NH+_ai_1999}.

\section{Observational results}
\label{s:obs results}
We analyze here the spectrum of the secondary hot core Sgr~B2(N2) at the position ($\alpha, \delta$)$_{\rm J2000}$= ($17^{\rm h}47^{\rm m}19.86^{\rm s}, -28^\circ22'13.4''$) \citep[][]{Belloche16}. The degree of spectral line confusion is lower toward Sgr~B2(N2) thanks to its narrower line widths ($FWHM \sim 5$~km~s$^{-1}$). The column densities of CH$_3$CN, C$_2$H$_5$CN, C$_2$H$_3$CN, and HC$_3$N extracted from the EMoCA survey have already been reported in \citet{Belloche16}. They result from a detailed modeling of the entire spectrum under the assumption of local thermodynamic equilibrium  (LTE), which is valid here given the high densities \citep[$> 10^7$~cm$^{-3}$, see][]{bonfand19}, and the calculation includes transitions from vibrationally excited states and isotopologs. 
For each investigated molecule, a synthetic spectrum is produced using the software Weeds \citep{Maret11} which takes into account the line opacities and the finite resolution of the observations in the radiative transfer calculation. The spectrum of each molecule is modeled with five free parameters: the size of the emission assumed to be Gaussian, the column density, the temperature, the line width, and the velocity offset with respect to the assumed systemic velocity of the source. These parameters are adjusted until a good fit to the observed spectrum is achieved, as evaluated by visual inspection. Blends with lines of other species already included in the model are naturally taken into account in this procedure. The source size is measured by fitting a two-dimensional Gaussian to integrated intensity maps of transitions that are found to be relatively free of contamination on the basis of the synthetic spectra. Population diagrams are built a posteriori for species that have detected lines over a sufficiently large range of upper-level energies. The column densities of CH$_3$CN, C$_2$H$_5$CN, C$_2$H$_3$CN, and HC$_3$N
are listed in Table~\ref{t:coldens} as reported in our previous study. Because the transitions in the vibrational ground states of CH$_3$CN and HC$_3$N are optically thick, the column densities of both species were derived from an analysis of transitions within vibrationally excited states but they correspond to the total column density of the molecules. They are consistent with the column densities obtained for the isotopologs, including their vibrational ground state, after accounting for the $^{12}$C/$^{13}$C isotopic ratio that characterizes Sgr~B2(N) \citep[see][]{Belloche16}.

\subsection{Detection of CH$_3$NC and HCCNC}
\label{s:obs_detections}

\begin{table*}[!ht]
 \begin{center}
 \caption{
 Parameters of our best-fit LTE model of alkyl cyanides and isocyanides, and related species, toward Sgr~B2(N2).
}
 \label{t:coldens}
 \vspace*{-1.2ex}
 \begin{tabular}{lcrccccccr}
 \hline\hline
 \multicolumn{1}{c}{Molecule} & \multicolumn{1}{c}{Status\tablefootmark{a}} & \multicolumn{1}{c}{$N_{\rm det}$\tablefootmark{b}} & \multicolumn{1}{c}{Size\tablefootmark{c}} & \multicolumn{1}{c}{$T_{\mathrm{rot}}$\tablefootmark{d}} & \multicolumn{1}{c}{$N$\tablefootmark{e}} & \multicolumn{1}{c}{$F_{\rm vib}$\tablefootmark{f}} & \multicolumn{1}{c}{$\Delta V$\tablefootmark{g}} & \multicolumn{1}{c}{$V_{\mathrm{off}}$\tablefootmark{h}} & \multicolumn{1}{c}{$\frac{N}{N_{\rm ref}}$\tablefootmark{i}} \\ 
  & & & \multicolumn{1}{c}{\scriptsize ($''$)} & \multicolumn{1}{c}{\scriptsize (K)} & \multicolumn{1}{c}{\scriptsize (cm$^{-2}$)} & & \multicolumn{1}{c}{\scriptsize (km~s$^{-1}$)} & \multicolumn{1}{c}{\scriptsize (km~s$^{-1}$)} &  \\ 
 \hline
 CH$_3$CN, $\varv_8=1$$^\star$ & d & 20 &  1.4 &  170 &  2.2 $\times$ 10$^{18}$ & 1.00 & 5.4 & -0.5 &    1 \\ 
 CH$_3$NC, $\varv=0$ & t & 2 &  1.4 &  170 &  1.0 $\times$ 10$^{16}$ & 1.45 & 5.4 & -0.5 & 0.0047 \\ 
 CH$_3$NC, $\varv=1$ & n & 0 &  1.4 &  170 &  1.0 $\times$ 10$^{16}$ & 1.45 & 5.4 & -0.5 & 0.0047 \\ 
\hline 
 C$_2$H$_5$CN, $\varv=0$$^\star$ & d & 154 &  1.2 &  150 & 6.2 $\times$ 10$^{18}$ & 1.38 & 5.0 & -0.8 &    1 \\ 
 C$_2$H$_5$NC, $\varv=0$ & n & 0 &  1.2 &  150 & $<$ 1.5 $\times$ 10$^{15}$ & 1.47 & 5.0 & -0.8 & $<$ 0.00024 \\ 
\hline 
 C$_2$H$_3$CN, $\varv=0$$^\star$ & d & 44 &  1.1 &  200 & 4.2 $\times$ 10$^{17}$ & 1.00 & 6.0 & -0.6 &    1 \\ 
 C$_2$H$_3$NC, $\varv=0$ & n & 0 &  1.1 &  200 & $<$ 3.0 $\times$ 10$^{15}$ & 1.49 & 6.0 & -0.6 & $<$ 0.0071 \\ 
\hline 
 HC$_3$N, $\varv_7=1$$^\star$ & d & 6 &  1.3 &  170 & 3.5 $\times$ 10$^{17}$ & 1.44 & 5.0 & -0.7 &    1 \\ 
 HCCNC, $\varv=0$ & t & 2 &  1.3 &  170 & 5.1 $\times$ 10$^{14}$ & 1.55 & 5.0 & -0.7 & 0.0015 \\ 
 HNC$_3$, $\varv=0$ & n & 0 &  1.3 &  170 & $<$ 6.6 $\times$ 10$^{13}$ & 1.65 & 5.0 & -0.7 & $<$ 0.00019 \\ 
 HC$_3$NH$^+$, $\varv=0$ & n & 0 &  1.3 &  170 & $<$ 5.8 $\times$ 10$^{14}$ & 1.65 & 5.0 & -0.7 & $<$ 0.0017 \\ 
\hline 
 \end{tabular}
 \end{center}
 \vspace*{-2.5ex}
 \tablefoot{The parameters reported for CH$_3$CN, C$_2$H$_5$CN, C$_2$H$_3$CN, and HC$_3$N are taken from \citet{Belloche16}.
 \tablefoottext{a}{d: detection, t: tentative detection, n: non-detection.}
 \tablefoottext{b}{Number of detected lines \citep[conservative estimate, see Sect.~3 of][]{Belloche16}. One line of a given species may mean a group of transitions of that species that are blended together.}
 \tablefoottext{c}{Source diameter (\textit{FWHM}).}
 \tablefoottext{d}{Measured or assumed rotational temperature.}
 \tablefoottext{e}{Total column density of the molecule.}
 \tablefoottext{f}{Correction factor that was applied to the column density to account for the contribution of vibrationally excited states, in the cases where this contribution was not included in the partition function of the spectroscopic predictions.}
 \tablefoottext{g}{Linewidth (\textit{FWHM}).}
 \tablefoottext{h}{Velocity offset with respect to the assumed systemic velocity of Sgr~B2(N2), $V_{\mathrm{lsr}} = 74$ km~s$^{-1}$.}
 \tablefoottext{i}{Column density ratio, with $N_{\rm ref}$ the column density of the previous reference species marked with a $\star$.}
 }
 \end{table*}

\begin{figure*}[!h]
\centerline{\resizebox{0.85\hsize}{!}{\includegraphics[angle=0]{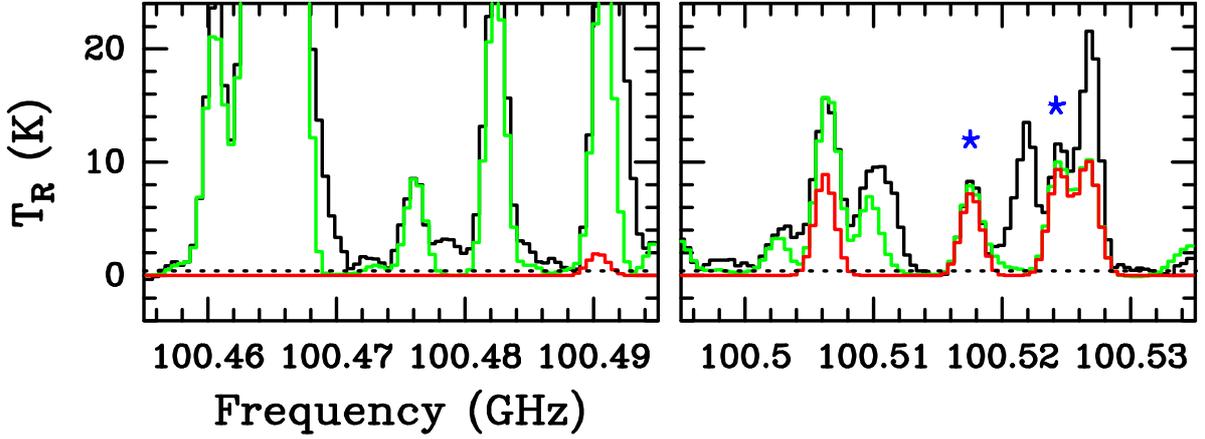}}}
\caption{Transitions of CH$_3$NC, $\varv = 0$ covered by our ALMA 
survey. The best-fit LTE synthetic spectrum of CH$_3$NC, $\varv = 0$ 
is displayed in red and overlaid on the observed spectrum of Sgr~B2(N2) shown 
in black. The green synthetic spectrum contains the contributions of all 
molecules identified in our survey so far, including the species shown in red. 
The central frequency and width are indicated in MHz below each panel. The y-axis is labeled in effective radiation temperature scale. The dotted line indicates the $3\sigma$ 
noise level. The lines counted as detected in Table~\ref{t:coldens} are marked with a blue star.
}
\label{f:spec_ch3nc_ve0}
\end{figure*}

We report here the tentative detection of CH$_3$NC toward Sgr~B2(N2). 
Figure~\ref{f:spec_ch3nc_ve0} shows the rotational transitions of the vibrational ground state of this molecule covered by the EMoCA survey. Two of them are detected (at 100518~MHz and 100524~MHz) without contamination from other species and well reproduced by our LTE model in red, which gives us confidence in the detection of CH$_3$NC. The line at 100524~MHz is considered as detected because it is closely associated with a peak in the observed spectrum (both in terms of velocity and
intensity). The parameters of the detected lines are listed in Table~\ref{t:list_r-nc}. The LTE model assumes the same parameters as for CH$_3$CN, with only the column density left as a free parameter. The other transitions of CH$_3$NC are consistent with the observed spectrum but blended with other species. Figure~\ref{f:spec_ch3nc_ve1} shows the rotational transitions from within this molecule's vibrationally excited state $\varv_8 = 1$ covered by the survey. The model shown in red assumes the same parameters as for the ground state. It is consistent with the observed spectrum, but all transitions are to some degree blended with lines from other (identified or unidentified) species and so a secure identification of this vibrational state cannot be made. The parameters of the best-fit LTE model of CH$_3$NC are reported in Table~\ref{t:coldens}.

\begin{figure*}[!h]
\centerline{\resizebox{\hsize}{!}{\includegraphics[angle=0]{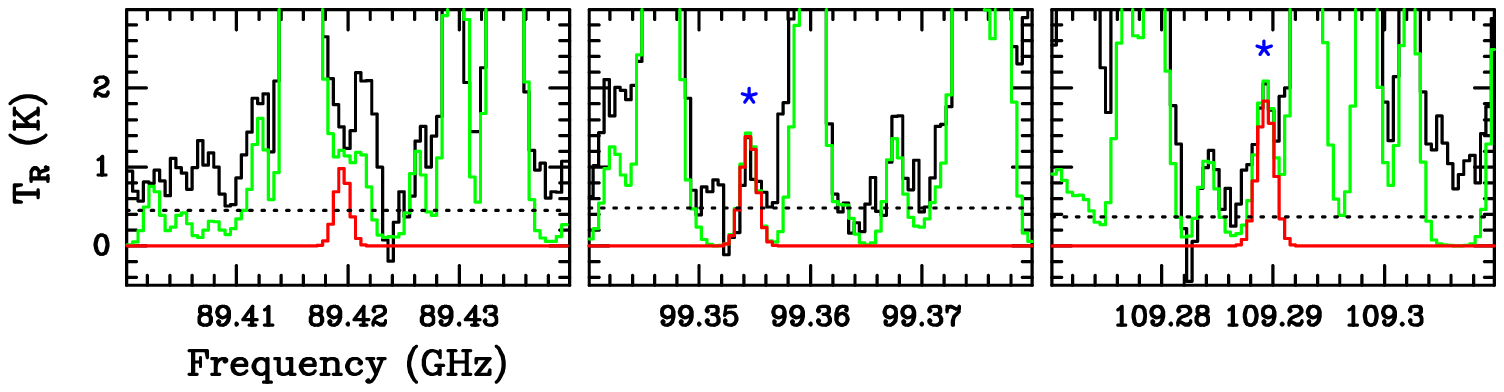}}}
\caption{Same as Fig.~\ref{f:spec_ch3nc_ve0} but for HCCNC, $\varv = 0$.
}
\label{f:spec_hccnc_ve0}
\end{figure*}

We also report the tentative detection of HCCNC toward Sgr~B2(N2). For the modeling, we assumed the same parameters as for HC$_3$N, except for the column density that was left as a free parameter. Out of the three lines of HCCNC covered by our survey, two are detected and well-reproduced by our model (see Fig.~\ref{f:spec_hccnc_ve0}), which gives us confidence in the detection of HCCNC. Like for CH$_3$NC, the second line at 109289~MHz is considered as 
detected because it is closely associated with a peak in the observed spectrum 
(both in terms of velocity and intensity). The third transition is consistent with the observed spectrum but blended with emission from other species. The parameters of the detected lines are listed in Table~\ref{t:list_r-nc}. The parameters of the best-fit LTE model of HCCNC are reported in Table~\ref{t:coldens}. We also searched for transitions from
within the vibrationally excited states $\varv_5 = 1$, $\varv_6 = 1$, and $\varv_7 = 1$ assuming the same parameters as for the ground state, but none are detected. Their predicted peak intensities are below the sensitivity limit of the EMoCA survey.

\begin{figure}[!h]
\centerline{\includegraphics[angle=0,width=\linewidth]{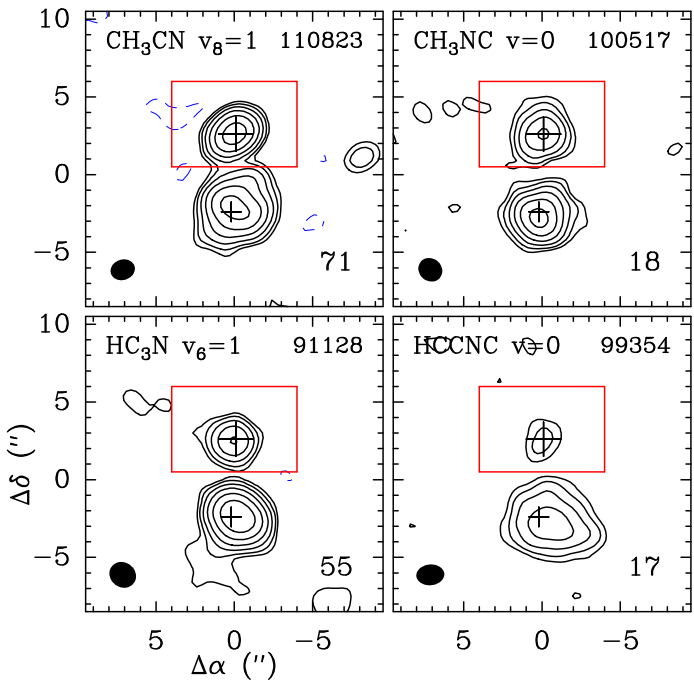}}
\caption{Integrated intensity maps of CH$_3$CN, CH$_3$NC, HC$_3$N, and HCCNC. In each panel, the name of the molecule followed by the vibrational state of the line is written in the top left corner, the line frequency in MHz is given in the top right corner, the rms noise level $\sigma$ in mJy~beam$^{-1}$~km~s$^{-1}$ is written in the bottom right corner, and the beam (HPBW) is shown in the bottom left corner. The black contour levels start at 3$\sigma$ and then increase geometrically by a factor of two at each step. The blue, dashed contours show the $-$3$\sigma$ level. The large and small crosses indicate the positions of the hot molecular cores Sgr~B2(N2) and Sgr~B2(N1), respectively. Because of the variation in systemic velocity across the field, the assignment of the detected emission to each molecule is valid only for the region around Sgr~B2(N2), highlighted with the red box.}
\label{f:maps}
\end{figure}

Examples of integrated intensity maps of rotational lines of CH$_3$CN, 
CH$_3$NC, HC$_3$N, and HCCNC are displayed in Fig.~\ref{f:maps}. The integration
was performed around the systemic velocity of Sgr~B2(N2). In all cases, the 
emission is centrally peaked on the hot core.

\subsection{Upper limits for C$_2$H$_5$NC, C$_2$H$_3$NC, HNC$_3$, and 
HC$_3$NH$^+$}
\label{s:obs_upperlimits}

We searched for C$_2$H$_5$NC, C$_2$H$_3$NC, HNC$_3$, and HC$_3$NH$^+$ toward Sgr~B2(N2) but did not detect these species. For a spectral line survey with thousands of lines detected, it is standard practice to model the emission and to use this model to derive upper limits. For the modeling, we assumed the same parameters as for C$_2$H$_5$CN, C$_2$H$_3$CN, HC$_3$N, and HC$_3$N, respectively, except for the column densities that were left as free 
parameters. The models used to obtain upper limits to their column densities are displayed in red in Figs.~\ref{f:spec_c2h5nc_ve0}--\ref{f:spec_hc3nhp_ve0}
and the upper limits are listed in Table~\ref{t:coldens}.

\section{Chemical modeling} \label{sec:modeling}
In this paper, we use the chemical kinetics code \textsc{MAGICKAL} \citep{garrod13}. The basis of the chemical network is taken from \citet{garrod17}, with the above-mentioned inclusion of \ce{CH3NC} chemistry first presented by  \citet{calcutt18}. The model also uses the grain-surface back-diffusion correction of \citet{willis17}. The model simulates a fully coupled gas-phase-, grain-surface-, and ice-mantle chemistry under time-dependent physical conditions appropriate to the source under consideration (see Section 5.2).

\subsection{Chemical network}
The chemical network for this study had to be expanded considerably to include isocyanide-related chemistry. This section is divided into subsections by molecule. Appendix B contains a more comprehensive list of reactions included in this updated model, as well as binding energies and enthalpies of formation for grain-surface species of note. Here we only focus on the most important reactions.

\subsubsection{\ce{CH3NC}}
The incorporation of \ce{CH3NC} into our chemical network was discussed in some detail by \citet{calcutt18}, but it is summarized here. \ce{CH3NC} is formed primarily through the radiative association of \ce{CH3+} and \ce{HCN}, which produces two isomers, \ce{CH3CNH+} and \ce{CH3NCH+}, in a ratio of 85:15 \citep{defrees85} due to unimolecular isomerization. These isomers can then recombine with electrons to produce \ce{CH3CN} and \ce{CH3NC} , respectively. We note that the protonated form of both the cyanide and isocyanide can be formed from proton transfer reactions with species such as \ce{H3O+}, but they are not assumed to be formed with enough internal energy to isomerize in that case. No isomerization is assumed to occur as a result of recombination. This schematic is shown below:
\begin{equation}
    \ce{CH3+ + HCN -> CH3CNH+/CH3NCH+ + h$\nu$}
    \label{eq:1}
\end{equation}
\begin{equation}
    \ce{CH3CNH+/CH3NCH+ + e- -> CH3CN/CH3NC + H}
    \label{eq:2}
\end{equation}
For both isomers, $\sim$40\% of recombinations produce \ce{CH3CN} and \ce{CH3NC}, respectively. The remaining $\sim$60\% produce more fragmented species. These branching ratios are taken from \citet{loison14}, based on laboratory work from \citet{plessis12}. There is currently no known efficient grain-surface formation path for \ce{CH3NC}. Due to the lack of efficient pathways for interconversion between the isomers, the point of divergence in the chemical networks for \ce{CH3CN} and \ce{CH3NC} therefore occurs with Reaction \ref{eq:1}.

Reaction with abundant positive ions (e.g., \ce{C+, H3+, H+}) is the primary destruction pathway for \ce{CH3NC} at T < 100~K. At higher temperatures, ion-molecule reactions are still important, but reaction with atomic hydrogen becomes the dominant destruction pathway for \ce{CH3NC} following:
\begin{equation}
    \ce{CH3NC + H -> CH3 + HCN}
    \label{eq:3}
\end{equation}
The activation energy barrier of this reaction is not known experimentally. The standard models presented in this paper use a barrier of 1200~K, which is assumed from the reaction of \ce{H} and \ce{HNC} \citep{graninger14}. This barrier was varied in several model tests to see what effect it would have on the chemistry of the isocyanides, and \ce{CH3NC} in particular. Recent theoretical work by \citet{Nguyen19} shows that \ce{CH3NC} does react with \ce{H} on surfaces with barriers similar to the standard value used here of 1200~K, though it is important to note that without experimental measurements, this value is still uncertain.

Binding energies for \ce{CH3CN} and \ce{CH3NC} on amorphous water ice are used, and are taken from \citet{bertin17}. The values are 6150~K for \ce{CH3CN} and 5686~K for \ce{CH3NC}.

\subsubsection{\ce{C2H5NC}}
Ethyl isocyanide has, to our knowledge, not been incorporated into any astrochemical networks until now. As such, the chemical network for this species had to be constructed from the ground up. In most cases, reactions were implemented based on analogous processes for \ce{C2H5CN}. 

In the models presented here, \ce{C2H5NC} is formed primarily through hydrogenation reactions on the surfaces and in the ice mantles of dust particles, specifically via the following:
\begin{equation}
    \ce{H + CH2CH2NC -> C2H5NC}
    \label{eq:4}
\end{equation}
\begin{equation}
    \ce{H + CH3CHNC -> C2H5NC}
    \label{eq:5}
\end{equation}
These large isocyanide radicals also had to be added to our chemical network, as they were not involved in any reactions before the introduction of ethyl isocyanide. \ce{CH2CH2NC} is formed through the following two reactions on grain surfaces and in grain ice mantles:
\begin{equation}
    \ce{CH2 + CH2NC -> CH2CH2NC}
    \label{eq:6}
\end{equation}
\begin{equation}
    \ce{H + C2H3NC -> CH2CH2NC}
    \label{eq:7}
\end{equation}
The first reaction (Reaction \ref{eq:6}) is the primary means by which the \ce{CH2CH2NC} radical is formed, and has no barrier. The second reaction is a secondary channel by which \ce{CH2CH2NC} can be formed, though it has an activation energy barrier of 1320~K, which we assume based on the analogous reaction with \ce{C2H3CN}. \ce{CH3CHNC} is formed solely through hydrogen addition to vinyl isocyanide:
\begin{equation}
    \ce{H + C2H3NC -> CH3CHNC}.
    \label{eq:8}
\end{equation}
This reaction has a barrier of 619~K, also taken from the analogous cyanide process. 

Another minor formation path for \ce{C2H5NC} that occurs on grains is the following reaction:
\begin{equation}
    \ce{CH3 + CH2NC -> C2H5NC}.
    \label{eq:9}
\end{equation}
Reaction \ref{eq:9} is significantly less efficient than Reactions \ref{eq:4} and \ref{eq:5} because of the need for mobility of the heavier \ce{CH3} radical. We note that \ce{CH2NC} is formed primarily in the gas phase from reactions of \ce{CH3CN+} with electrons and CO molecules, as well as by recombination of \ce{CH3NCH+}. 

Standard destruction mechanisms for \ce{C2H5NC} were also included. These include photo-dissociation and cosmic-ray-induced (photo-)dissociation. However, the most efficient destruction mechanism for \ce{C2H5NC} is ion-molecule reactions with abundant gas-phase ions.

Grain-surface binding energies for \ce{C2H5NC} and related radicals were chosen to mimic those of the corresponding cyanides, considering the lack of experimental data on these species. These values can be found in Table \ref{t:physical_quantities}. We note that there is no reaction of \ce{C2H5NC} with \ce{H}, which would be analogous to Reaction \ref{eq:3}. Since there has been, to the authors' knowledge, no experimental or theoretical work done on this reaction, it was decided not to extrapolate the barrier used in Reaction \ref{eq:3} to the larger  \ce{C2H5NC}.

\subsubsection{\ce{C2H3NC}}
Vinyl isocyanide was incorporated into our chemical network. This is the first time this species has been included in an astrochemical network, to the authors' knowledge. A similar strategy was followed with vinyl isocyanide as is explained for ethyl isocyanide, in that reactions were selected as analogs to the cyanide isomer, vinyl cyanide. 

There are important formation reactions for \ce{C2H3NC} in both the gas phase and on grains in our network. On grains, it is formed through hydrogenation of the \ce{C2H2NC} radical, through the following reaction:
\begin{equation}
    \ce{H + C2H2NC -> C2H3NC}.
    \label{eq:10}
\end{equation}
\ce{C2H2NC} is formed on grains from hydrogenation of \ce{HCCNC}. The formation of \ce{HCCNC}  at low temperatures is dominated by the dissociative recombination of \ce{C3H2N+} (via a rearrangement of the heavy atoms of the carbon backbone -CCCN to -CCNC), while at higher temperatures (greater than $\sim$26K in the models presented below), production is dominated by the reaction:
\begin{equation}
    \ce{H + HCNCC -> HCCNC + H}.
    \label{eq:11}
\end{equation}
\ce{HCNCC} is primarily formed from dissociative recombination of \ce{C3H2N+}. We note that \ce{C3H2N+} is formed from the reaction of \ce{C3NH+} with \ce{H2} in the gas phase. \ce{C3NH+} is formed from the reaction of \ce{H2} with \ce{C3N+}. These reactions were already present in previous networks.

In the gas phase, \ce{C2H3NC} is formed in one of the primary recombination channels of protonated ethyl isocyanide (\ce{C2H6NC+}):
\begin{equation}
    \ce{C2H6NC+ + e- -> C2H3NC + H2 + H}.
    \label{eq:12}
\end{equation}
Thus the chemistry of ethyl and vinyl isocyanide is linked. Approximately 40\% of recombinations are assumed to go through this channel, which is consistent with the experiments of \citet{vigren12}.

Similar to methyl and ethyl isocyanide discussed previously, standard destruction mechanisms for vinyl isocyanide are included as well. These include photo-dissociation, cosmic ray-photon induced dissociation, and ion-molecule reactions. Binding energies are chosen to be equivalent to vinyl cyanide, in the absence of experimental data, and are shown in Table \ref{t:physical_quantities}. 

\subsubsection{\ce{HC3N}, \ce{HCCNC}}
Although both cyanoacetylene and isocyanoacetylene have been studied in detail previously \citep[e.g.,][]{hebrard09,woon09}, a review of their basic chemistry is presented here. \ce{HC3N} has a few primary gas-phase formation pathways. At lower temperatures, it is formed primarily through the following reaction:
\begin{equation}
    \ce{C + CH2CN -> HC3N + H},
    \label{eq:13}
\end{equation}
while at higher temperatures ($\gtrsim$27K in the models presented below) the following reaction takes over:
\begin{equation}
    \ce{N + C3H3 -> HC3N + H2}.
    \label{eq:14}
\end{equation}
\ce{C2H} + \ce{HCN} is also a viable formation path, though it is secondarily important. There are minor grain-surface formation pathways as well, namely hydrogenation of \ce{C3N} and the reaction of atomic N with \ce{C3H3}. Moreover, \ce{HC3N} is destroyed on the grains by further hydrogenation to \ce{C2H2CN}, which has an activation energy barrier of 1710~K, and in the gas via standard ion-molecule destruction routes. 

The chemistry of \ce{HCCNC} is less complex. It has two primary gas-phase formation pathways. It is formed from Reaction \ref{eq:11} and the following reaction:
\begin{equation}
    \ce{C3H2N+ + e- -> HCCNC + H}.
    \label{eq:15}
\end{equation}
Standard ion-molecule destruction routes are also included. Grain-surface chemistry for \ce{HCCNC} is minimal, as there are no known formation routes, and, similarly to \ce{HC3N}, it is destroyed by hydrogenation to \ce{C2H2NC}, with the same barrier. 

\subsection{Physical model}
For the chemical modeling in this paper, we introduce a new way of incorporating the physical profile of astronomical sources. Traditional astrochemical models of hot cores consist of two stages of physical evolution; these are described in detail by \citet{garrod17}. First, the molecular cloud undergoes a cold collapse to some specified gas density. This is followed by a static warm-up to a dust and gas temperature of 400 K. This warm-up can be tuned to occur at different rates depending on the source, with the warm-up rate believed to be roughly correlated with the mass of the source \citep{garrod06}. 

However, there are some issues with a modeling approach like this, as discussed by \citet{coutens18}, who applied it to a low-mass source. In two-stage chemical models, the maximum gas density in the simulation is reached before any warm-up has occurred. This provides an inaccurate physical picture of star formation, particularly in cases where extreme densities are reached in the central core ($\sim$10$^{9}$ cm$^{-3}$, based on the results from \citealt{coutens18}). A more accurate physical depiction of these sources includes a density gradient coupled with a temperature gradient, whereby colder regions at the outer edge of the cloud are less dense than the warm regions in the interior of the source. Treating the physics more accurately will lead to a more accurate chemical treatment of the source.

In this paper, we have incorporated a new method of physical modeling into {\em MAGICKAL}. Instead of following the canonical two-stage approach to modeling hot cores, we have introduced a single-stage modeling approach. This method can be thought of as following an infalling parcel of gas through a physical profile, which dictates the physical conditions of the chemical model.

First, a spatial density and temperature profile is obtained from observations of a source, in this case Sgr B2(N2). The profiles have the form shown in equations 16 and 17, where $n_{0}$ and $T_{0}$ are the density and temperature at radius $r_{0}$, and $\alpha$ and $q$ correspond to indices for the power law for density and temperature, respectively. The physical profile culminates in a  final density, $n_{f}$. Here, $T_0$, $n_0$, and $r_0$ are determined from observations of Sgr B2(N2) taken as part of the EMoCA survey \citep{bonfand17,bonfand19}, while $\alpha$ and $q$ are assumptions for hot-core sources ($\alpha$: \citealt{shu77}, $q$: \citealt{terebey93}). To determine $T_0$, we assume the rotational temperature measured for COMs at $r_0$ to be equal to the dust temperature at that radius and this dust temperature to result from the radiative heating by the protostar. 
\begin{equation}
    n = n_{0} (\frac{r_{0}}{r})^{\alpha}
    \label{eq:density}
,\end{equation}

\begin{equation}
    T = T_{0} (\frac{r_{0}}{r})^{q}
    \label{eq:temp}
.\end{equation}
For the chemical modeling, we have made some changes to these profiles. We have chosen to give the density profile a minimum gas-phase density of $10^{4}$ cm$^{-3}$. This is to represent the background density of the Sgr B2(N2) region, since it is unlikely that lower densities would be reached. Moreover, implementing this density floor gives us a good fit (within a factor of two) to the observational \ce{H2} column density. The density profile is otherwise unchanged from the assumed power-law shape as discussed earlier. In other words, Equation 16 is assumed until a radius at which the density falls to $10^{4}$ cm$^{-3}$, at which point the density is assumed to remain constant at this value at larger radii. The visual extinction is then re-computed from this new density profile with the following standard relation \citep{bohlin78}:
\begin{equation}
    A_v = \frac{3.1}{5.8 \times 10^{21}} N_H
    \label{eq:av}
.\end{equation}
This density minimum leads to a higher extinction at lower temperatures than in the unaltered profile. We note that in the chemical model, the dust temperature in the outer envelope of the source is calculated from the visual extinction, as in \citet[; Equation 17]{garrod11}, until that temperature and the temperature given by the observational profile cross. At this point, the temperature is determined by Equation \ref{eq:temp}. The collapse is stopped once a temperature of 400~K is reached, as that is the highest temperature at which our chemical network is reliable. Relevant parameters for the chemical modeling are shown in Table \ref{t:phys_params}.

\begin{table}
 \begin{center}
 \caption{Physical parameters used in chemical model.}
 \label{t:phys_params}
 \begin{tabular}{ll}
 \hline\hline\\[-1.0em]
 \multicolumn{1}{l}{Parameter} & \multicolumn{1}{l}{Value}
 \\ \hline\\[-1.0em]
    $r_{0}$ & 6010 au \\ 
    $n_{0}$ & 1.54 $\times$ 10$^{7}$ cm$^{-3}$ \\
    $T_{0}$ & 135.5 K \\
    $\alpha$ & 1.5 \\
    $q$ & 0.38 \\
    $n_{f}$ & 2.74 $\times$ 10$^{9}$ cm$^{-3}$ \\ \hline
 \end{tabular}
 \end{center}
 \tablefoot{$r_0$: Reference radius used in Eqs. 16-17. Here, $n_0$ is the density of total hydrogen at $r_0$; $T_0$ is the dust temperature at $r_0$; $\alpha$ is the exponent in Equation 16; $q$ is the exponent in Equation 17; and $n_f$ is the final density of the total hydrogen for the models. }
 \end{table}

These physical parameters are then used to compute a full physical profile of the source using a simple free-fall collapse model. This profile is assumed to remain static through time, while a parcel of gas freefalls inward, experiencing physical conditions dictated by the profile. This approach has the advantage of monitoring the observed profiles throughout in the absence of any historical information about the profiles, while still allowing a more accurate progression in $T$ and n$_H$ for the chemical modeling. A cubic spline interpolation is used to calculate the physical conditions between time points.  

\subsection{Cosmic-ray ionization}
There is mounting evidence that the cosmic-ray ionization rate ($\zeta$) in diffuse clouds may be significantly higher than the canonical value of 1.3 x 10$^{-17}$ s$^{-1}$ typically assumed in astrochemical models \citep{indriolo07,gerin10}. Diffuse clouds are not thought to be the only sources to experience this enhancement, however. In particular, the central molecular zone (CMZ) displays a very high abundance of H$_{3}^{+}$, which has been theorized to be caused by a very high $\zeta$ \citep{lepetit16}. These authors determined a $\zeta$ on the order of 10$^{-14}$ s$^{-1}$ for the diffuse medium of the CMZ, which is where Sgr B2(N2) is located. \citet{indriolo15} also determined a very high $\zeta$ towards the diffuse medium in the Galactic Centre using {\it Herschel} observations to derive values $>10^{-15}$ s$^{-1}$. Therefore, it makes sense to investigate models in which $\zeta$ is higher than the canonical value. 

In addition to this higher $\zeta$, it is not physically accurate to assume that $\zeta$ will be constant throughout the source. In fact, $\zeta$ will vary with the column density in the region \citep{rimmer12}. To this end, we ran models using our new single-stage physical profile where $\zeta$ varies throughout the source, using the following equation:
\begin{equation}
    \zeta = 3.9 \times 10^{-16}(A_{v})^{-0.6} + 10^{-17} s^{-1}
    \label{eq:zeta}
.\end{equation}
This is based on Equation 10 from \citet{rimmer12} with the $N_H$-to-$A_v$ conversion from \citet[; our Equation 18]{bohlin78}. We note that Equation \ref{eq:zeta} diverges from plausible values at low $A_v$. For the modeling in this paper, the lowest $A_v$ is 2 mag, and as such this divergence does not present an issue.

\section{Modeling results} \label{sec:results}
\subsection{Standard model and \ce{H + CH3NC} barrier}
We ran several different chemical models in this study in order to investigate the effect of different chemical and physical parameters. We named them Model 1 through Model 7. Table \ref{t:model_legend} shows a legend for easy reference. 
\begin{table}
 \begin{center}
 \caption{Legend for chemical modeling presented in this study.}
 \label{t:model_legend}
 \begin{tabular}{ll}
 \hline\hline\\[-1.0em]
 \multicolumn{1}{l}{Model} & \multicolumn{1}{l}{Description}
 \\ \hline\\[-1.0em]
    Model 1 & Standard single-stage model; \\ & constant $\zeta=\num{1.3e-17}$ $s^{-1}$ \\ & 1200~K barrier for \ce{H + CH3NC} \\ 
    Model 2 & Model 1, but with 3000~K barrier for \ce{H + CH3NC}  \\
    Model 3 & Model 1, with $A_V$-dependent $\zeta$ (low) \\
    Model 4 & Model 1, with $A_V$-dependent $\zeta$ (medium) \\
    Model 5 & Model 1, with $A_V$-dependent $\zeta$ (high)  \\ 
    Model 6 & Model 1, with $\zeta=\num{1e-16}$ s$^{-1}$ \\
    Model 7 & Model 1, with $\zeta=\num{3e-14}$ s$^{-1}$ \\\hline
 \end{tabular}
 \end{center}
 \end{table}
 
The results for Model 1 are discussed first. This model includes the collapse and warm-up phases combined into a single stage. The standard $\zeta$ is used (\num{1.3e-17} s$^{-1}$) and is held constant throughout the model run. The chemical network has been updated as discussed in Sect. 5. This model will serve as a standard point of comparison throughout this study. 

\begin{figure*}[htb]
    \centering
    \begin{tabular}{cc}
         \includegraphics[width=.46\textwidth]{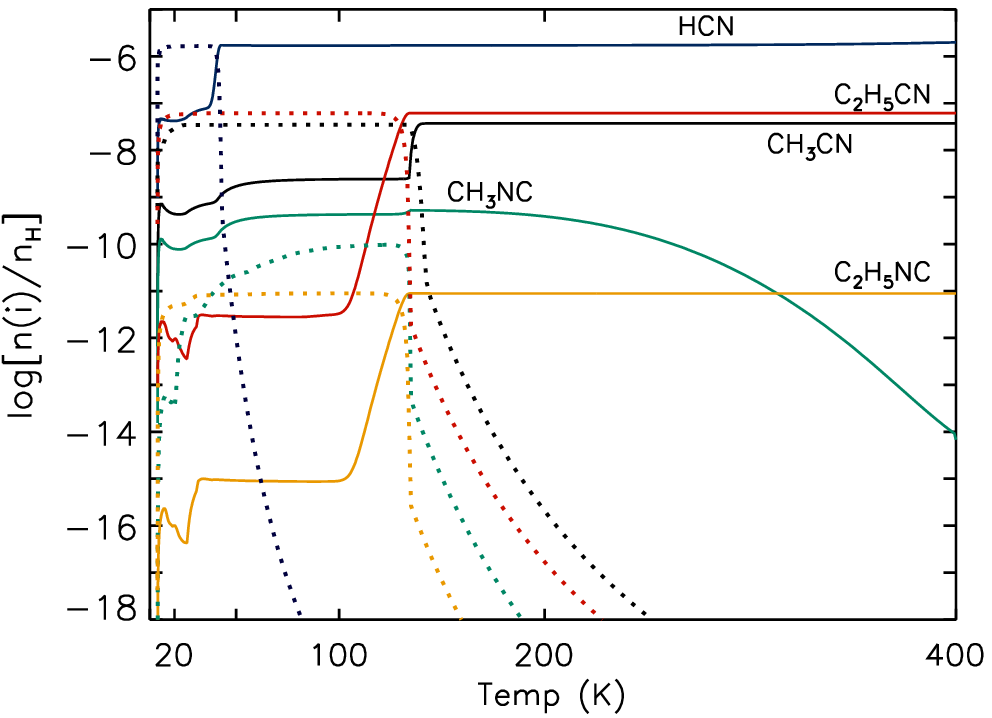} &
         \includegraphics[width=.46\textwidth]{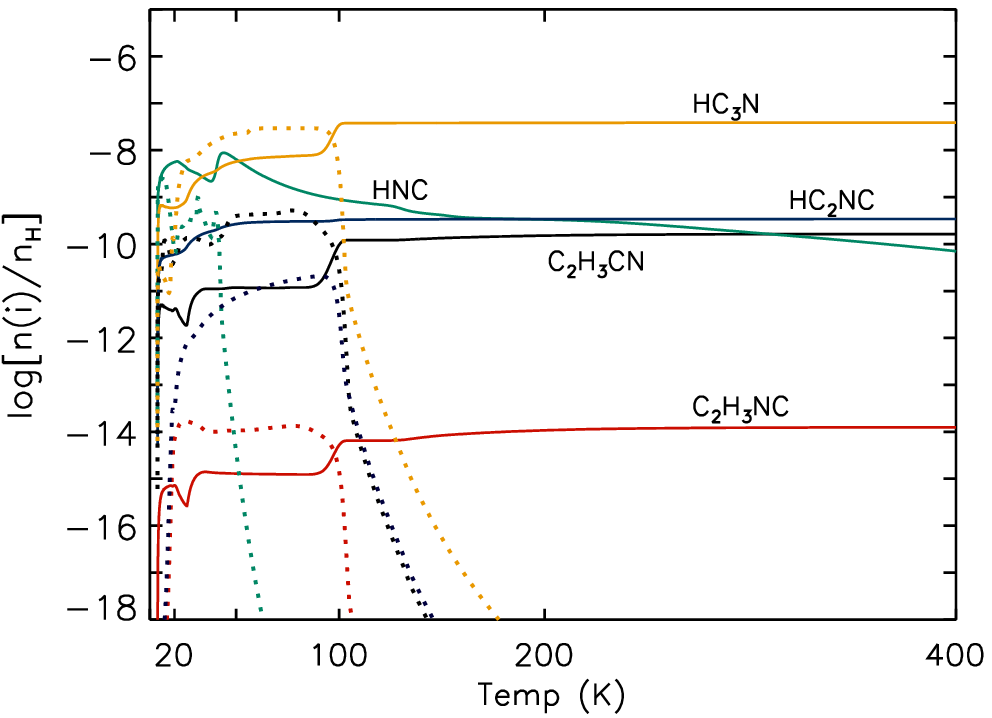} \\
    \end{tabular}
    \caption{Abundances of cyanides and isocyanides in the standard model presented in this paper (Model 1). Dashed lines correspond to grain abundances, while solid lines correspond to gas-phase abundances. Left panel: \ce{HCN, CH3CN, CH3NC, C2H5CN, C2H5NC}. Right panel: \ce{HNC, C2H3CN, C2H3NC, HC3N, HC2NC}.}
    \label{fig:mod1}
\end{figure*}

We begin by discussing the final chemical abundances in the model, as they provide a good point of reference when comparing between models, and the fractional abundance of the molecules we are studying  usually do not change much once they desorb from the grains. It can be seen from Figure \ref{fig:mod1} that \ce{HCN} is the most abundant of the molecules we are interested in, with a final fractional abundance of $\sim$\num{2e-6} with respect to total hydrogen. HCN is followed by \ce{HC3N}, \ce{C2H5CN}, and \ce{CH3CN} with fractional abundances $>$ 10$^{-8}$. We find that \ce{HCCNC}, \ce{HNC}, and \ce{C2H3CN} comprise the next most abundant molecules, with final abundances on the order of 10$^{-10}$. \ce{C2H5NC} follows with a fractional abundance of $\sim$10$^{-11}$, while \ce{C2H3NC} and \ce{CH3NC} finish with very low fractional abundances, at $\sim$10$^{-14}$. 

\ce{CH3NC} is a particularly interesting molecule, as although it finishes with a very low fractional abundance, it has a much higher {peak} abundance value of $\sim$\num{5e-10}. This dramatic decrease after desorption from the grain surfaces is a result of destruction in the gas phase by reaction with atomic hydrogen, via Reaction \ref{eq:3}. Although the vast majority of hydrogen in the model is contained in \ce{H2}, there is still a very high gas-phase abundance of \ce{H} to react with.

\begin{figure}[htb]
    \centering
    \includegraphics[width=.46\textwidth]{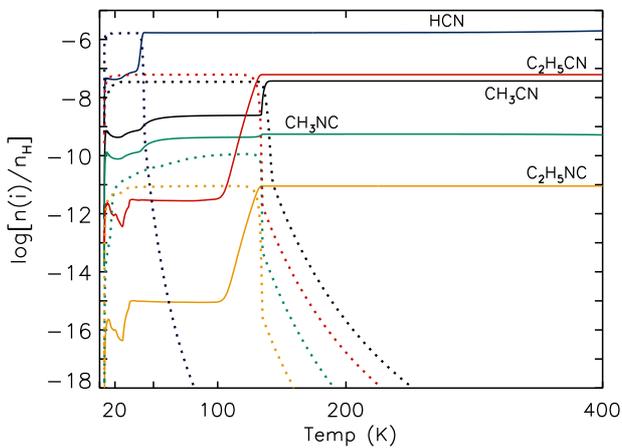}
    \caption{Abundances of cyanides and isocyanides in the model with a barrier of 3000~K for the reaction of \ce{H + CH3NC} (Model 2).}
    \label{fig:mod2}
\end{figure}

As discussed earlier, there is some uncertainty in the activation energy barrier of Reaction \ref{eq:3}. The standard version of our chemical network uses an activation energy barrier of 1200~K for this reaction based on the reaction of H + HNC discussed by \citet{graninger14}. We note that this reaction is not as important for \ce{HNC} as it is for \ce{CH3NC}, as there are other, barrierless paths for the destruction of that molecule, such as reaction with \ce{C}.
We ran Model 2 in order to test the impact of this reaction, whereby the barrier for this process was varied. A value of 3000~K was chosen, and the results of this are shown in Figure \ref{fig:mod2}. 
The only molecule that is affected by this change is \ce{CH3NC}. Instead of the abundance falling off steadily after desorbing into the gas-phase, it stays at its peak value of $\sim$\num{5e-10}, which is due to the fact that the 3000~K barrier is not overcome at an appreciable rate. Trials were run varying the barrier to 5000 and 10000~K as well, but no difference was noted, as the barrier becomes totally insurmountable without considering tunneling. 

\begin{figure}[!htb]
    \centering
    \includegraphics[width=\linewidth]{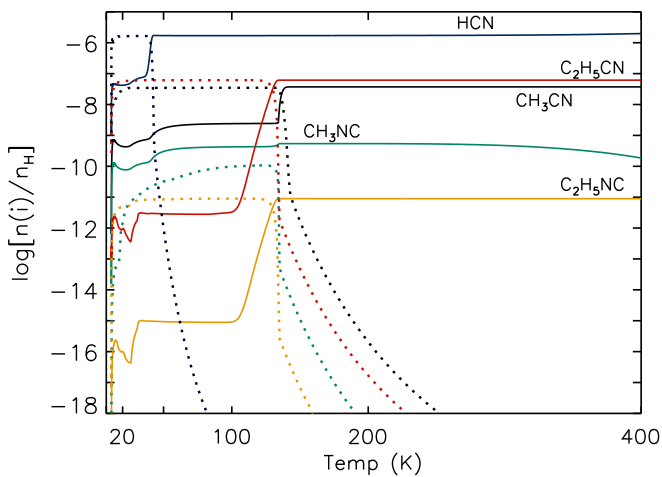}
    \caption{Abundances of cyanides and isocyanides in the model with a barrier of 2000~K for the reaction of \ce{H + CH3NC}.}
    \label{fig:2000K_barrier}
\end{figure}

A final test model with a barrier of 2000~K was chosen to study the effect of an intermediate barrier value between Models 1 and 2. The results for a select group of molecules from this model are shown in Figure \ref{fig:2000K_barrier}, and it can be seen that the abundance of \ce{CH3NC} begins to fall off at the end of this model, when the temperature becomes sufficient to overcome the kinetic barrier. For the remaining models in this paper, the value of 1200~K is used, because that is closest to theoretical estimates.

\subsection{Comparison to old physical model}
\begin{figure*}[htb]
    \centering
    \begin{tabular}{cc}
         \includegraphics[width=.46\textwidth]{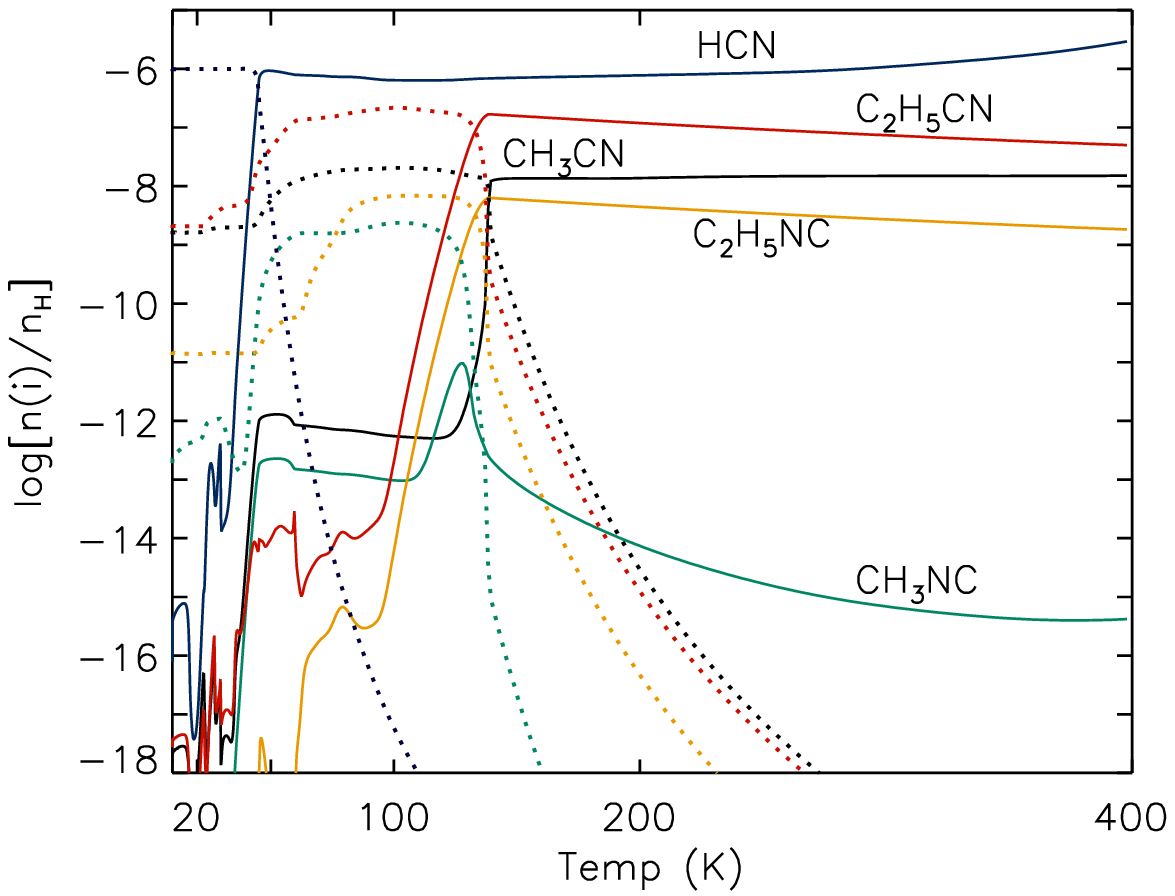} &
         \includegraphics[width=.46\textwidth]{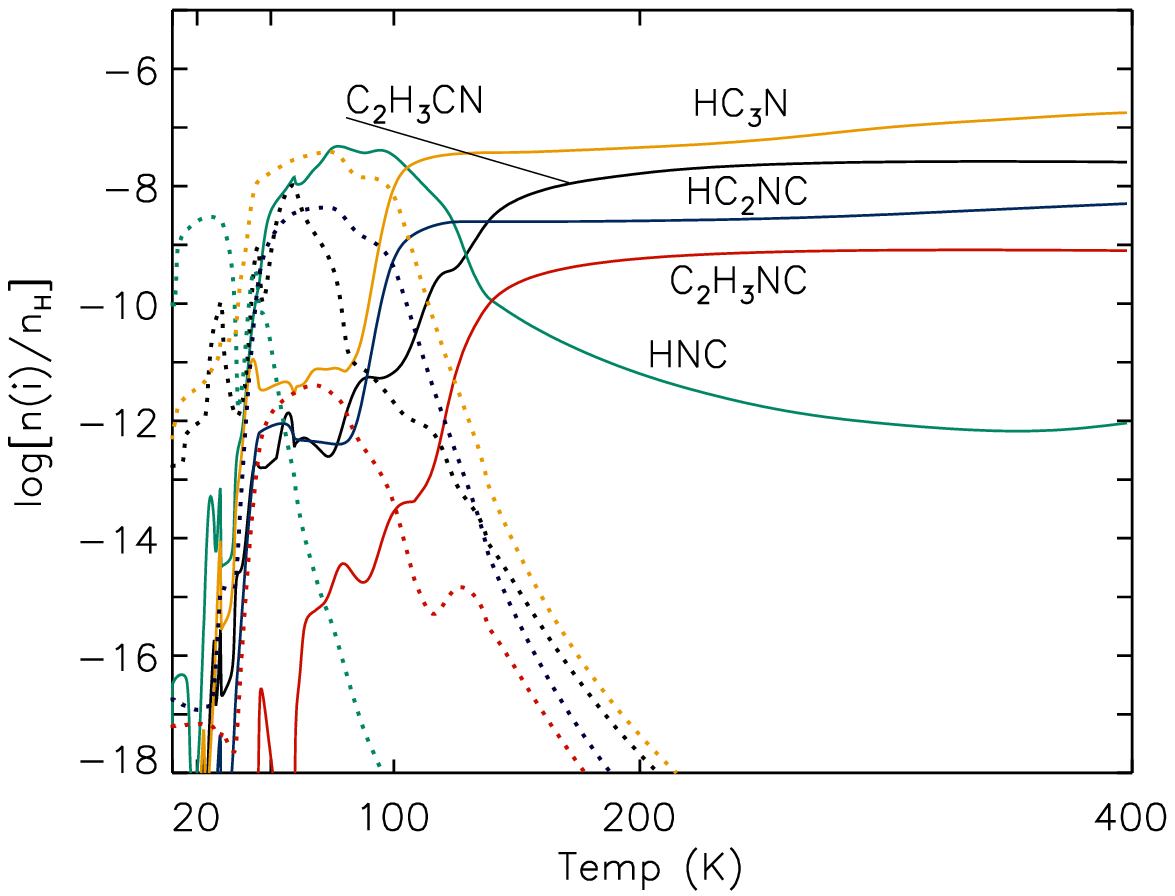} \\
    \end{tabular}
    \caption{Abundances of cyanides and isocyanides in the standard two-phase hot-core chemical model. Dashed lines correspond to grain abundances, while solid lines correspond to gas-phase abundances. Left panel: \ce{HCN, CH3CN, CH3NC, C2H5CN, C2H5NC}. Right panel: \ce{HNC, C2H3CN, C2H3NC, HC3N, HC2NC}.}
    \label{fig:old_model}
\end{figure*}

Since the focus of this paper is chemical modeling using the single-stage physical model of Sgr~B2(N2), it is instructive to compare the results of this model to that of a standard two-phase hot-core model. Figure \ref{fig:old_model} shows the fractional abundances of the same cyanide and isocyanide species as displayed in Figure \ref{fig:mod1}. This model uses the same initial and final density as the one-stage model used throughout the paper. We assume an intermediate warm-up timescale for this model, as discussed in \citet{garrod13}. The warm-up phase of this model reaches 400~K in \num{2.85e5} years. This is slightly faster than the single-stage model, which reaches 400~K in \num{4.327e5} years. One can see some significant differences compared with the new physical model. In the left panel, the peak and final abundances of \ce{HCN, C2H5CN, and CH3CN} are not affected significantly. However, \ce{C2H5NC} shows an abundance two orders of magnitude lower in the new physical model. Similar effects are observed for the chemical species plotted in the right panel. \ce{C2H3CN} is two orders of magnitude lower in the new physical model as well. Perhaps most pronounced is \ce{C2H3NC}, which has a final abundance of approximately five orders of magnitude higher in the old physical model. In fact, the abundance shown in this model indicates that \ce{C2H3NC} should be detectable in this source, which it does not appear to be.

\begin{figure*}[htb]
    \centering
    \begin{tabular}{cc}
         \includegraphics[width=.46\textwidth]{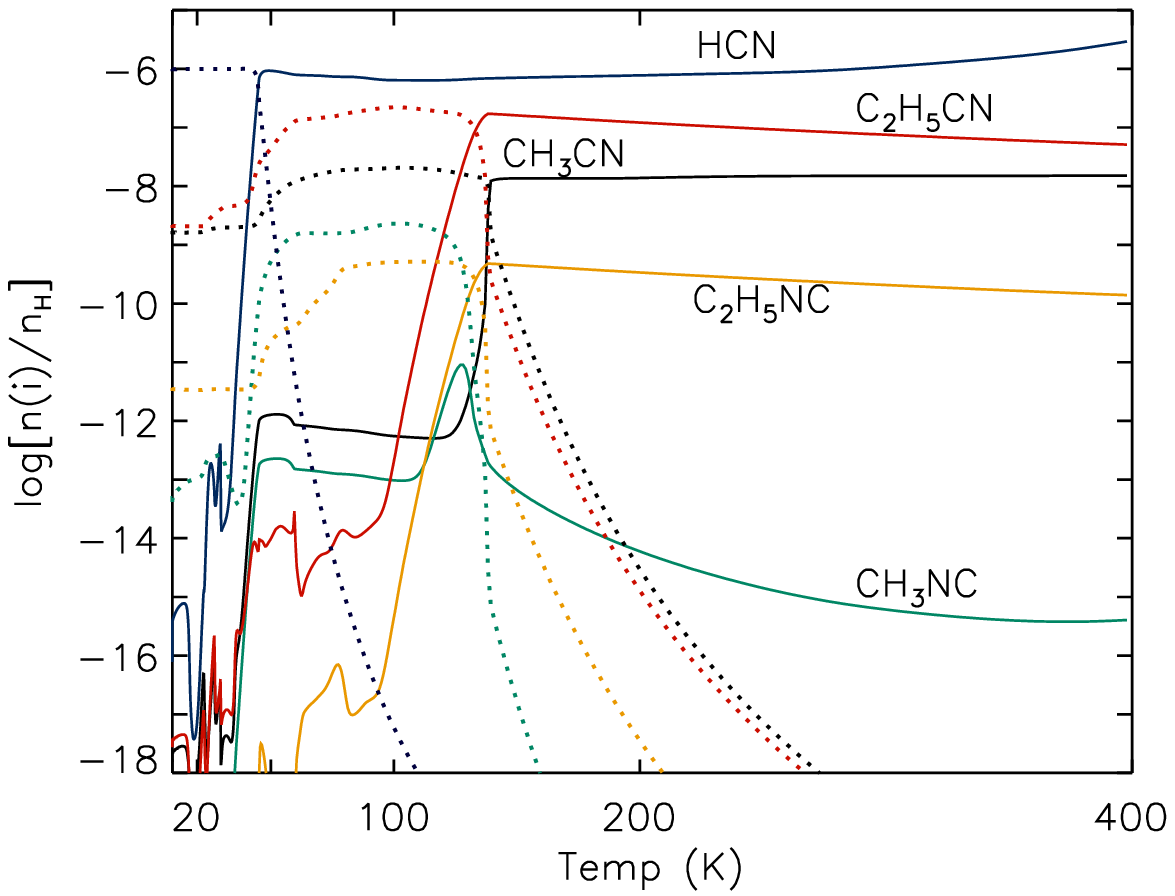} &
         \includegraphics[width=.46\textwidth]{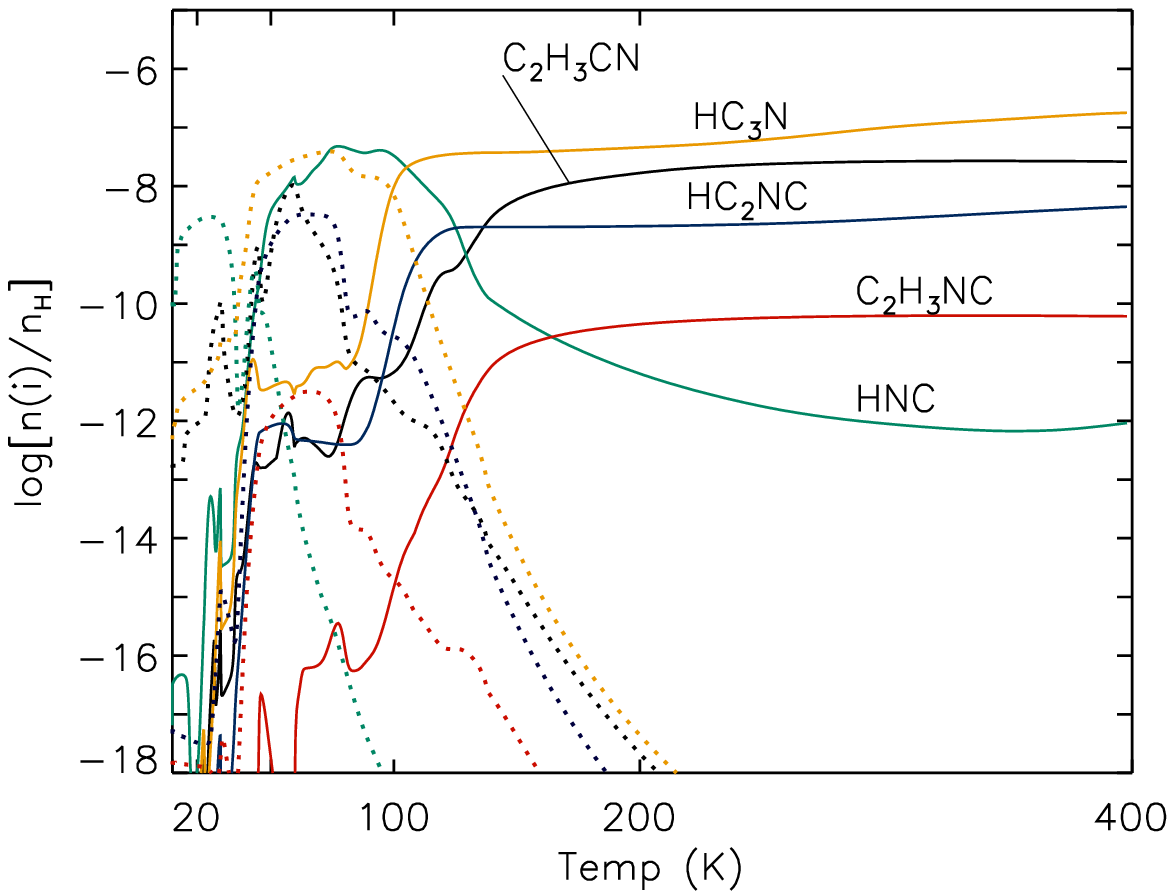} \\
    \end{tabular}
    \caption{Abundances of cyanides and isocyanides in the old physical model with \ce{H + HCCNC -> HCN + C2H} (Eq. \ref{eq:hccnc}) added. Left panel: \ce{HCN, CH3CN, CH3NC, C2H5CN, C2H5NC}. Right panel: \ce{HNC, C2H3CN, C2H3NC, HC3N, HC2NC}.}
    \label{fig:hccnc_1}
\end{figure*}

\begin{figure*}[htb]
    \centering
    \begin{tabular}{cc}
         \includegraphics[width=.46\textwidth]{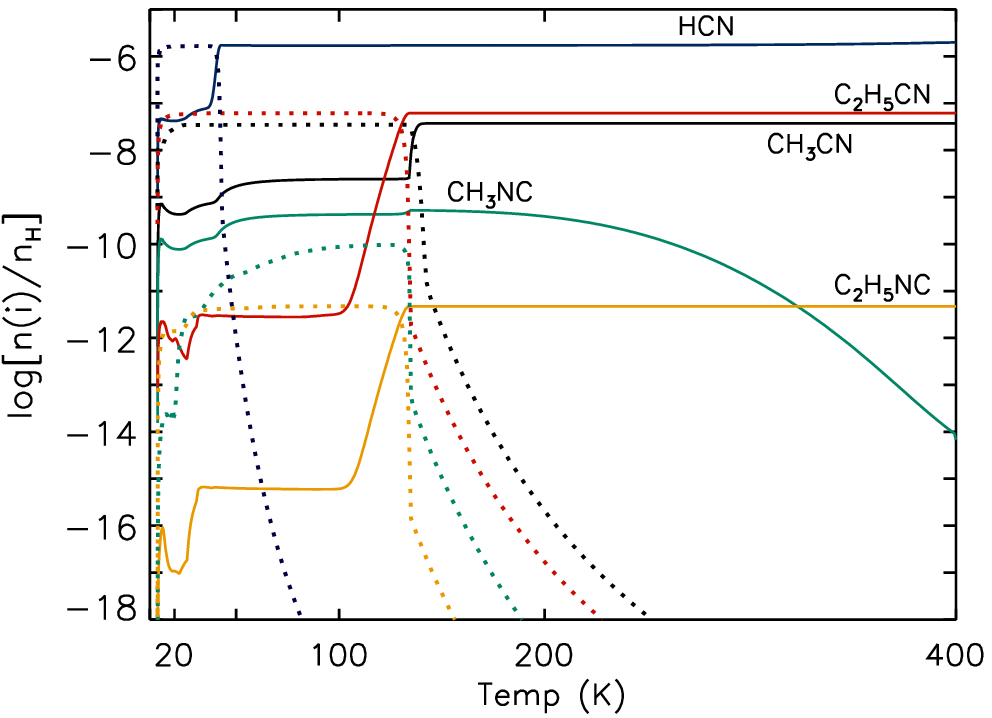} &
         \includegraphics[width=.46\textwidth]{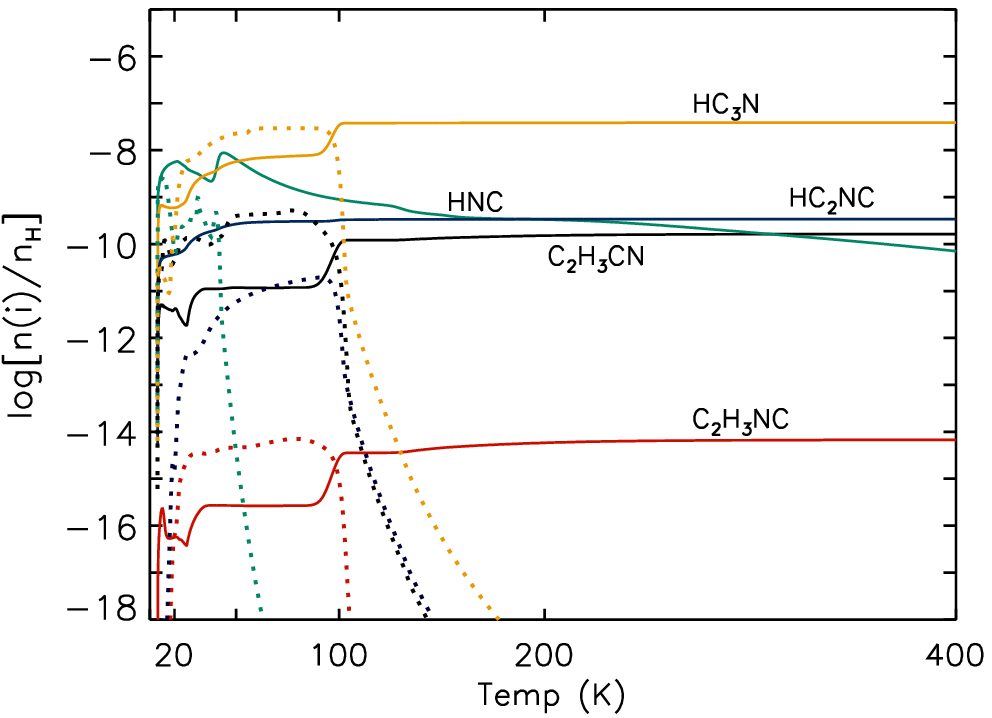} \\
    \end{tabular}
    \caption{Abundances of cyanides and isocyanides in the standard model presented in this paper (Model 1), with \ce{H + HCCNC -> HCN + C2H} (Eq. \ref{eq:hccnc}) added. Left panel: \ce{HCN, CH3CN, CH3NC, C2H5CN, C2H5NC}. Right panel: \ce{HNC, C2H3CN, C2H3NC, HC3N, HC2NC}.}
    \label{fig:hccnc_2}
\end{figure*}

These significant chemical differences are directly related to the physical differences between the two modeling approaches. Although the total warm-up timescales are within a factor of two of each other, the time spent in key temperature ranges is significantly different between the models due to the different dynamics. For example, in the new physical model, the time that it takes the warm-up to traverse from 50~K to 80~K is $\sim$ 500 years, whereas the old physical model takes $\sim$ 29000 years. In this temperature range, \ce{HCCNC}, which is formed in the gas phase, freezes out onto the grains, and is subsequently hydrogenated to \ce{C2H3NC} with an activation energy barrier of 1700~K. \ce{C2H3NC} is then hydrogenated to \ce{C2H5NC} via Reactions \ref{eq:8} and \ref{eq:5}. In the old physical models, this process has significantly more time to take effect before \ce{HCCNC} desorbs, thus leading to increased abundances of \ce{C2H3NC} and \ce{C2H5NC}. To illustrate the effect of these processes, we ran additional models that incorporate an alternate reaction of \ce{HCCNC} with \ce{H}, analogous to Reaction \ref{eq:3}, shown below:

\begin{equation}
    \label{eq:hccnc}
    \ce{H + HCCNC -> HCN + C2H}.
\end{equation}
This reaction is given the same barrier as Reaction \ref{eq:3} (1200~K). In this model, most reactions between H and \ce{HCCNC} will proceed via this path, since it has a lower activation energy barrier than simple hydrogenation. Thus, the two physical models should begin to converge on the abundances of \ce{C2H3NC} and \ce{C2H5NC}. Figures \ref{fig:hccnc_1} and \ref{fig:hccnc_2} illustrate this. The final abundances of these species in the old physical model are lower than in Figure \ref{fig:old_model}, and the abundances in the new physical model are virtually unchanged from Figure \ref{fig:mod1}. The reason the models do not converge perfectly is that there is still a small fraction of \ce{HCCNC} that is getting converted to \ce{C2H3NC} in the old physical models, and there is significantly more time for this process to occur. A very similar process is responsible for the decreased abundance of \ce{C2H3CN} in the new physical model, as this molecule is predominantly formed through hydrogenation of \ce{HC3N} which is subject to the same timescale effects as hydrogenation of \ce{HCCNC}.

Other differences are noted as well. In general, the cold-phase chemistry of the cyanides and isocyanides also appears to be much more efficient in the new physical model. In Figure \ref{fig:mod1} the ice-phase abundances of all of the molecules reach a peak at very low temperature ($\sim$20~K), whereas in the old models (Figure \ref{fig:old_model}), the peak values for these same molecules are not reached until much later ($\sim$50~K). This seems to be a result of the amount of time spent at low temperatures in the models. The new physical model takes significantly longer to go from 15~K to 20~K ($\sim$\num{6.3e4} years), whereas the old physical model spends only $\sim$ \num{1.2e4} years in this temperature range. This longer timescale for collapse leads to peak ice abundances being reached at lower temperatures, which has complex effects on the chemistry. The new physical model used throughout this paper is more directly relevant to the source structure of Sgr~B2(N2), and a more accurate representation of hot-core dynamics in general.

\subsection{Cosmic-ray ionization rate}

\begin{figure*}[htb]
    \centering
    \begin{tabular}{cc}
    \includegraphics[width=.46\textwidth]{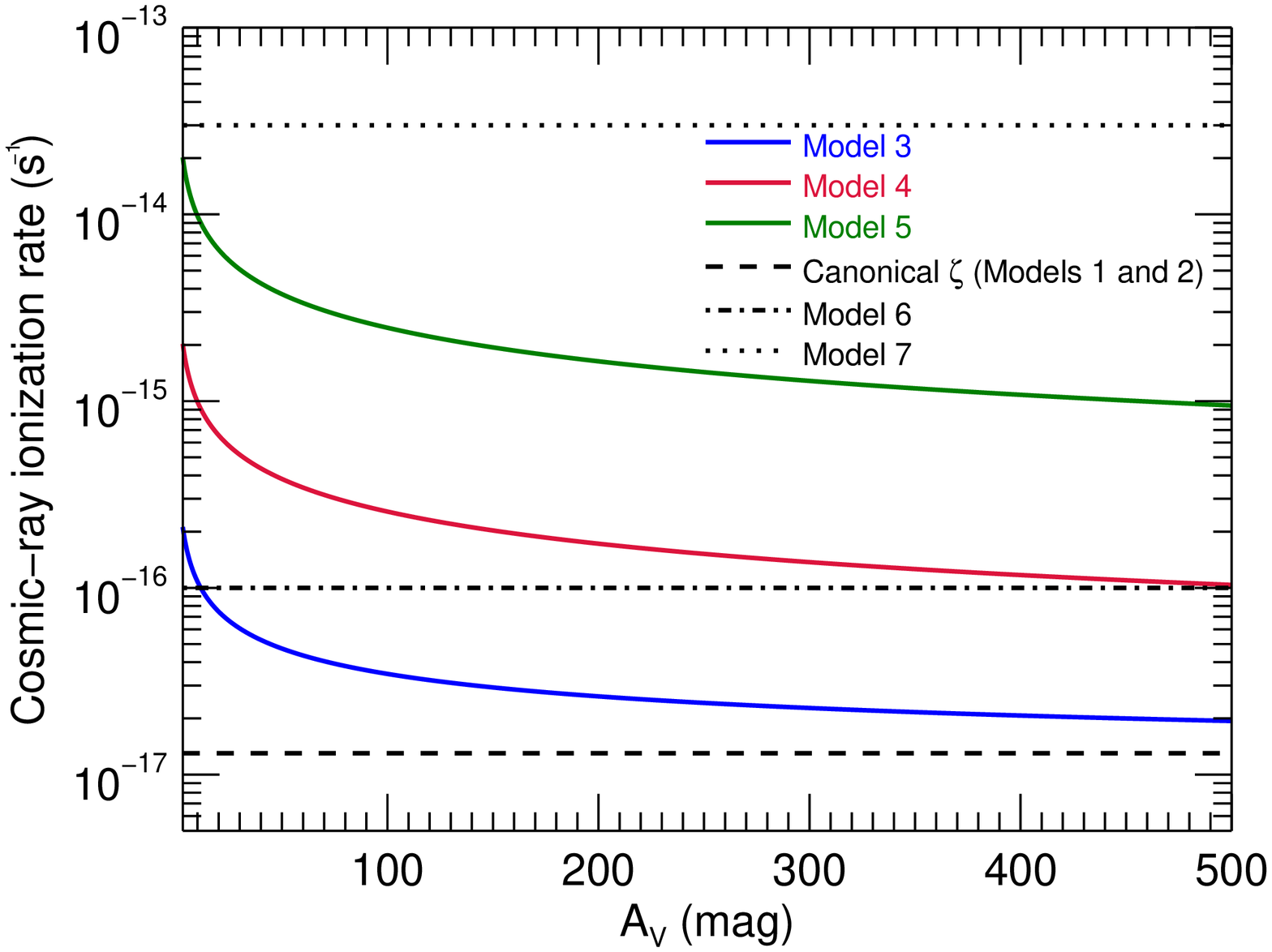} & \includegraphics[width=.46\textwidth]{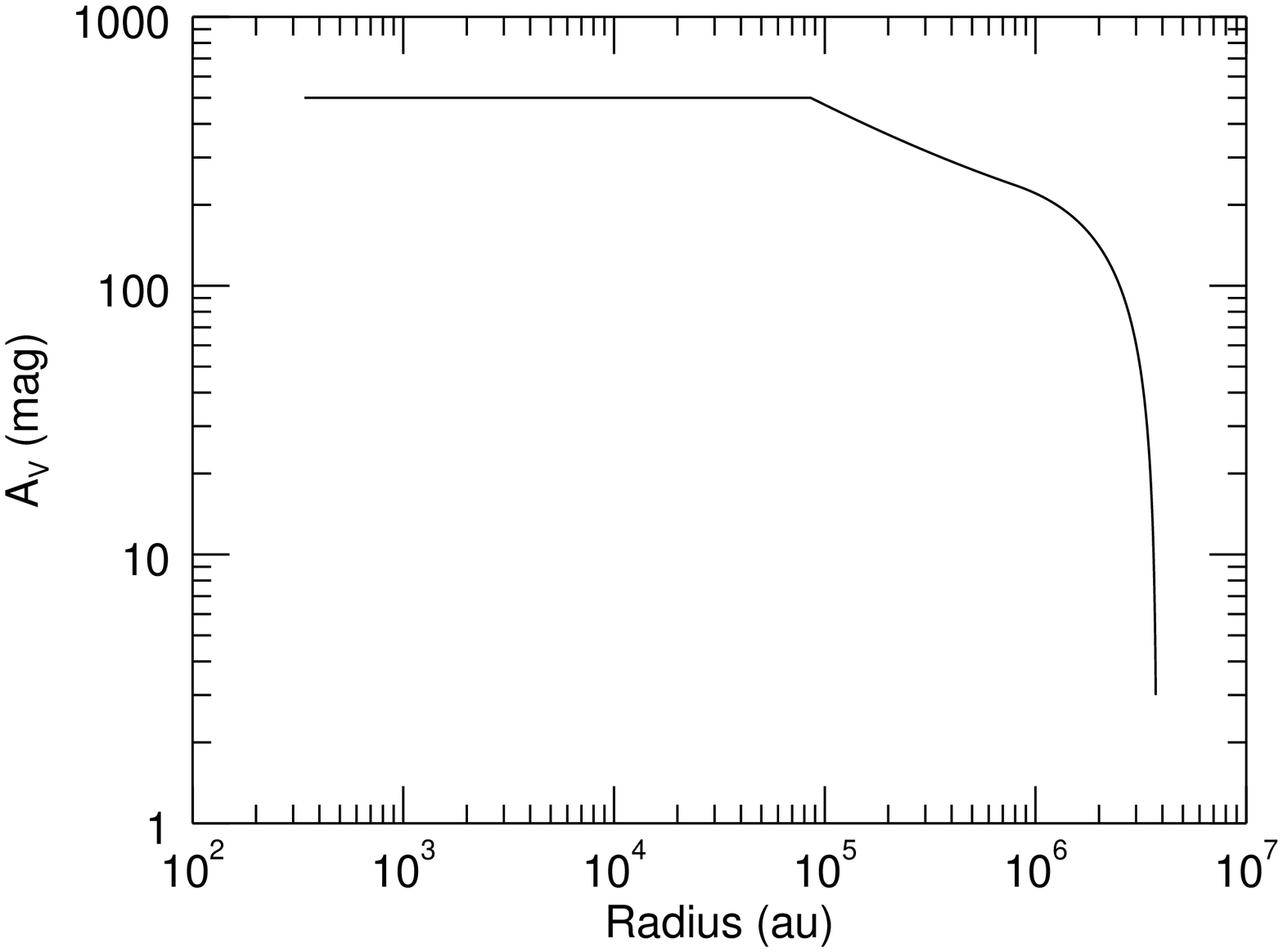} \\
    \end{tabular}
    \caption{$\zeta$ profiles for each model as a function of $A_V$. The dashed line shows the reference value that is typically assumed in hot-core models (\num{1.3e-17} $s^{-1}$). Recent observational constraints have placed $\zeta$ for the diffuse medium around Sgr B2 at 10$^{-15}$-10$^{-14}$ s$^{-1}$. The panel on the right shows the $A_V$ profile as a function of radius.}
    \label{fig:zeta_av}
\end{figure*}

To test the effect of varying $\zeta$ on the fractional abundances of cyanides and isocyanides, we ran three chemical models using Equation \ref{eq:zeta} to vary $\zeta$ throughout the source. The magnitude of the $A_V$ multiplier was changed in each run. The chosen values of the multiplier were \num{3.9e-16} (Model 3), \num{3.9e-15} (Model 4), and \num{3.9e-14} (Model 5). As a further test of the effect of $\zeta$ on hot-core chemistry, we also ran two additional models (Models 6 and 7). These models exhibit a constant $\zeta$ throughout the source at higher values than the canonical rate (\num{1e-16} s$^{-1}$ for Model 6 and  \num{3e-14} s$^{-1}$ for Model 7). These values were chosen to correspond to a value within the range of variable $\zeta$ values in Models 3-5, and to one value above that range. The values of the cosmic-ray ionization rate as a function of $A_V$ for each model are shown in Figure \ref{fig:zeta_av}. The radial profile of $A_V$ is also shown, for illustrative purposes. 

\begin{figure*}[htb]
    \centering
    \begin{tabular}{cc}
         \includegraphics[width=.46\textwidth]{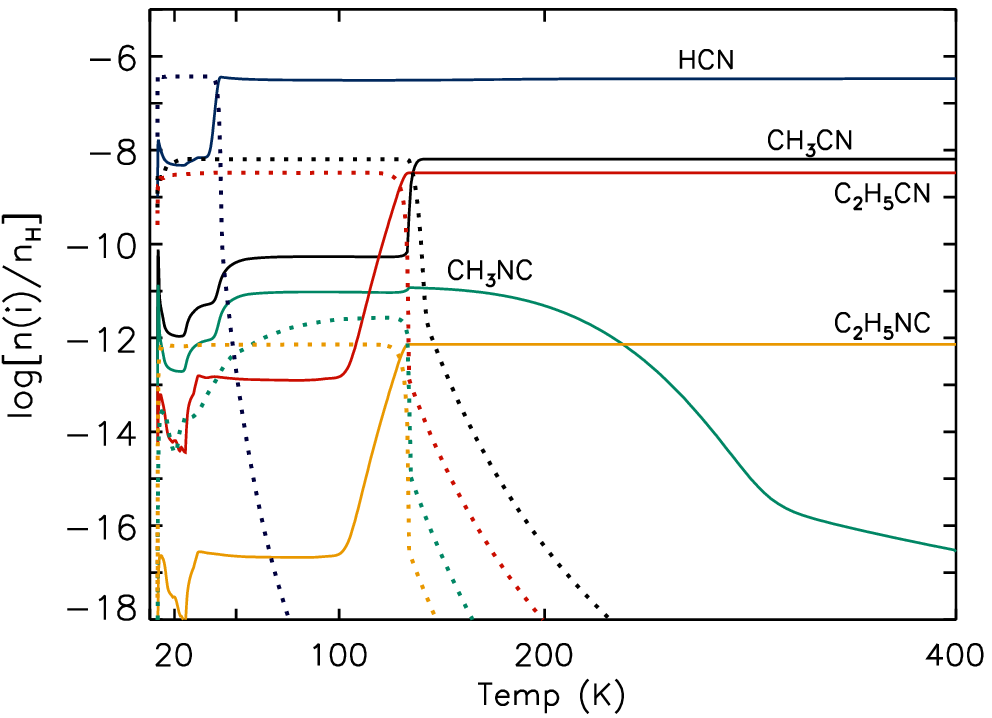} &
         \includegraphics[width=.46\textwidth]{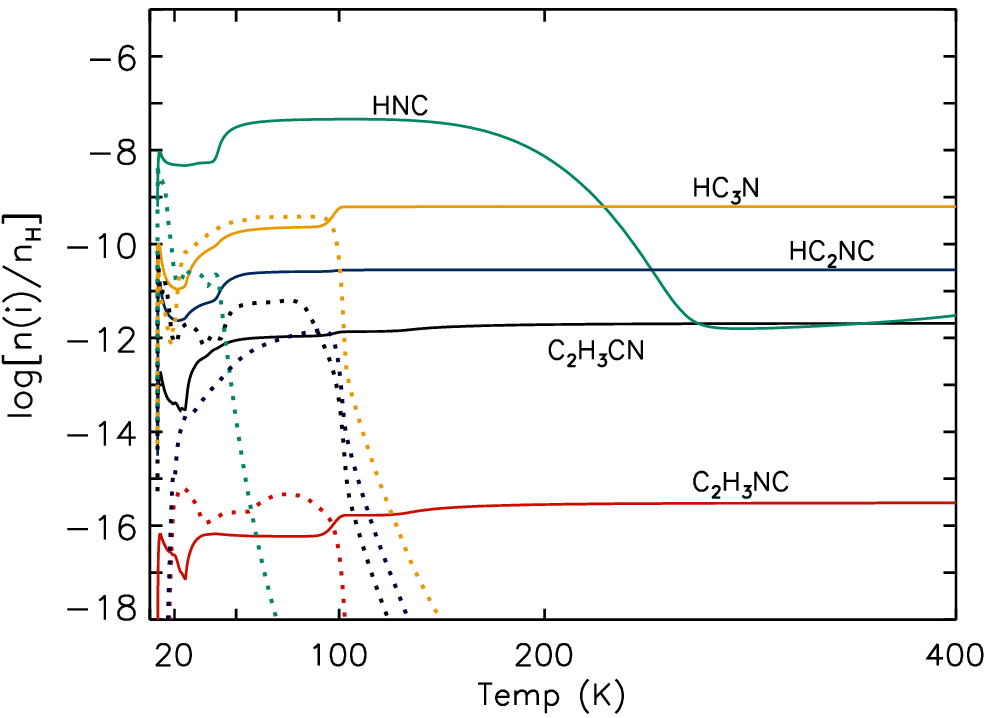} \\
    \end{tabular}
    \caption{Abundances of cyanides and isocyanides in Model 3, which has an $A_V$-dependent $\zeta$ shown in Figure \ref{fig:zeta_av}. Left panel: \ce{HCN, CH3CN, CH3NC, C2H5CN, C2H5NC}. Right panel: \ce{HNC, C2H3CN, C2H3NC, HC3N, HC2NC}.}
    \label{fig:mod3}
\end{figure*}

\begin{figure*}[htb]
    \centering
    \begin{tabular}{cc}
         \includegraphics[width=.46\textwidth]{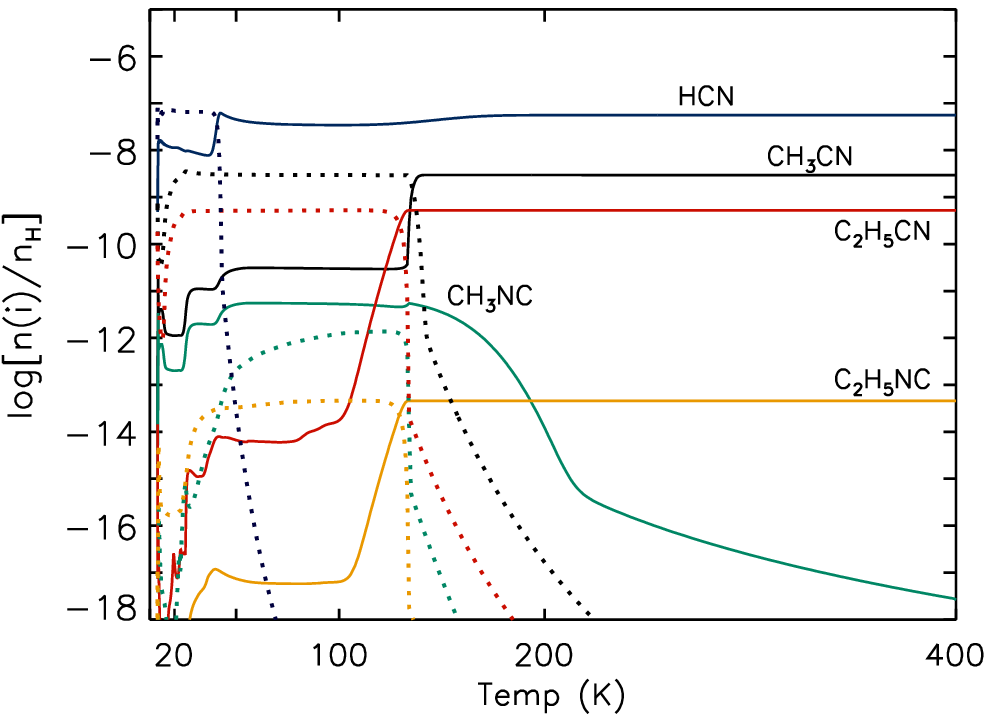} &
         \includegraphics[width=.46\textwidth]{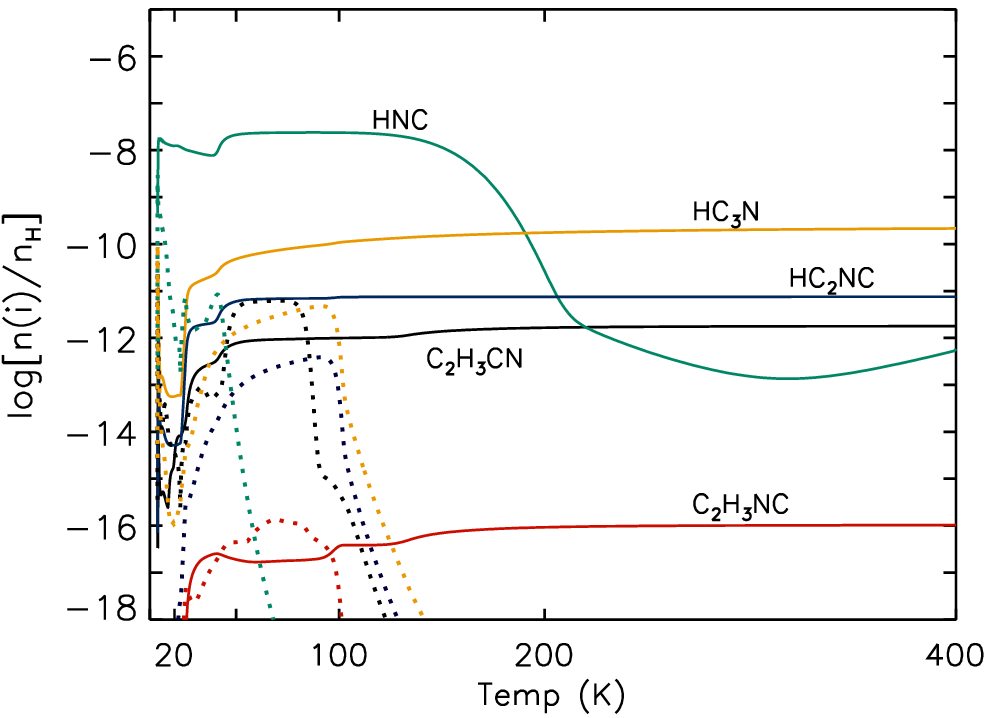} \\
    \end{tabular}
    \caption{Abundances of cyanides and isocyanides in Model 4, which has an $A_V$-dependent $\zeta$ shown in Figure \ref{fig:zeta_av}. Left panel: \ce{HCN, CH3CN, CH3NC, C2H5CN, C2H5NC}. Right panel: \ce{HNC, C2H3CN, C2H3NC, HC3N, HC2NC}.}
    \label{fig:mod4}
\end{figure*}

\begin{figure*}[htb]
    \centering
    \begin{tabular}{cc}
         \includegraphics[width=.46\textwidth]{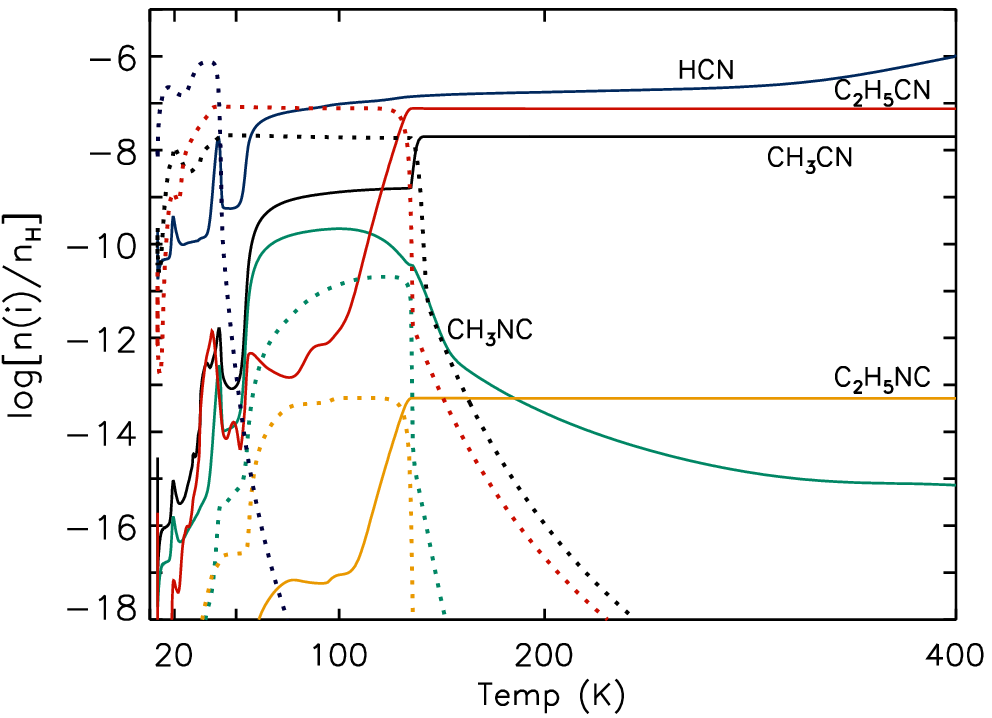} &
         \includegraphics[width=.46\textwidth]{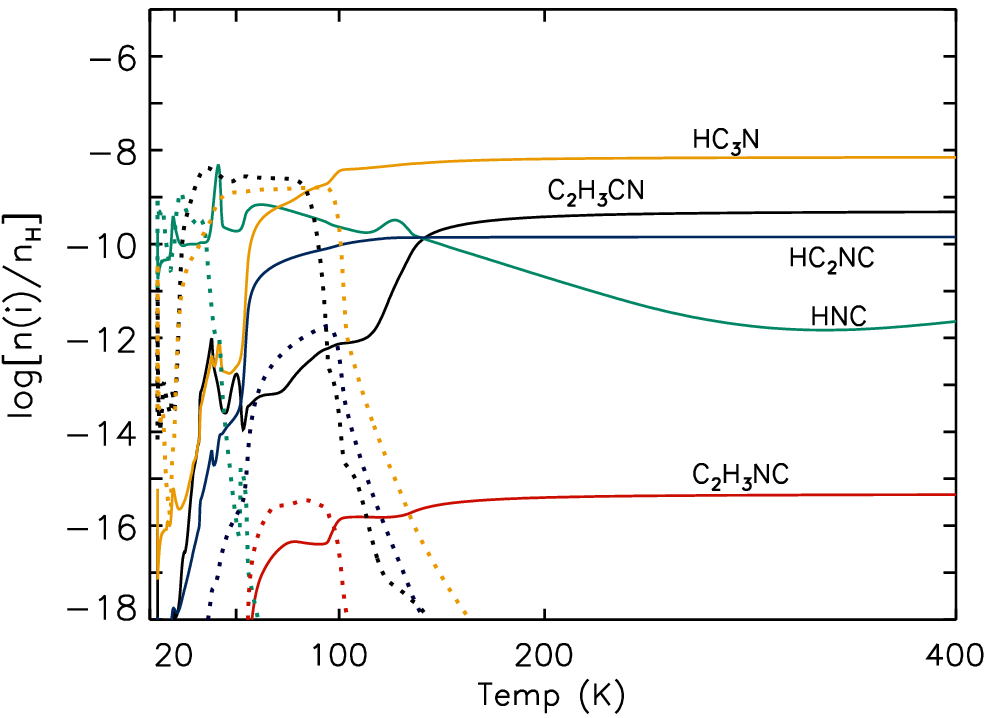} \\
    \end{tabular}
    \caption{Abundances of cyanides and isocyanides in Model 5, which has an $A_V$-dependent $\zeta$ shown in Figure \ref{fig:zeta_av}. Left panel: \ce{HCN, CH3CN, CH3NC, C2H5CN, C2H5NC}. Right panel: \ce{HNC, C2H3CN, C2H3NC, HC3N, HC2NC}.}
    \label{fig:mod5}
\end{figure*}
Figure \ref{fig:mod3} displays the fractional abundance profiles for Model 3. Model 3 has a variable $\zeta$ governed by Equation \ref{eq:zeta} that varies from \num{2.1e-16} s$^{-1}$ at the outer edge of the source to \num{1.9e-17} s$^{-1}$ at the inner edge of the source. This results in a higher cosmic-ray ionization rate throughout the entirety of the source than what is exhibited in Model 1. 

 In general, the shapes of the abundance profiles of the molecules are not affected in a significant way. This is true for all molecules in Figure \ref{fig:mod3}, except for \ce{HNC}. With a higher $\zeta$, the abundance of \ce{HNC} remains at its peak value for much longer than in Model 1. It only begins to decrease at $\sim$150~K. The reason for this is the much lower abundance of atomic \ce{C} in Model 3 as compared to Model 1, which is a result of ionization by cosmic rays, as well as reactions with species (such as \ce{O2}) that are produced in larger quantities in high-$\zeta$ environments. Also of note is that \ce{HNC} becomes the second-most abundant cyanide or isocyanide in terms of peak abundance. It reaches a high initial abundance due to the fact that its formation is dependent on recombination of larger ions, which are produced in greater abundance with a higher $\zeta$.
 
The increased and variable $\zeta$ impacts the peak and final abundances of all . In all cases, with the exception of the {peak} abundance of \ce{HNC}, the peak and final abundances of all molecules are decreased in Model 3 with respect to Model 1. The peak abundance of \ce{HNC} is actually higher in Model 3 by a factor of approximately five, while the final abundance is lower, as discussed previously. One other major difference in Model 3 is that \ce{CH3CN} becomes more abundant than \ce{C2H5CN}. This is a result of the fact that \ce{C2H5CN} is more readily destroyed by cosmic rays than \ce{CH3CN}, which is due to higher dissociation rates. 

Figure \ref{fig:mod4} displays the results for a model with higher variable $\zeta$ (Model 4). Model 4 has a $\zeta$ that varies from $\sim$\num{2.0e-15} s$^{-1}$ to $\sim$\num{1.0e-16} s$^{-1}$. The general shapes of the fractional abundance profiles are not changed significantly when going from Model 3 to Model 4. However, the values of peak and final abundances for all molecules are decreased.

A few molecules exhibit very slight changes in peak and final abundance with this increase in $\zeta$. \ce{C2H3NC} decreases by less than a factor of three, for example. Changing the values of $\zeta$ seems to have varying impacts on different types of molecules, however. \ce{HCN} and \ce{C2H5CN} both decrease by a factor of about six in Model 4, while the peak abundance of \ce{C2H5NC} falls even lower, decreasing by over an order of magnitude relative to Model 3.

Model 5, shown in Figure \ref{fig:mod5}, has a $\zeta$ that varies from $\sim$\num{2.0e-14} s$^{-1}$ to $\sim$\num{9.5e-16} s$^{-1}$. The chemical behavior displayed in Model 5 is in some cases significantly different from lower-$\zeta$ models. Relative to Model 4, all final abundances are actually higher in this model. The only peak abundance that decreases is that of \ce{HNC}, which decreases by a factor of approximately five. Many of these increases are quite significant as well. \ce{C2H5CN} and \ce{C2H3CN} increase by two orders of magnitude relative to Model 4, and in fact have higher peak and final abundances in Model 5 than in Model 3 as well. \ce{HCN} exhibits a similar increase.

In the case of \ce{HCN}, this increase manifests at higher temperatures ($>$200~K), due to a greater abundance of the \ce{CN} radical in the gas phase. \ce{CN} can react with species such as \ce{H2} and \ce{NH3} to produce \ce{HCN}. For \ce{CH3CN}, the abundance increase occurs at much earlier times, on the grains. At early temperatures ($<$20~K), the abundance of \ce{CH3} is much higher on grains in Model 5. This is because there is more \ce{OH} as well, which can abstract hydrogen atoms from \ce{CH4} to form \ce{CH3}. The larger amount of \ce{OH} is due to increased cosmic-ray dissociation rates of larger species such as \ce{H2O and C2H5OH}. Therefore, \ce{CH3} can react with \ce{CN} on grains to form \ce{CH3CN}. This is the same reason for the increase in peak and final abundance of \ce{C2H5CN}, as \ce{CH3} can also react with \ce{CH2CN} to produce \ce{C2H5CN}.

\begin{figure*}[htb]
    \centering
    \begin{tabular}{cc}
         \includegraphics[width=.46\textwidth]{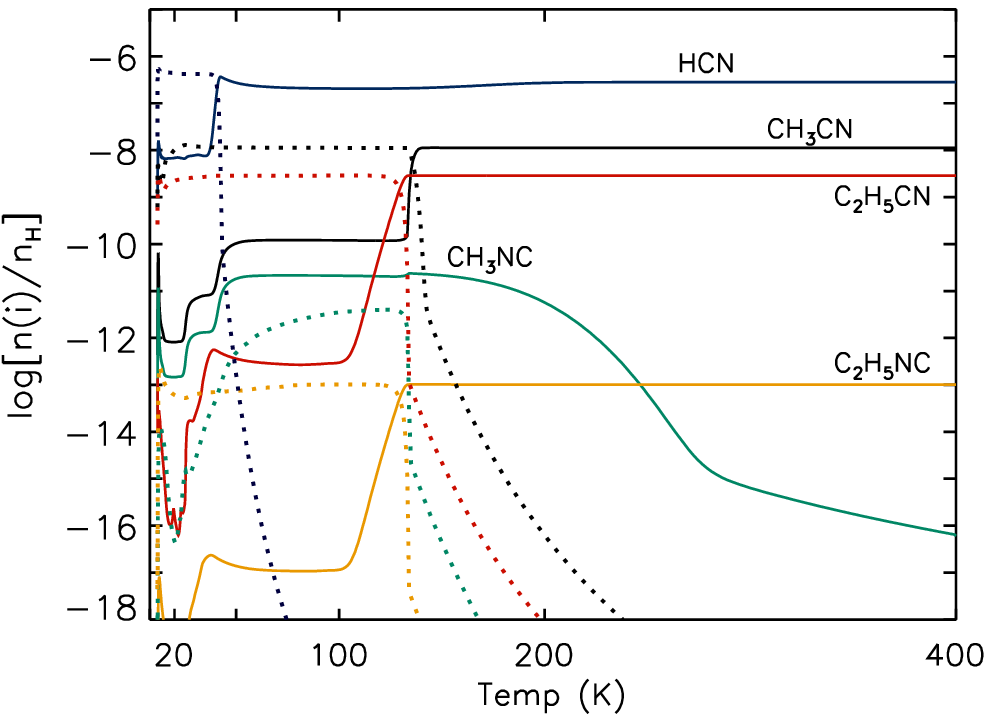} &
         \includegraphics[width=.46\textwidth]{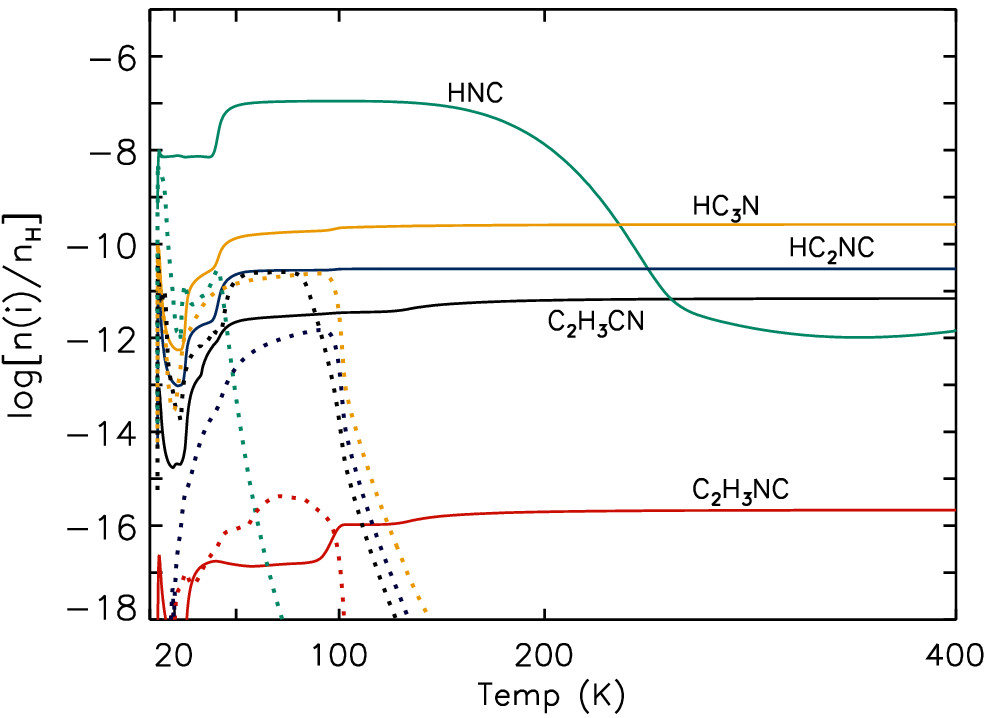} \\
    \end{tabular}
    \caption{Abundances of cyanides and isocyanides in Model 6, with a constant $\zeta$ of \num{1e-16} s$^{-1}$. Left panel: \ce{HCN, CH3CN, CH3NC, C2H5CN, C2H5NC}. Right panel: \ce{HNC, C2H3CN, C2H3NC, HC3N, HC2NC}.}
    \label{fig:mod6}
\end{figure*}

The results of Model 6 are shown in Figure \ref{fig:mod6}. The abundance profiles for Model 6 are intermediate between Models 3 and 4, but slightly more similar to those of Model 4 (Figure \ref{fig:mod4}), which has a minimum $\zeta$ of $\sim$\num{1e-16} s$^{-1}$, the same value as Model 6. However, the absolute values of the abundances are different. In most cases, the peak and final abundance values in Model 6 are greater than those in Model 4. In many cases, they are about five times higher in Model 6. There are some exceptions to this behavior. The final abundance of \ce{CH3NC} and the peak abundance of \ce{HNC3} (not shown on the plot) are over an order of magnitude higher in Model 6, whereas the peak and final abundance of \ce{HC3N} is actually slightly lower in Model 6.

\begin{figure*}[htb]
    \centering
    \begin{tabular}{cc}
         \includegraphics[width=.46\textwidth]{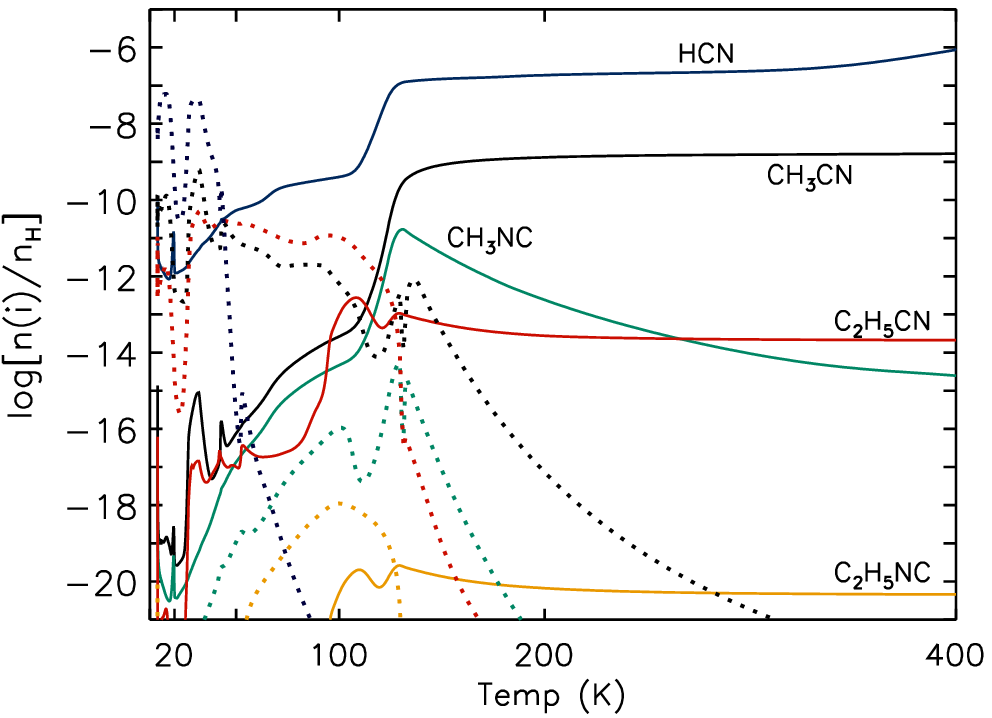} &
         \includegraphics[width=.46\textwidth]{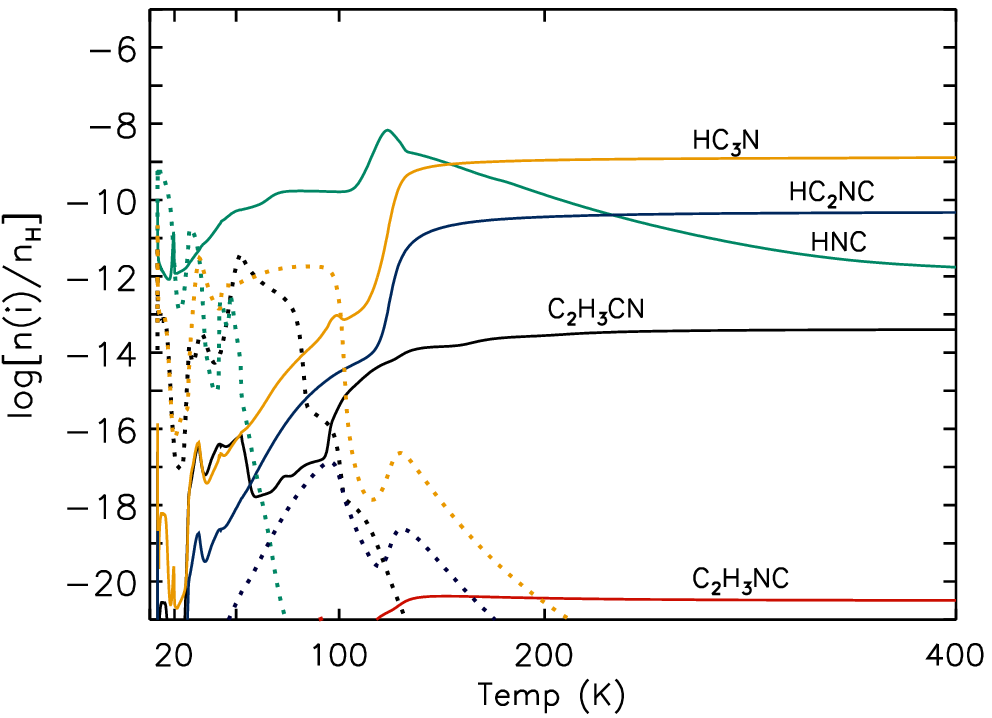} \\
    \end{tabular}
    \caption{Abundances of cyanides and isocyanides in Model 7, with a constant $\zeta$ of \num{3e-14} s$^{-1}$. Left panel: \ce{HCN, CH3CN, CH3NC, C2H5CN, C2H5NC}. Right panel: \ce{HNC, C2H3CN, C2H3NC, HC3N, HC2NC}.}
    \label{fig:mod7}
\end{figure*}

The results from Model 7 are shown in Figure \ref{fig:mod7}. This model exhibits the highest $\zeta$ in our study, and as such it is useful to compare it to Model 5 (our other high-$\zeta$ model), as well as Model 6. In this model, the peak and final abundances of most molecules decrease from those in Model 5. The only exception to this is the peak abundance of \ce{HNC}, which increases slightly from that in Model 5. All other species exhibit both peak and final abundance decreases. Some of these are quite substantial. For example, \ce{C2H5CN, C2H3CN, C2H5NC, and C2H3NC} all decrease by several orders of magnitude in Model 7. The increased flux of cosmic rays throughout the cloud serves to destroy large molecules much faster than they are able to be formed. This model appears to be a very poor fit to observations, especially considering the predicted undetectable abundance of \ce{C2H5CN} at $\sim$\num{2e-14}. The abundance profiles of molecules exhibit large changes as well, particularly in the low-temperature regions. 

The behavior of Model 7 with respect to Model 6 is also quite complex. Some molecules exhibit increases in peak and final abundance, while others exhibit decreases. \ce{HCN} and \ce{HCCNC} exhibit a modest (factor of $\sim$2) increase in peak and final abundances, while \ce{C2H3CN} and \ce{HC3NH+} exhibit larger increases. This agrees with the results for the variable-$\zeta$ models, in that these molecules appear to exhibit higher abundances with increasing cosmic-ray flux. All other species exhibit decreases in peak abundance when going from Model 6 to Model 7.
\subsection{Comparison of chemical modeling to observations and spectral modeling}
There are a few methods for making a comparison between our astrochemical models and observational column densities. We discuss two of them below. The first is a simple comparison of fractional abundance values. Up until now, we have focused on peak and final abundances, but for a comparison with observational data, we are more interested in the abundances of molecules at their observationally determined excitation temperatures (given in Table \ref{t:coldens}). Table \ref{t:tex_abuns} shows the fractional abundance for each molecule of interest at its observationally determined excitation temperature for each model. 

\begin{table*}
 \begin{center}
 \caption{Abundances relative to total hydrogen at the observationally-determined excitation temperature for all molecules of interest in each model.}
 \label{t:tex_abuns}
 \vspace*{-2.0ex}
 \begin{tabular}{llllllll}
 \hline\hline\\[-1.0em]
 \multicolumn{1}{l}{Molecule} & \multicolumn{1}{l}{Model 1} & \multicolumn{1}{l}{Model 2} & \multicolumn{1}{l}{Model 3} & \multicolumn{1}{l}{Model 4} & \multicolumn{1}{l}{Model 5} & \multicolumn{1}{l}{Model 6} & \multicolumn{1}{l}{Model 7}
 \\ \hline
    \ce{CH3CN} & \num{3.7e-8} & \num{3.7e-8} & \num{6.4e-9} & \num{3.0e-9} & \num{1.9e-8} & \num{1.1e-8} & \num{1.1e-9}\\ 
    \ce{CH3NC} & \num{4.8e-10} & \num{5.5e-10} & \num{9.2e-12} & \num{9.7e-13} & \num{1.2e-13} & \num{1.6e-11} & \num{1.1e-12}\\ 
    \ce{C2H5CN} & \num{6.1e-8} & \num{6.1e-8} & \num{3.3e-9} & \num{5.2e-10} & \num{7.7e-8} & \num{2.9e-9} & \num{5.7e-14}\\
    \ce{C2H5NC} & \num{8.9e-12} & \num{9.0e-12} & \num{7.3e-13} & \num{4.6e-14} & \num{5.2e-14} & \num{1.0e-13} & \num{1.4e-20}\\
    \ce{C2H3CN} & \num{1.5e-10} & \num{1.5e-10} & \num{1.9e-12} & \num{1.6e-12} & \num{3.8e-10} & \num{6.3e-12} & \num{2.8e-14}\\
    \ce{C2H3NC} & \num{1.1e-14} & \num{1.1e-14} & \num{2.8e-16} & \num{9.2e-17} & \num{4.0e-16} & \num{2.0e-16} & \num{3.6e-21}\\
    \ce{HC3N} & \num{3.8e-8} & \num{3.8e-8} & \num{6.3e-10} & \num{1.6e-10} & \num{6.1e-9} & \num{2.5e-10} & \num{9.9e-10}\\
    \ce{HCCNC} & \num{3.4e-10} & \num{3.4e-10} & \num{2.8e-11} & \num{7.5e-12} & \num{1.4e-10} & \num{3.0e-11} & \num{2.9e-11}\\
    \ce{HNCCC} & \num{1.6e-13} & \num{1.6e-13} & \num{2.4e-16} & \num{6.0e-17} & \num{1.2e-15} & \num{3.2e-16} & \num{2.6e-15}\\
    \ce{HC3NH+} & \num{6.4e-13} & \num{6.4e-13} & \num{2.9e-15} & \num{3.3e-15} & \num{5.6e-13} & \num{5.5e-15} & \num{4.5e-12}\\ \hline
 \end{tabular}
 \end{center}
 \vspace*{-2.5ex}
 \end{table*}
\begin{table*}
 \begin{center}
 \caption{Fractional abundance ratios relative to total hydrogen for models at the observationally-determined excitation temperature for each species, as well as the observational column density ratios.}
 \label{t:ratios_comparison}
 \vspace*{-2.0ex}
 \begin{tabular}{lllllllll}
 \hline\hline\\[-1.0em]
 \multicolumn{1}{c}{Ratio} & \multicolumn{1}{c}{Observations} &
 \multicolumn{1}{c}{\hspace*{-2ex} Model 1} &
 \multicolumn{1}{c}{\hspace*{-2ex} Model 2} &
 \multicolumn{1}{c}{\hspace*{-2ex} Model 3} &
 \multicolumn{1}{c}{\hspace*{-2ex} Model 4} &
 \multicolumn{1}{c}{\hspace*{-2ex} Model 5} &
 \multicolumn{1}{c}{\hspace*{-2ex} Model 6} &
 \multicolumn{1}{c}{\hspace*{-2ex} Model 7} \\
 \hline\\[-1.0em]
 CH$_3$NC/CH$_3$CN & \num{4.7e-3} & \hspace*{-2ex} \num{1.3e-2} & \hspace*{-2ex} \num{1.5e-2} & \hspace*{-2ex} \num{1.4e-3} & \hspace*{-2ex} \num{3.3e-4} & \hspace*{-2ex} \num{6.3e-6} & \hspace*{-2ex} \num{1.4e-3} & \hspace*{-2ex} \num{9.8e-4}\\[0.2ex]
 C$_2$H$_5$NC/C$_2$H$_5$CN & $< \num{2.4e-4}$ & \hspace*{-2ex} \num{1.4e-4} & \hspace*{-2ex} \num{1.5e-4} & \hspace*{-2ex} \num{2.2e-4} & \hspace*{-2ex} \num{8.8e-5} & \hspace*{-2ex} \num{6.8e-7} & \hspace*{-2ex} \num{3.6e-5} & \hspace*{-2ex} \num{2.4e-7}\\[0.2ex]
 C$_2$H$_3$NC/C$_2$H$_3$CN & $<$ \num{7.1e-3} & \hspace*{-2ex} \num{7.1e-5} & \hspace*{-2ex} \num{7.2e-5} & \hspace*{-2ex} \num{1.5e-4} & \hspace*{-2ex} \num{5.6e-5} & \hspace*{-2ex} \num{1.0e-6} & \hspace*{-2ex} \num{3.1e-5} & \hspace*{-2ex} \num{1.3e-7}\\[0.2ex]
 HCCNC/HC$_3$N & \num{1.5e-3} & \hspace*{-2ex} \num{8.9e-3} & \hspace*{-2ex} \num{8.9e-3} & \hspace*{-2ex} \num{4.5e-2} & \hspace*{-2ex} \num{4.7e-2} & \hspace*{-2ex} \num{2.3e-2} & \hspace*{-2ex} \num{1.2e-1} & \hspace*{-2ex} \num{2.9e-2}\\[0.2ex]
 \hline\\[-1.0em]
 HNC$_3$/HC$_3$N & $< \num{1.9e-4}$ & \hspace*{-2ex} \num{4.3e-6} & \hspace*{-2ex} \num{4.3e-6} & \hspace*{-2ex} \num{3.8e-7} & \hspace*{-2ex} \num{3.7e-7} & \hspace*{-2ex} \num{2.0e-7} & \hspace*{-2ex} \num{1.3e-6} & \hspace*{-2ex} \num{2.6e-6}\\[0.2ex]
 HC$_3$NH$^+$/HC$_3$N & $<$\num{1.7e-3} & \hspace*{-2ex} \num{1.7e-5} & \hspace*{-2ex} \num{1.7e-5} & \hspace*{-2ex} \num{4.7e-6} & \hspace*{-2ex} \num{3.7e-7} & \hspace*{-2ex} \num{2.0e-7} & \hspace*{-2ex} \num{2.2e-5} & \hspace*{-2ex} \num{4.6e-3}\\[0.2ex]
 \hline
 \end{tabular}
 \end{center}
 \vspace*{-2.5ex}
 \tablefoot{Mod 1 - standard model. Mod 2 - 3000K barrier model. Mod 3 - Variable $\zeta$ - low. Mod 4 - Variable $\zeta$ - med. Mod 5 - Variable $\zeta$ - high. Mod 6 - Med. constant $\zeta$. Mod 7 - High constant $\zeta$.}
 \end{table*}

Since there are many uncertainties associated with our chemical models, as well as the observational data, comparing raw fractional abundance values is not very robust. Thus, it is generally better to compare fractional abundance {ratios} of two related molecules, to assess the model agreement with observations. Table \ref{t:ratios_comparison} shows the fractional abundance ratios for each model at the  observational excitation temperature of the corresponding species, as well as the observational column density ratios. By this method, all models produce ratios of \ce{C2H5NC}:\ce{C2H5CN}, \ce{C2H3NC}:\ce{C2H3CN}, and \ce{HNC3}:\ce{HC3N}, consistent with the observational upper limits. The behavior of the \ce{HC3NH+}:\ce{HC3N} ratio is more complicated. In this case, all models except Model 7 are consistent with the observational upper limit.

The two tentatively detected ratios, \ce{CH3NC}:\ce{CH3CN} and \ce{HCCNC}:\ce{HNC3}, are not exactly reproduced in any of the models. For the \ce{CH3NC}:\ce{CH3CN} ratio, Models 1 and 2 have roughly the same value, exhibiting a ratio almost an order of magnitude too high, meaning either an overproduction of the isocyanide or an underproduction of the normal cyanide. As mentioned in Sect. 6.1, the only difference between Models 1 and 2 is the value of the barrier in Reaction \ref{eq:3}.  Model 3 and Model 6 reproduce the observational ratio within a factor of three, the best fit out of any of the models. Meanwhile, Models 4, 5, and 7 do a much poorer job of reproducing this ratio, with Models 4 and 7 being an order of magnitude too low, and Model 5 being almost three orders of magnitude too low. This seems to indicate that higher $\zeta$ values push this ratio too low, as there is a relatively clear trend observed among the models for which $\zeta$ is changed where this is the case. Within both the variable and constant $\zeta$ models, higher values of $\zeta$ lead to lower values of the \ce{CH3NC}:\ce{CH3CN} ratio.

The \ce{HCCNC}:\ce{HC3N} ratio exhibits different behavior. Models 1 and 2 exhibit ratios that are of the same order of magnitude as the observations, but a factor of approximately four too high. Models 3-5 are about an order of magnitude too high, caused by a large decrease in the abundance of \ce{HC3N} with respect to Models 1 and 2. \ce{HCCNC} also exhibits a decrease in abundance, but it is not as large as \ce{HC3N}. Models 6 and 7, with higher constant $\zeta$ values, exhibit the highest \ce{HCCNC}:\ce{HC3N} ratios, with Model 6 being almost two orders of magnitude too high and Model 7 being over an order of magnitude too high.

Models 2, 4, and 5 do not appear to be particularly good fits to the observational ratios. Model 6 is a reasonably good fit for the ratio of \ce{CH3NC}:\ce{CH3CN}, but it is not a good fit for the \ce{HCCNC}:\ce{HC3N} ratio. Model 7 is a better fit, but it does not actually have observable abundances of \ce{CH3NC} at the observationally determined excitation temperature (a value of \num{5.7e-13} with respect to total hydrogen at T = 170~K), and so it cannot be considered a good fit either. Models 1 and 3 appear to be the best fits. Model 1 reproduces the \ce{HCCNC}:\ce{HC3N} ratio better than Model 3, while Model 3 does a better job with \ce{CH3NC}:\ce{CH3CN}. However, when looking at both ratios together, Model 1 reproduces the observational data slightly better than Model 3. 

Inspecting fractional abundance profiles from chemical models is a very effective tool to determine important chemical pathways and chemical behaviors as a function of physical parameters. However, observations of star-forming sources provide only column densities of molecules averaged over the telescope beam. Therefore, it is not precisely accurate to compare abundance values for a single point in a chemical model to those averaged over an entire telescope beam. Another method of comparing our models with observations is to take the fractional abundances of the molecules calculated in the chemical models and model their radiative transfer using observational physical profiles. Details of the radiative transfer model used here are given in \citet{garrod13}, and a brief summary of the procedure is given here. 

\begin{figure*}[htb]
    \centering
    \begin{tabular}{cc}
         \includegraphics[width=.46\textwidth]{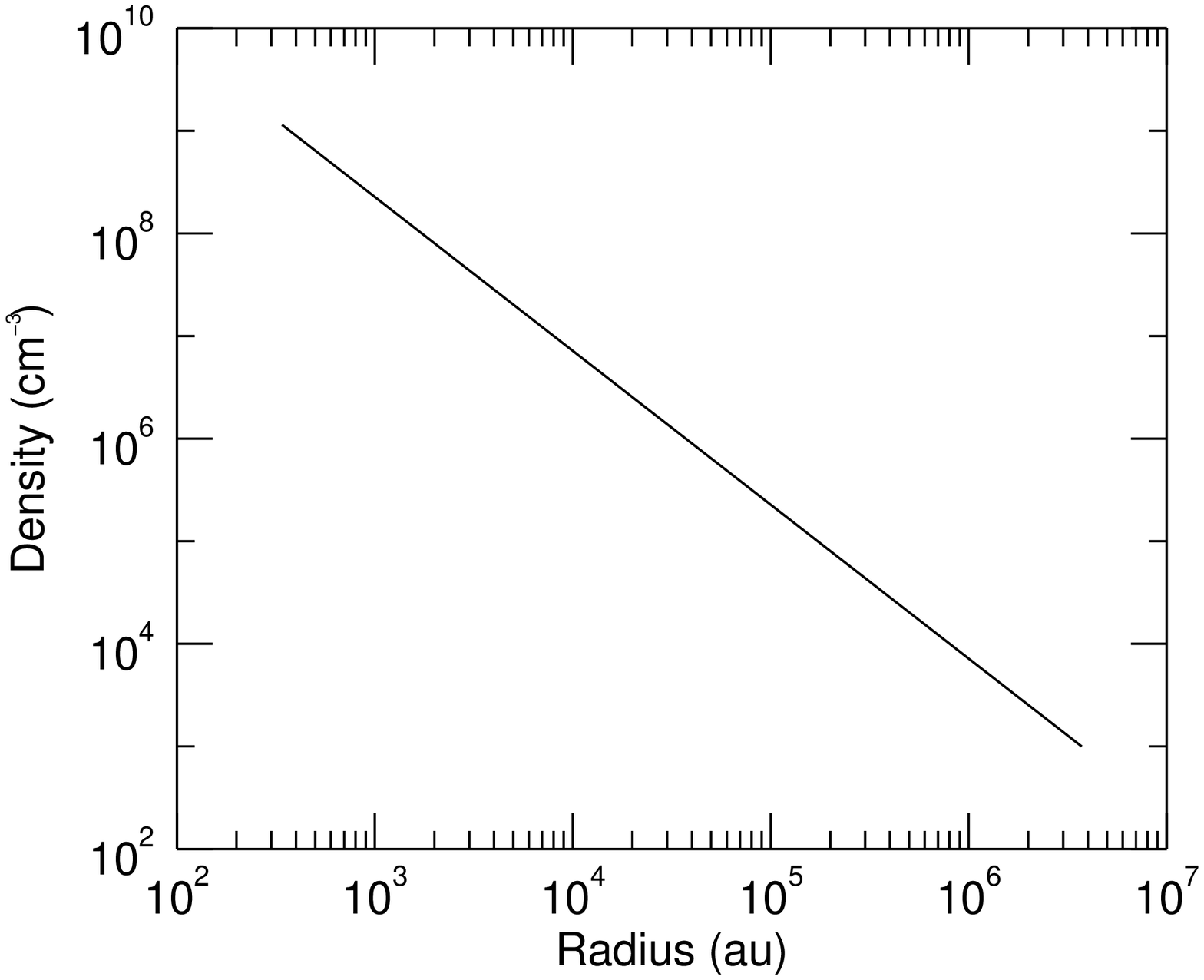} &
         \includegraphics[width=.46\textwidth]{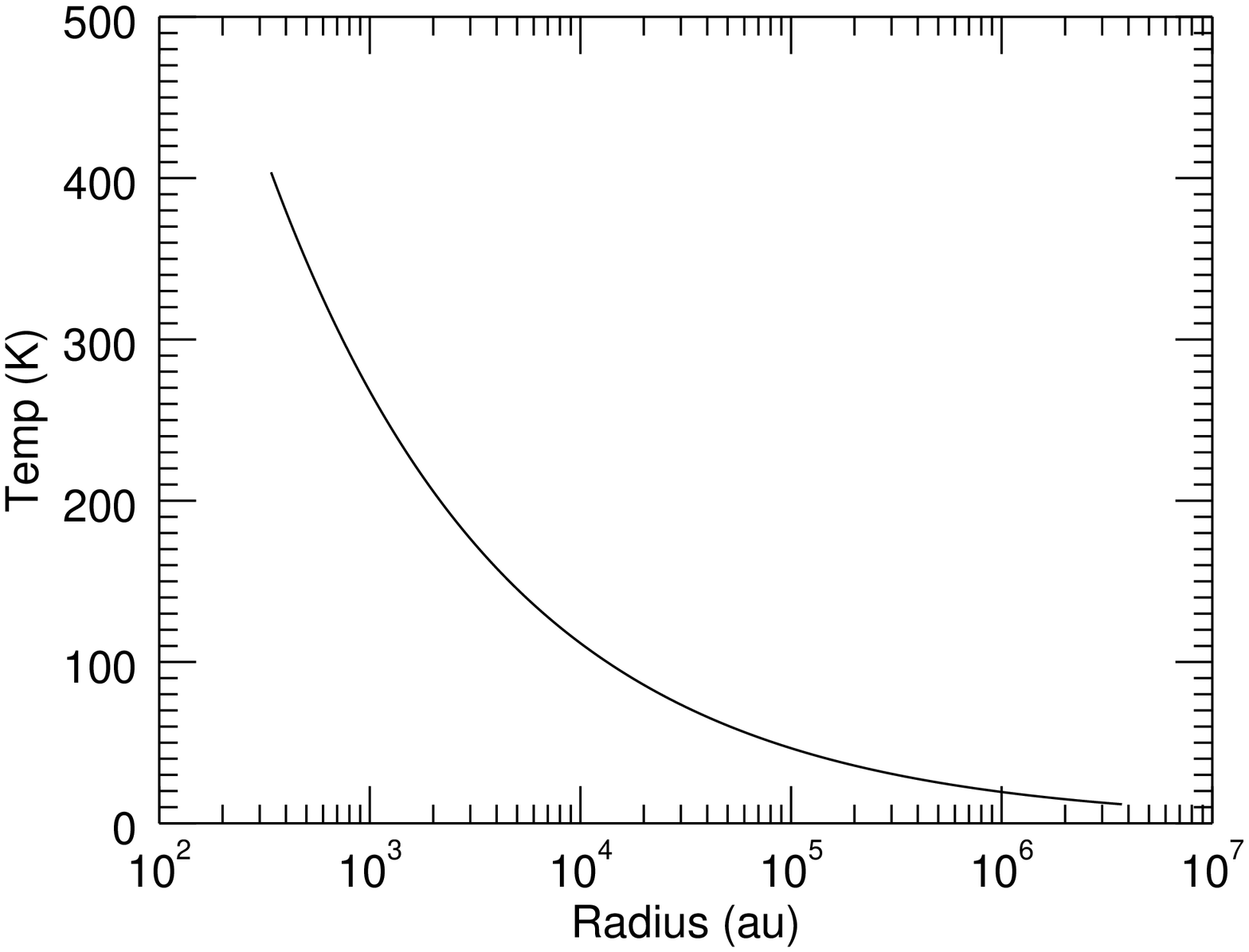} \\
    \end{tabular}
    \caption{Density and temperature profiles used in the radiative transfer calculations for the modeling data. The density is that of total hydrogen.} 
    \label{fig:profiles}
\end{figure*}

Since we are interested in modeling observations of Sgr B2(N2), we use the observational physical profiles discussed in Sect. 5.2 for our radiative transfer model. The density and temperature profiles that are used are shown in Figure \ref{fig:profiles}. The fractional abundance profiles for each molecule are mapped onto the corresponding temperature and density profiles. As a result, we produce a spherically symmetric model of Sgr B2(N2), in which the abundance of each molecule is defined at each point. Local absorption and emission coefficients are then calculated. Since Sgr B2(N2) should be well within LTE conditions \citep{Belloche16}, we then use the LTE approximation to calculate radiative transfer along lines of sight, thus producing emission maps for each molecule in each frequency channel.

These emission maps are then convolved with a Gaussian beam in order to produce simulated spectra for each line. In this case, we model all lines within the frequency range of the EMoCA survey for each molecule. The beam width for the convolution is chosen to replicate the ALMA beam for the observations outlined in \citet{Belloche16}.

We then use the simulated spectra to produce rotational diagrams for each molecule studied here, following the method of \citet{goldsmithlanger}. We correct for the optical depth of lines by fitting the wings of each spectral line to a Gaussian. We then use the ratio of the area of the fitted Gaussian to the area of the optically thick spectral line to determine the optical depth correction factor, $c_\tau$. We integrate the emission of each molecule only for a radius of $\sim$\num{8e4} au, which is the maximum recoverable scale of ALMA in the EMoCA survey. This limit is also enforced in the line-of-sight integrations that produce the raw emission maps.

\begin{figure*}[htb]
    \centering
    \begin{tabular}{cc}
         \includegraphics[width=.46\textwidth]{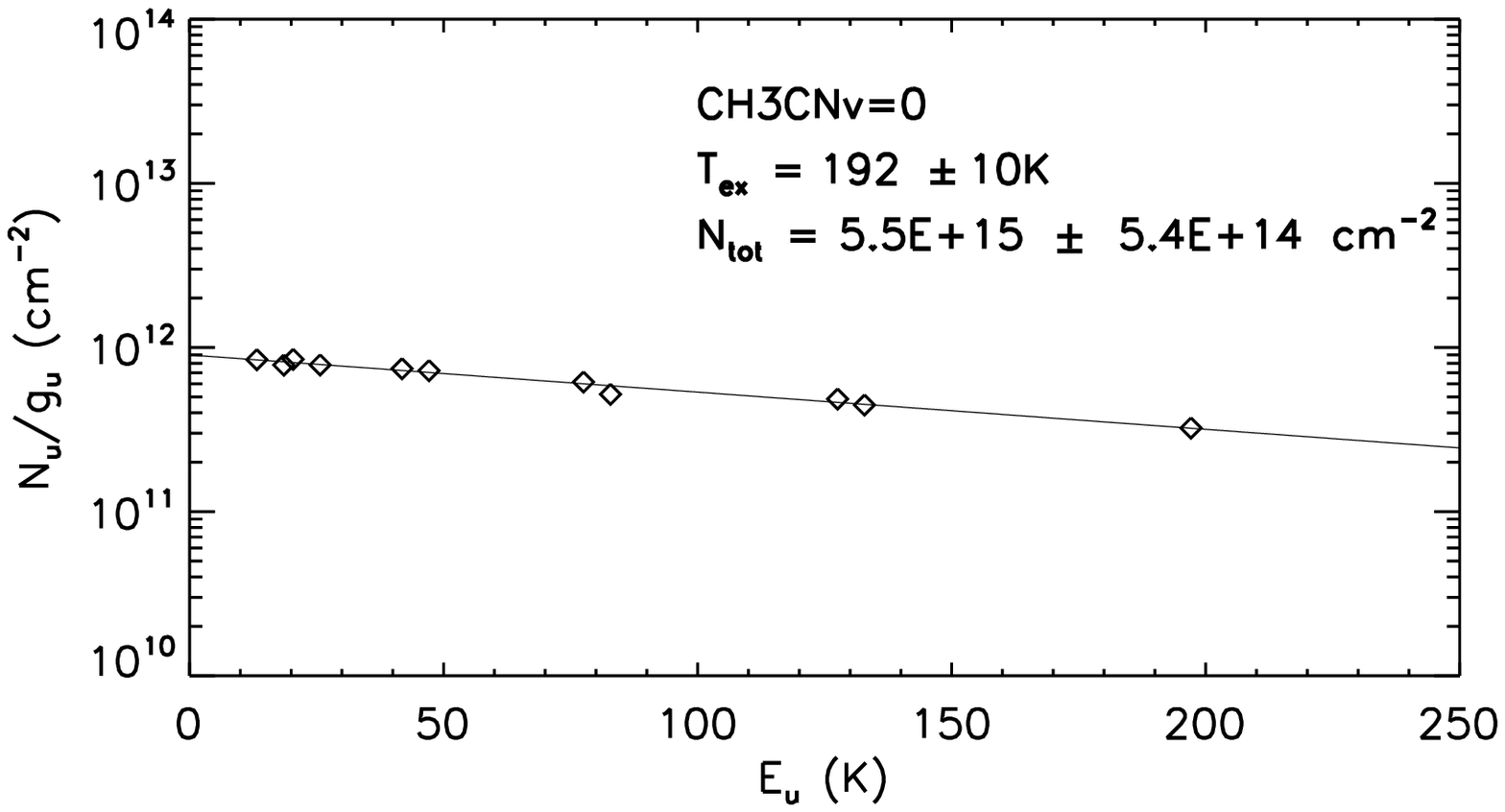} &
         \includegraphics[width=.46\textwidth]{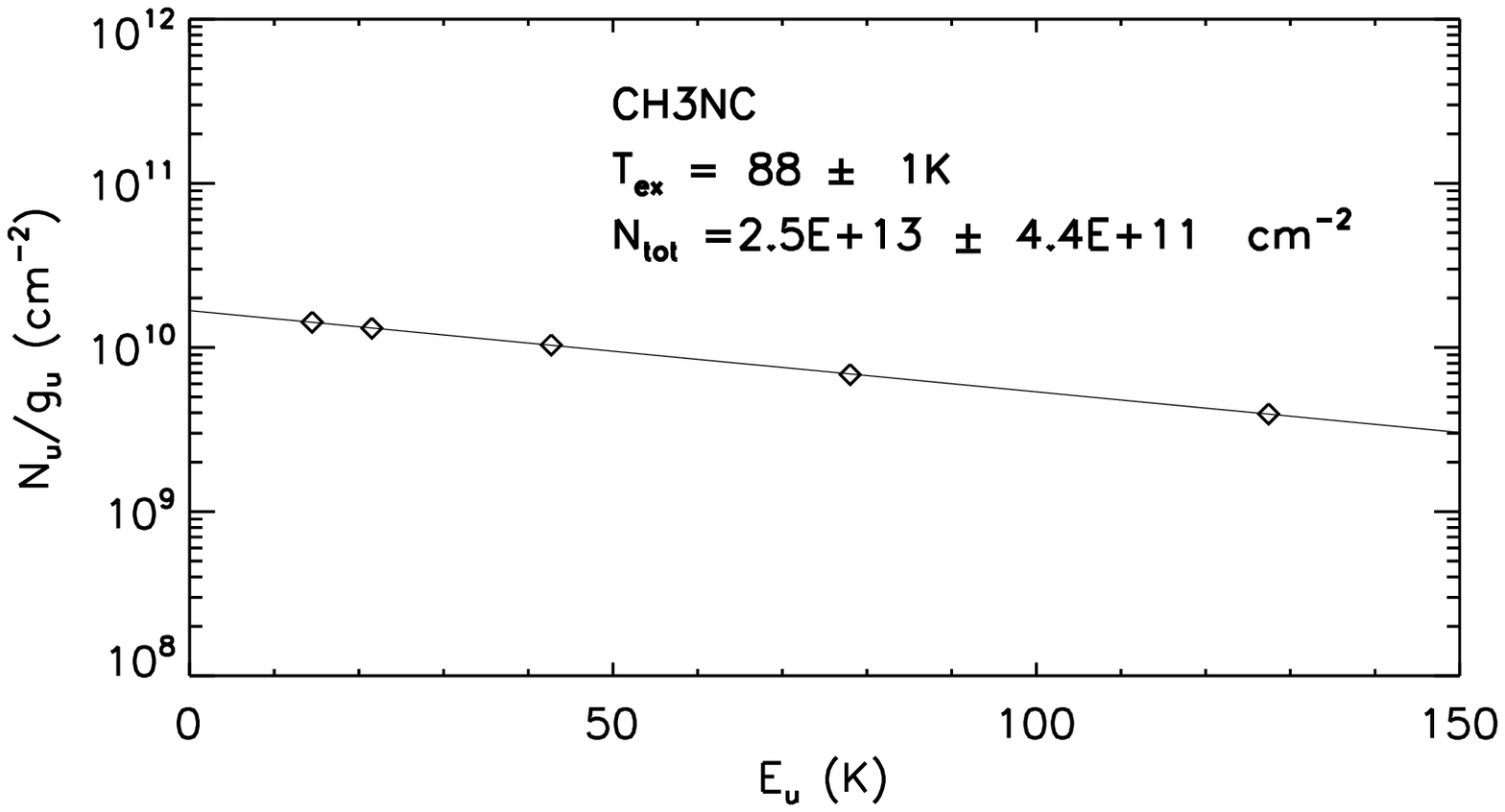} \\
         \includegraphics[width=.46\textwidth]{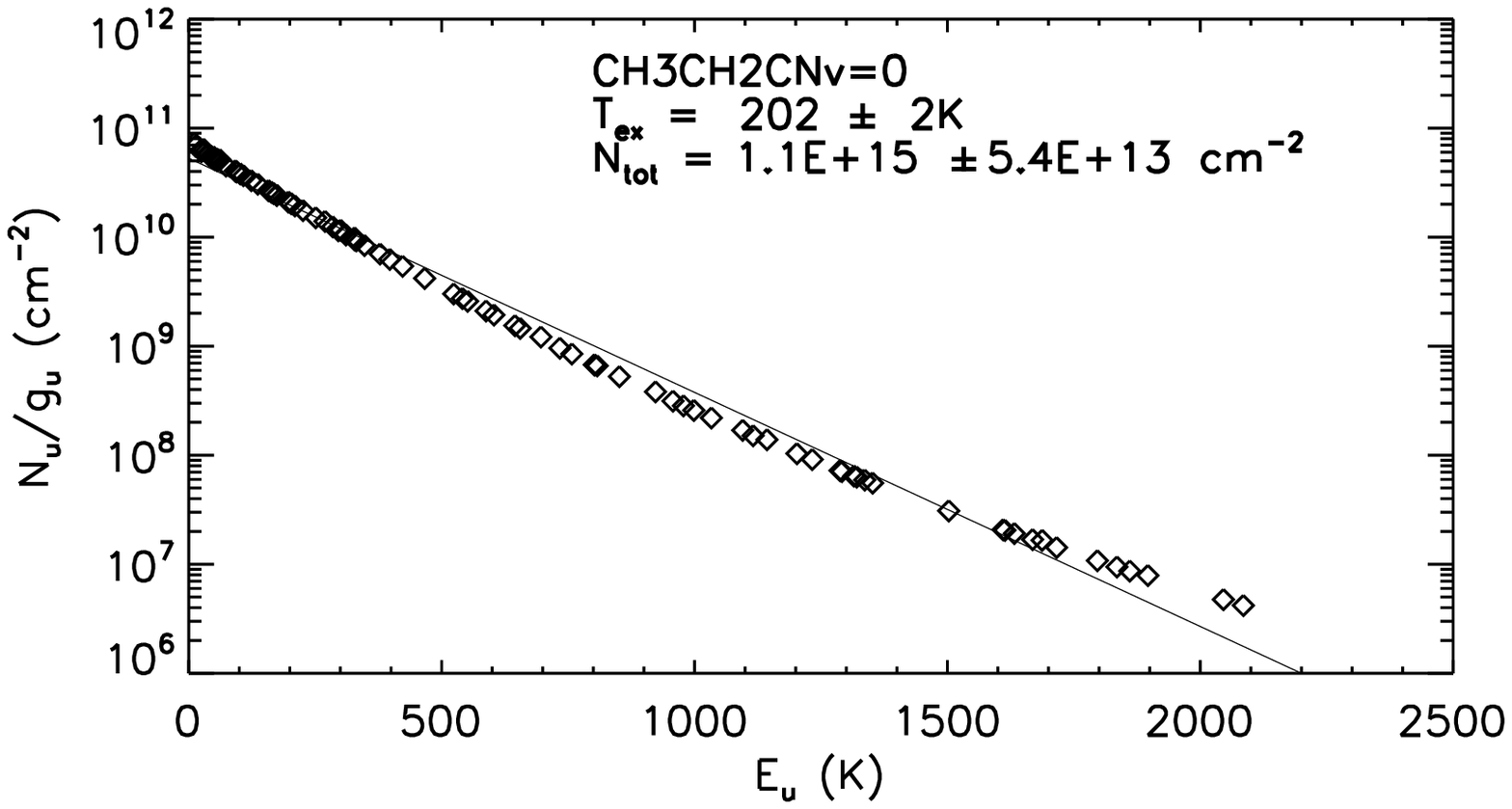} &
         \includegraphics[width=.46\textwidth]{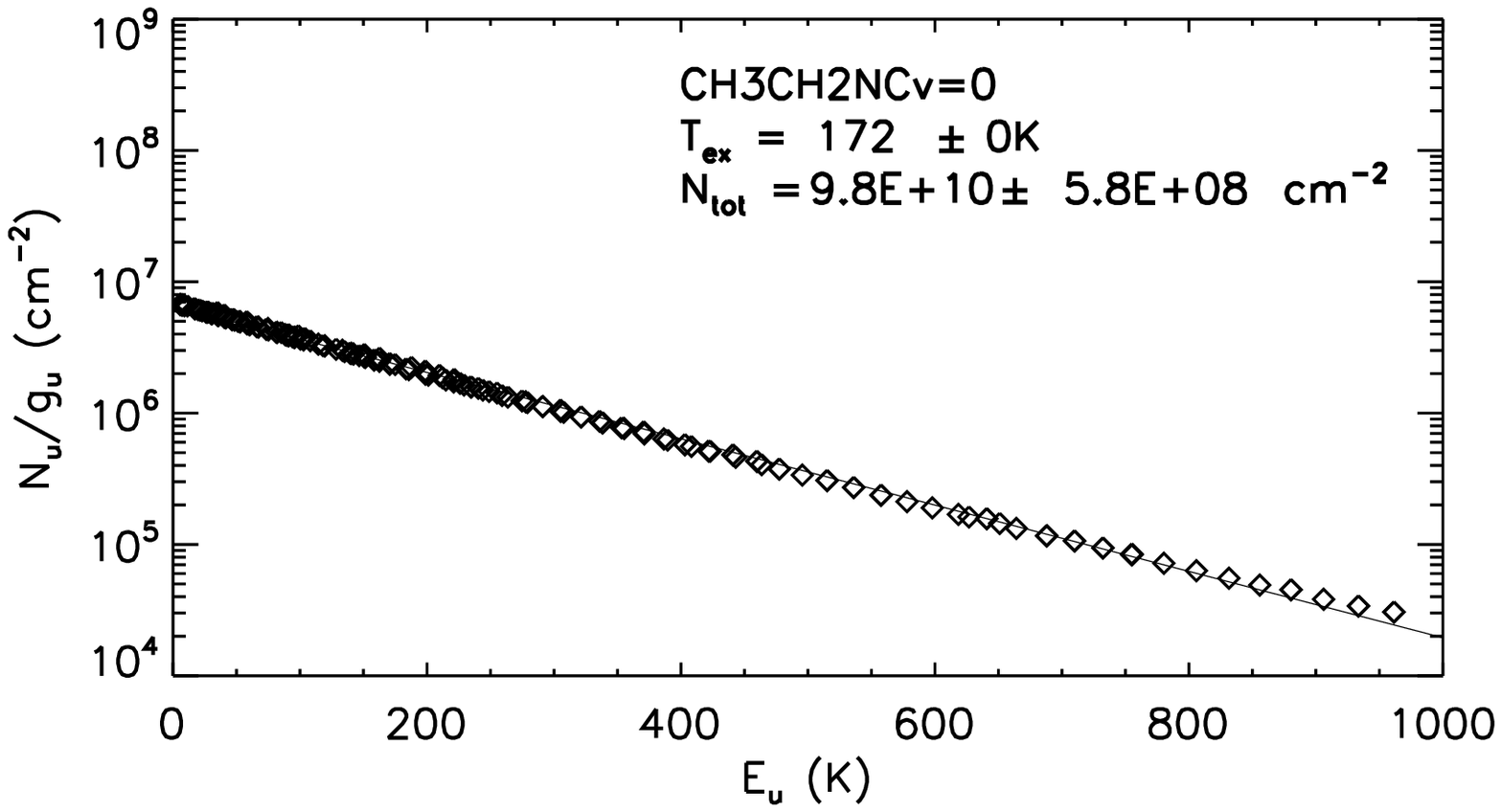} \\         \includegraphics[width=.46\textwidth]{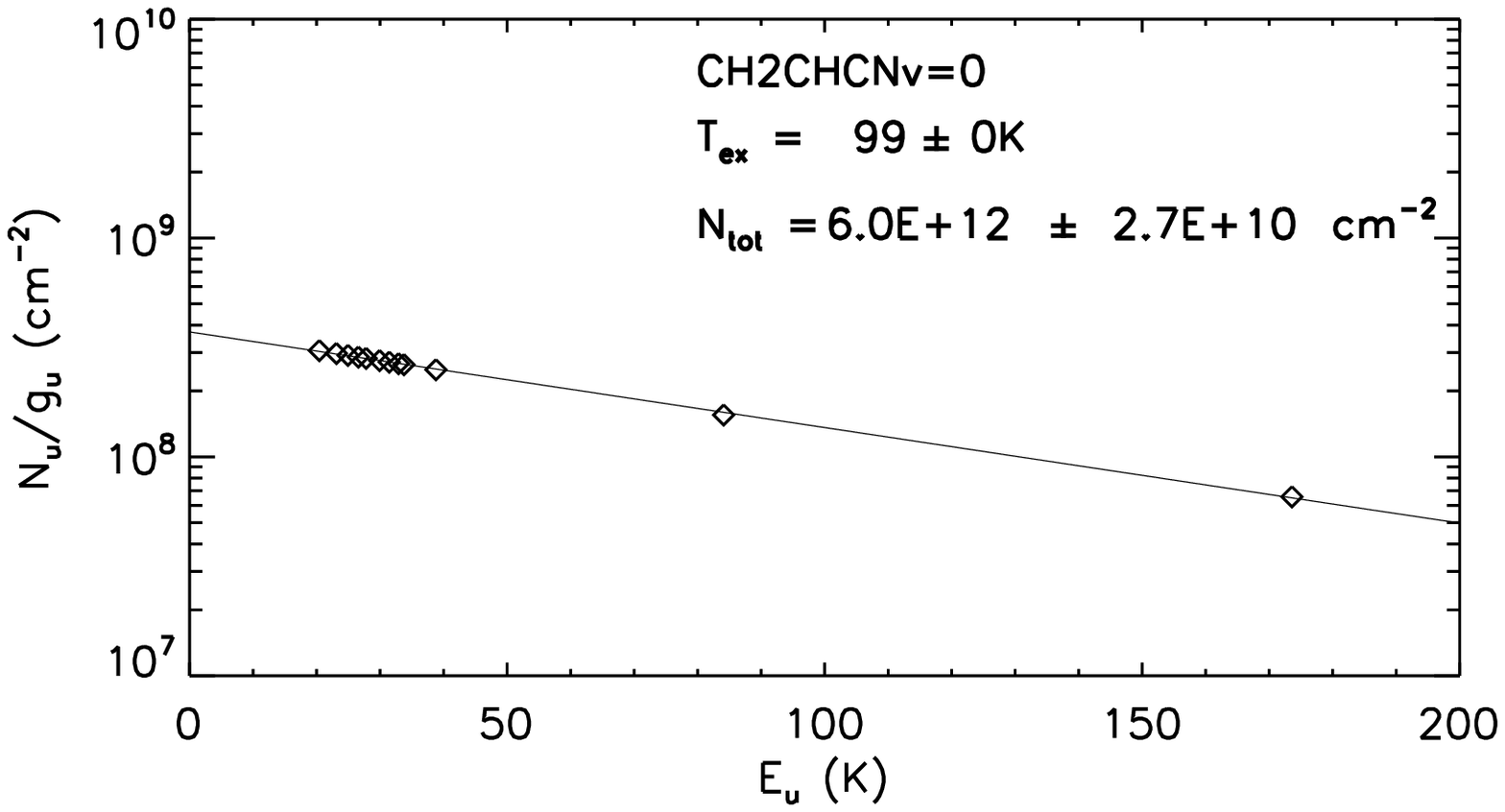} &
         \includegraphics[width=.46\textwidth]{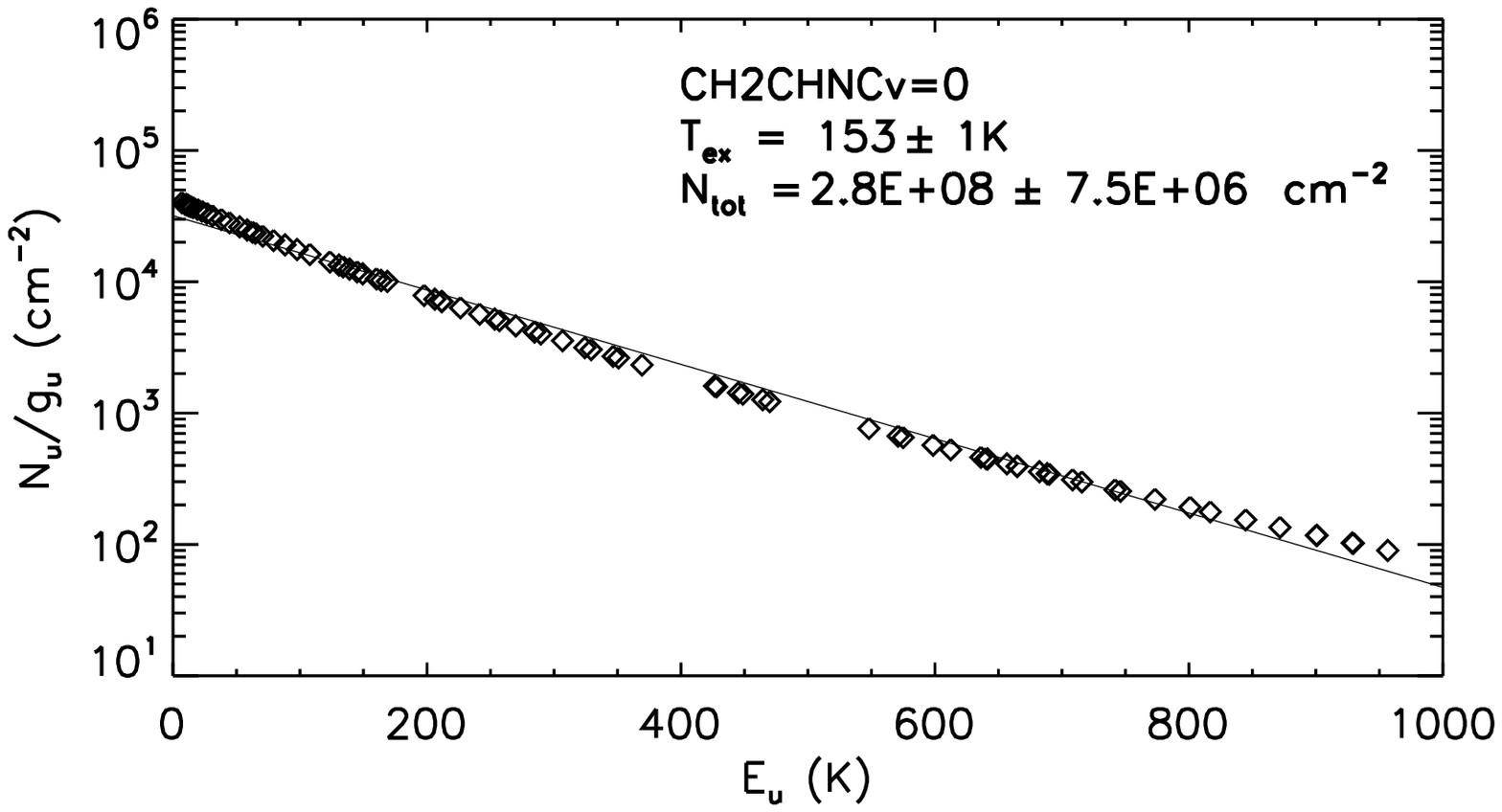} \\
         \includegraphics[width=.46\textwidth]{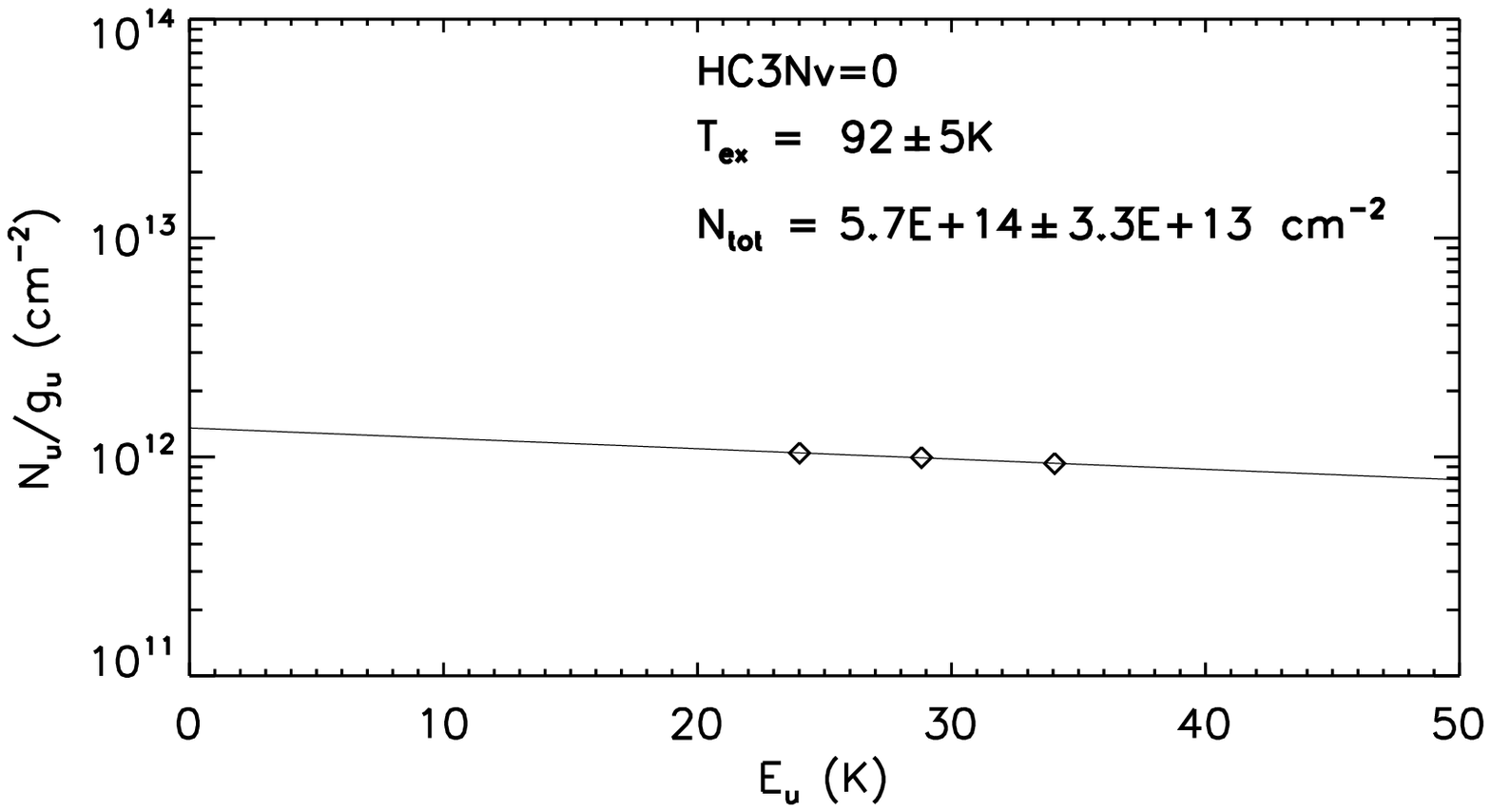} &
         \includegraphics[width=.46\textwidth]{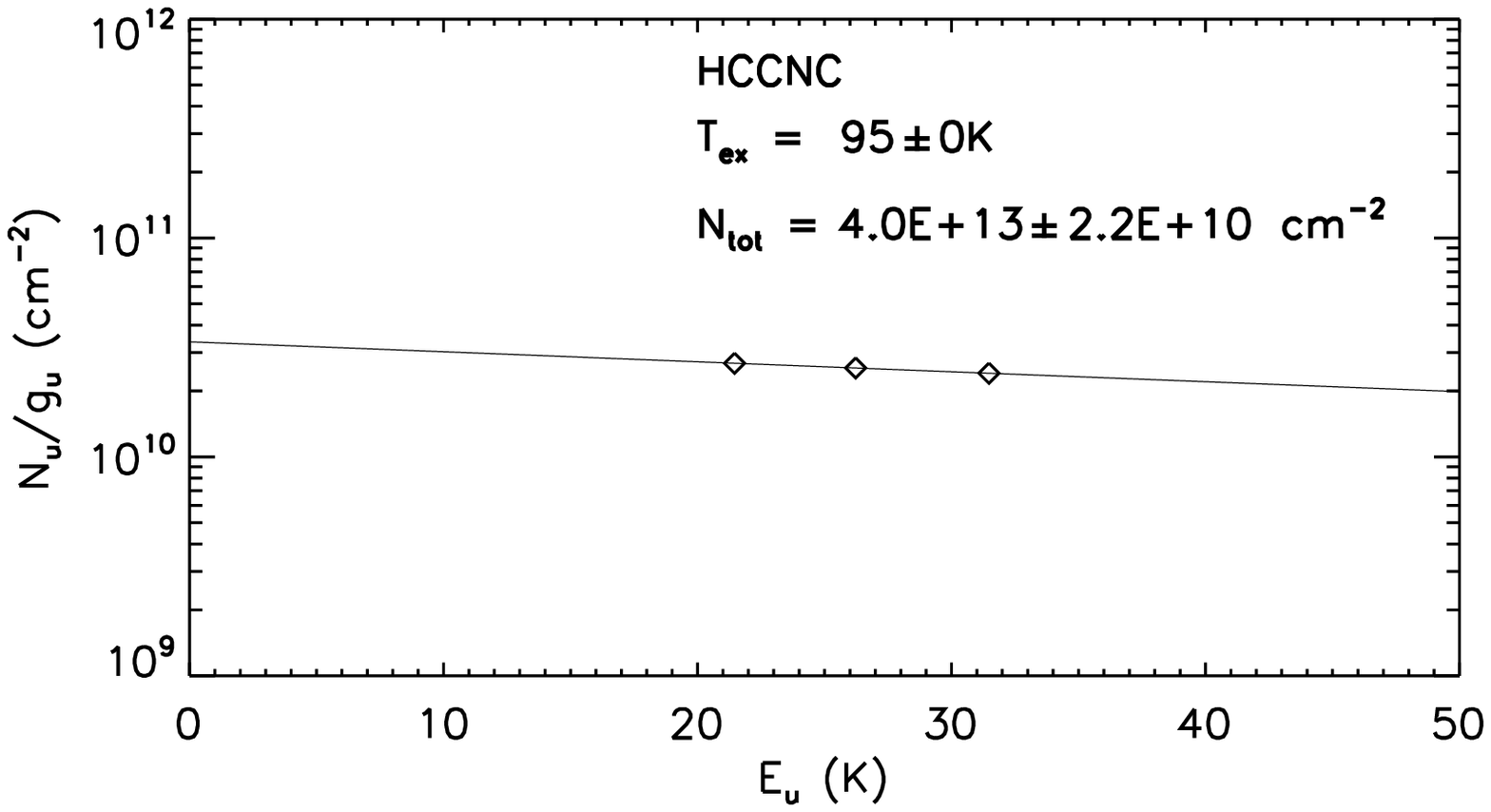} \\
         \includegraphics[width=.46\textwidth]{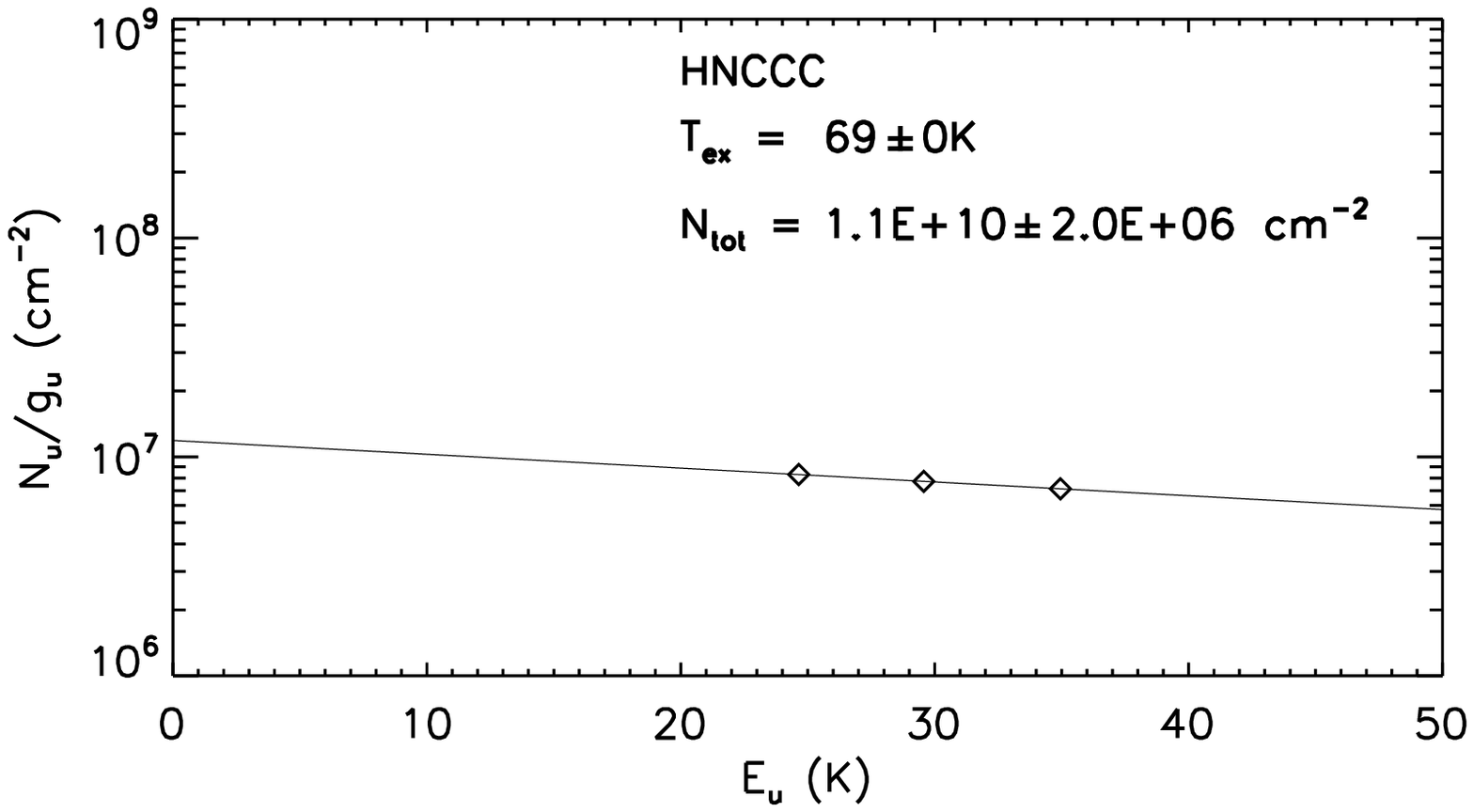} &
         \includegraphics[width=.46\textwidth]{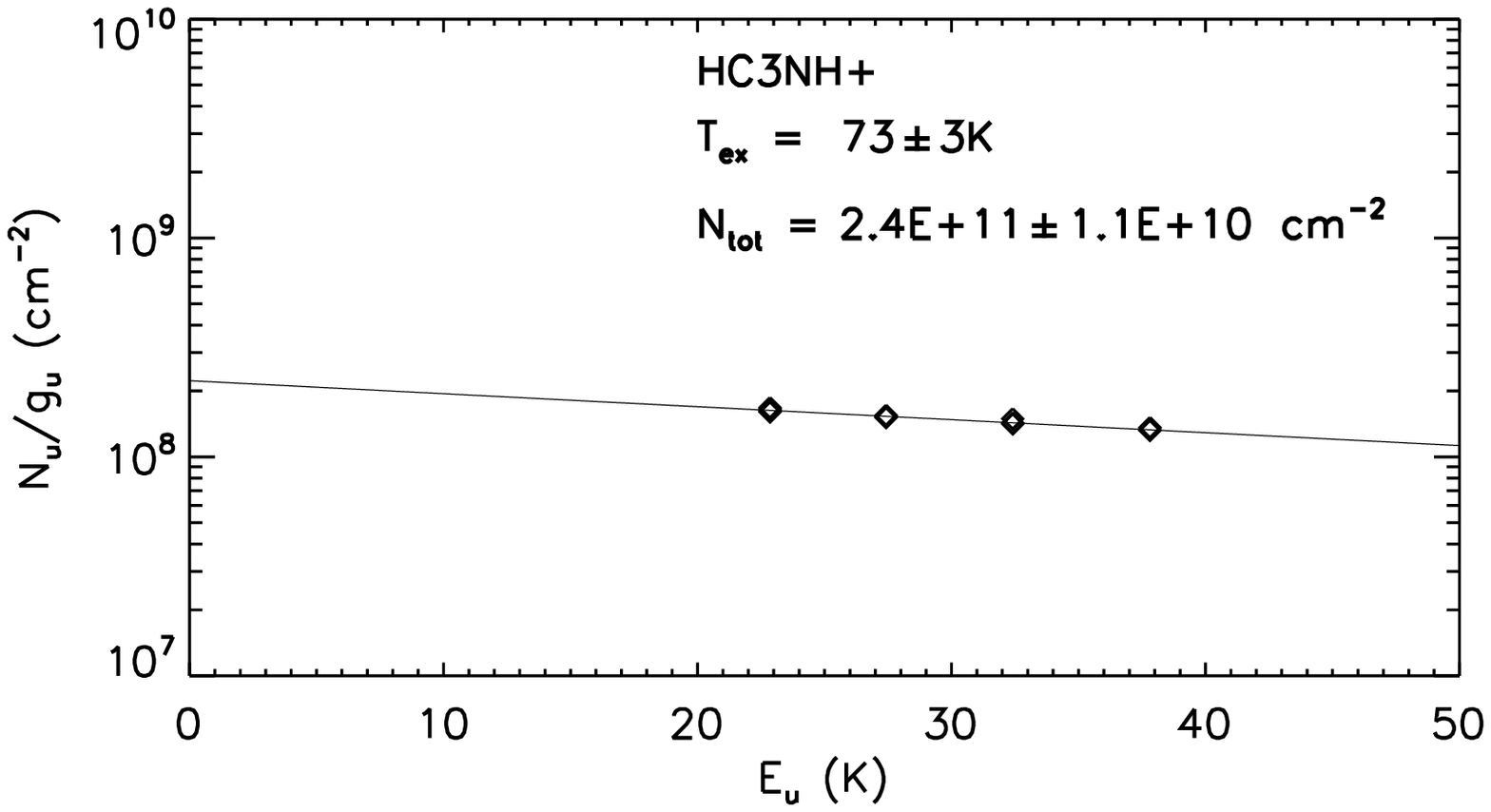} \\
         \end{tabular}
    \caption{Rotational diagrams for cyanides and isocyanides. We note that Model 4 was used to produce these diagrams.} 
    \label{fig:rot_diagrams}
\end{figure*}

From these rotational diagrams, excitation temperatures ($T_{Ex}$) and total column densities ($N_{TOT}$) can be determined for comparison to observations. Figure \ref{fig:rot_diagrams} shows the rotational diagrams that were produced. We note that the assumed desorption radius of the molecules studied here is on the order of the simulated beam size, and, as such, beam dilution effects are negligible. Table \ref{t:rt_coldens} shows the column densities obtained from each model presented here, along with the observational values, while Table \ref{t:rt_tex} shows the same for the excitation temperatures. We primarily focus on column densities going forward, both for consistency and because we feel that this comparison is more valid than simply comparing fractional abundances at one specific point. We note that column density ratios and $T_{Ex}$ values are plotted in Figures 12 and 13.

\begin{table*}
 \begin{center}
 \caption{Column densities (in cm$^{-2}$) for each molecule of interest in our observations and for each model, calculated using our radiative transfer model.}
 \label{t:rt_coldens}
 \vspace*{-2.0ex}
 \begin{tabular}{lllllllll}
 \hline\hline\\[-1.0em]
 \multicolumn{1}{l}{Molecule} & \multicolumn{1}{l}{Observations} & \multicolumn{1}{l}{Model 1} & \multicolumn{1}{l}{Model 2} & \multicolumn{1}{l}{Model 3} & \multicolumn{1}{l}{Model 4} & \multicolumn{1}{l}{Model 5} & \multicolumn{1}{l}{Model 6} & \multicolumn{1}{l}{Model 7}
 \\ \hline
    \ce{CH3CN} & \num{2.2e18} & \num{4.4e16} & \num{4.4e16} & \num{1.2e16} & \num{5.5e15} & \num{4.2e16} & \num{2.2e16} & \num{1.9e15}\\ 
    \ce{CH3NC} & \num{1.0e16} & \num{2.3e15} & \num{2.4e15} & \num{5.0e13} & \num{2.5e13} & \num{6.3e14} & \num{1.1e14} & \num{1.3e13}\\ 
    \ce{C2H5CN} & \num{6.2e18} & \num{1.1e17} & \num{1.1e17} & \num{6.8e15} & \num{1.1e15} & \num{1.4e17} & \num{6.0e15} & \num{3.3e11}\\
    \ce{C2H5NC} & $<$\num{1.5e15} & \num{1.9e13} & \num{1.9e13} & \num{1.5e12} & \num{9.8e10} & \num{1.1e11} & \num{2.2e11} & \num{7.9e4}\\
    \ce{C2H3CN} & \num{4.2e17} & \num{4.8e14} & \num{4.8e14} & \num{7.5e12} & \num{6.0e12} & \num{4.7e14} & \num{2.2e13} & \num{4.1e10}\\
    \ce{C2H3NC} & $<$\num{3.0e15} & \num{3.5e10} & \num{3.5e10} & \num{9.8e8} & \num{2.8e8} & \num{1.0e9} & \num{6.0e8} & \num{6.3e4}\\
    \ce{HC3N} & \num{3.5e17} & \num{5.0e16} & \num{5.0e16} & \num{1.2e15} & \num{5.7e14} & \num{6.5e15} & \num{6.6e14} & \num{6.0e14}\\
    \ce{HCCNC} & \num{5.1e14} & \num{1.7e15} & \num{1.7e15} & \num{1.5e14} & \num{4.0e13} & \num{5.6e14} & \num{1.9e14} & \num{4.9e13}\\
    \ce{HNCCC} & $<$\num{6.6e13} & \num{3.3e12} & \num{3.3e13} & \num{7.2e10} & \num{1.1e10} & \num{1.9e11} & \num{9.7e10} & \num{9.4e9}\\
    \ce{HC3NH+} & $<$\num{5.8e14} & \num{6.3e12} & \num{6.3e13} & \num{2.2e11} & \num{2.4e11} & \num{2.5e13} & \num{4.3e11} & \num{8.5e12}\\ \hline
 \end{tabular}
 \end{center}
 \vspace*{-2.5ex}
 \end{table*}
\begin{table*}
 \begin{center}
 \caption{Excitation temperatures (in K) for each molecule of interest in each model, calculated using our radiative transfer model.}
 \label{t:rt_tex}
 \vspace*{-2.0ex}
 \begin{tabular}{lllllllll}
 \hline\hline\\[-1.0em]
 \multicolumn{1}{l}{Molecule} & \multicolumn{1}{l}{Observations} & \multicolumn{1}{l}{Model 1} & \multicolumn{1}{l}{Model 2} & \multicolumn{1}{l}{Model 3} & \multicolumn{1}{l}{Model 4} & \multicolumn{1}{l}{Model 5} & \multicolumn{1}{l}{Model 6} & \multicolumn{1}{l}{Model 7}
 \\ \hline
    \ce{CH3CN} & \num{170} & \num{166} & \num{166} & \num{218} & \num{192} & \num{244} & \num{227} & \num{181}\\ 
    \ce{CH3NC} & \num{170} & \num{110} & \num{114} & \num{92} & \num{88} & \num{100} & \num{95} & \num{141}\\ 
    \ce{C2H5CN} & \num{150} & \num{207} & \num{207} & \num{203} & \num{202} & \num{208} & \num{203} & \num{157}\\
    \ce{C2H5NC} & \num{150} & \num{172} & \num{172} & \num{172} & \num{172} & \num{173} & \num{172} & \num{129}\\
    \ce{C2H3CN} & \num{200} & \num{121} & \num{121} & \num{100} & \num{99} & \num{171} & \num{103} & \num{157}\\
    \ce{C2H3NC} & \num{200} & \num{153} & \num{153} & \num{149} & \num{153} & \num{161} & \num{155} & \num{132}\\
    \ce{HC3N} & \num{170} & \num{51} & \num{51} & \num{35} & \num{92} & \num{19} & \num{36} & \num{37}\\
    \ce{HCCNC} & \num{170} & \num{95} & \num{95} & \num{96} & \num{95} & \num{114} & \num{120} & \num{186}\\
    \ce{HNCCC} & \num{170} & \num{75} & \num{75} & \num{64} & \num{69} & \num{88} & \num{66} & \num{154}\\
    \ce{HC3NH+} & \num{170} & \num{97} & \num{97} & \num{76} & \num{73} & \num{94} & \num{76} & \num{180}\\ \hline
 \end{tabular}
 \end{center}
 \vspace*{-2.5ex}
 \end{table*}

It is immediately noticeable that the column densities determined theoretically for most molecules are significantly lower than those determined observationally (Table \ref{t:coldens}). An exception is noted in \ce{HCCNC}, which is an order of magnitude higher in the radiative transfer model for Models 1 and 2, and is reproduced within a factor of a few for Models 3, 5, and 6.

The calculated excitation temperatures also exhibit some discrepancies in this radiative transfer model, though they are better reproduced than the column densities. \ce{CH3CN}, \ce{CH3NC}, \ce{C2H5CN}, \ce{C2H5NC}, \ce{C2H3CN}, \ce{C2H3NC}, and \ce{HCCNC} are reproduced within a factor of two in all models, and for many of these molecules the calculated $T_{Ex}$ agrees with the observational value within the error. However, the values for \ce{HC3N}, \ce{HNCCC}, and  \ce{HC3NH+} are significantly lower in most models than in the observations. This is most pronounced for \ce{HC3N}, which has a very low $T_{Ex}$ in most models (as low as 19~K in Model 5). However, Model 4 agrees within a factor of two with the observations, with a $T_{Ex}$ of 92~K. Generally speaking, \ce{HNC3} and \ce{HC3NH+} exhibit modest agreement with observations in models with higher $\zeta$, with Model 7 being the best for each species. Nevertheless, other models show low excitation temperatures for these molecules.

As mentioned previously, we are more interested in the ratios of related species. Table \ref{t:cd_ratios} contains the column density ratios determined from these radiative transfer calculations, along with the observational value. 
\begin{table*}
 \begin{center}
 \caption{Column density ratios for models, as well as the observational column density ratios.}
 \label{t:cd_ratios}
 \vspace*{-2.0ex}
 \begin{tabular}{lllllllll}
 \hline\hline\\[-1.0em]
 \multicolumn{1}{c}{Ratio} & \multicolumn{1}{c}{Observations} &
 \multicolumn{1}{c}{\hspace*{-2ex} Mod 1} &
 \multicolumn{1}{c}{\hspace*{-2ex} Mod 2} &
 \multicolumn{1}{c}{\hspace*{-2ex} Mod 3} &
 \multicolumn{1}{c}{\hspace*{-2ex} Mod 4} &
 \multicolumn{1}{c}{\hspace*{-2ex} Mod 5} &
 \multicolumn{1}{c}{\hspace*{-2ex} Mod 6} &
 \multicolumn{1}{c}{\hspace*{-2ex} Mod 7} \\
 \hline\\[-1.0em]
 CH$_3$NC/CH$_3$CN & \num{4.7e-3} & \hspace*{-2ex} \num{5.2e-2} & \hspace*{-2ex} \num{5.5e-2} & \hspace*{-2ex} \num{4.2e-3} & \hspace*{-2ex} \num{4.5e-3} & \hspace*{-2ex} \num{1.5e-2} & \hspace*{-2ex} \num{5.0e-3} & \hspace*{-2ex} \num{6.8e-3}\\[0.2ex]
 C$_2$H$_5$NC/C$_2$H$_5$CN & $< \num{2.4e-4}$ & \hspace*{-2ex} \num{1.7e-4} & \hspace*{-2ex} \num{1.7e-4} & \hspace*{-2ex} \num{2.2e-4} & \hspace*{-2ex} \num{8.9e-5} & \hspace*{-2ex} \num{7.9e-7} & \hspace*{-2ex} \num{3.7e-5} & \hspace*{-2ex} \num{2.4e-7}\\[0.2ex]
 C$_2$H$_3$NC/C$_2$H$_3$CN & $<$ \num{7.9e-5} & \hspace*{-2ex} \num{7.3e-5} & \hspace*{-2ex} \num{7.3e-5} & \hspace*{-2ex} \num{1.3e-4} & \hspace*{-2ex} \num{4.7e-5} & \hspace*{-2ex} \num{2.1e-6} & \hspace*{-2ex} \num{2.7e-5} & \hspace*{-2ex} \num{1.5e-6}\\[0.2ex]
 HCCNC/HC$_3$N & \num{1.5e-3} & \hspace*{-2ex} \num{3.4e-2} & \hspace*{-2ex} \num{3.4e-2} & \hspace*{-2ex} \num{1.2e-1} & \hspace*{-2ex} \num{7.0e-2} & \hspace*{-2ex} \num{8.6e-2} & \hspace*{-2ex} \num{2.9e-1} & \hspace*{-2ex} \num{8.2e-2}\\[0.2ex]
 \hline\\[-1.0em]
 HNC$_3$/HC$_3$N & $< \num{1.9e-4}$ & \hspace*{-2ex} \num{6.6e-5} & \hspace*{-2ex} \num{6.6e-5} & \hspace*{-2ex} \num{6.0e-5} & \hspace*{-2ex} \num{1.9e-5} & \hspace*{-2ex} \num{2.9e-5} & \hspace*{-2ex} \num{1.5e-4} & \hspace*{-2ex} \num{1.6e-5}\\[0.2ex]
 HC$_3$NH$^+$/HC$_3$N & $<$\num{1.7e-3} & \hspace*{-2ex} \num{1.3e-4} & \hspace*{-2ex} \num{1.3e-4} & \hspace*{-2ex} \num{1.8e-4} & \hspace*{-2ex} \num{4.2e-4} & \hspace*{-2ex} \num{3.8e-3} & \hspace*{-2ex} \num{6.5e-4} & \hspace*{-2ex} \num{1.4e-2}\\[0.2ex]
 \hline
 \end{tabular}
 \end{center}
 \vspace*{-2.5ex}
 \tablefoot{Mod 1 - standard model. Mod 2 - 3000K barrier model. Mod 3 - Variable $\zeta$ - low. Mod 4 - Variable $\zeta$ - med. Mod 5 - Variable $\zeta$ - high. Mod 6 - Med. constant $\zeta$. Mod 7 - High constant $\zeta$}
 \end{table*}
The column density results obtained from the radiative transfer modeling vary from model to model in most cases. For those column density ratios for which we only have an upper limit, there is generally good agreement with observations. For \ce{C2H5NC}:\ce{C2H5CN}, Models 1-7 all agree with the observationally determined upper limit of \num{2.4e-4}. All models also agree with the upper limit of \num{1.9e-4} for \ce{HNC3}:\ce{HC3N}. For \ce{C2H3NC}:\ce{C2H3CN}, Model 3 exhibits a column density ratio greater than the observational upper limit, while all other models are in agreement with the upper limit. For \ce{HC3NH+}:\ce{HC3N}, only Models 5 and 7 (those with very high $\zeta$) do not agree with the observational upper limit. 

It is perhaps more useful to compare the column density ratios of \ce{CH3NC}:\ce{CH3CN} and \ce{HCCNC}:\ce{HC3N}, since these pairs have observationally defined values of \num{4.7e-3} and \num{1.5e-3} respectively. \ce{CH3NC}:\ce{CH3CN} is remarkably well-produced by some of the radiative transfer models presented here. For example, Models 3, 4, 6, and 7 are all within a factor of two of the observational value, with Model 4 being almost an exact match. These results provide evidence that a $\zeta$ higher than the canonical value of \num{1.3e-17} s$^{-1}$ is needed to re-produce the \ce{CH3NC}:\ce{CH3CN} ratio for Sgr B2(N2). 

The agreement for \ce{HCCNC}:\ce{HC3N} is generally much poorer across the models than for \ce{CH3NC}:\ce{CH3CN}. All radiative transfer models overproduce this ratio by at least an order of magnitude, with Models 1 and 2 showing the best-fit value of \num{3.4e-2}, still a factor of approximately 23 too high. This systematic over-production of \ce{HCCNC} relative to \ce{HC3N} for all models, which was also seen in the fractional abundance values from Table \ref{t:tex_abuns}, likely means that there is something missing from our network in regards to the chemistry of these molecules. Since Model 4 provides such a good fit to \ce{CH3NC}:\ce{CH3CN}, and is consistent with all upper limits, we show the rotational diagrams for that model as a sample of the output from our radiative transfer modeling in Figure \ref{fig:rot_diagrams}.

\section{Discussion}
\label{s:discussion}

\subsection{\ce{H + CH3NC} reaction}
The results of Models 1 and 2 show that varying the barrier of Reaction \ref{eq:3} has a significant impact on the fractional abundance of \ce{CH3NC}. A lower barrier for this reaction means that the abundance of \ce{CH3NC} falls off at earlier times and temperatures (see Figure \ref{fig:mod1}). Higher values see a less significant decrease in abundance (Figures 2 and 3). The existence of this reaction also means that the abundance of \ce{CH3NC}, and the ratio of \ce{CH3NC}:\ce{CH3CN} is moderately temperature sensitive. In light of the recent quantum chemical calculations of \citet{Nguyen19} which show a barrier closer to that selected in Model 1, it is possible that the \ce{CH3NC}:\ce{CH3CN} ratio could be used as a diagnostic for temperature. The barrier of Reaction 3 does not appear to affect any other species studied here.

\subsection{Effects of changing $\zeta$}
A detailed investigation of the effects of changing $\zeta$ on the chemistry of complex cyanides, as well as other complex molecules, is beyond the scope of this paper. However, it is clear from the models presented here that the effect is complex and nonlinear. Models 3-7 are those models for which $\zeta$ is changed compared to Model 1. For the purposes of this comparison, we first focus on the three models with extinction-dependent $\zeta$ (Models 3, 4, and 5), as well as Model 1.

Most species experience relatively complex behavior with increasing $\zeta$. The exceptions to this are \ce{C2H5NC} and \ce{C2H3NC}, which have peak abundances that decrease monotonically with increasing $\zeta$. These two species are efficiently destroyed by cosmic rays, as well as smaller radicals on the grain surface that are produced in larger abundances due to higher $\zeta$, such as \ce{OH}.

All other species of interest exhibit nonmonotonic behavior with increasing $\zeta$. All species demonstrate lower final abundances when going from Model 1 to Model 3. Most species then exhibit still-lower final abundances when $\zeta$ is increased further, from Model 3 to Model 4. However, \ce{CH3CN} and \ce{HCCNC} remain relatively flat in both peak and final abundance, exhibiting no real change. 

Conversely, going from Model 4 to Model 5, most species exhibit enhancements in peak and final abundances, with the exceptions being the aforementioned \ce{C2H5NC} and \ce{C2H3NC}. In fact, \ce{CH3CN} actually exhibits its highest peak and final abundance out of any model in this study in Model 5. It is formed more efficiently on grains in this model, as a result of the availability of more \ce{CH3} and \ce{CN} radicals, as well as a larger abundance of \ce{CH2CN} on the grains.

Model 6, as mentioned above, exhibits remarkably similar abundance profile shapes to Model 4, though the peak and final abundances are different, and in many cases higher. This is perhaps unsurprising, as Model 6 has a constant $\zeta$ of \num{1e-16} s$^{-1}$, which is the lower bound of the $\zeta$-profile for Model 4 (Figure \ref{fig:zeta_av}). However, it is difficult to disentangle the effects of having an $A_V$-dependent cosmic-ray ionization rate from the absolute magnitude of the rate.

It is also interesting to investigate the chemical behavior in Model 7. Since Model 7 has the highest $\zeta$ of all models presented in this paper, we first compare the fractional abundances to those of Model 5, which has the next-highest average $\zeta$. Comparing Model 7 to Model 5, half of the species (\ce{HCN, CH3NC, C2H5CN, HC3N, and HCCNC}) demonstrate higher final abundances in Model 7, while the other half exhibit lower final abundances.

In addition to changing the peak and final abundances of species,  changing $\zeta$ can also affect the shapes of the abundance profiles for many species, as evidenced by the fact that models with higher $\zeta$ show that \ce{CH3NC} and \ce{HNC} drop off at lower temperatures than in the standard model (Model 1).

Another important result to note is that incorporating an $A_V$-dependent $\zeta$ profile into astrochemical models has an important impact on the chemistry of star-forming regions. Although Model 6 exhibits similarities to Model 4, the models with constant $\zeta$ throughout the source exhibit different behavior from those with extinction-dependent $\zeta$. This can be seen when inspecting both the fractional abundance ratios in Table \ref{t:tex_abuns} and the column density ratios in Table \ref{t:cd_ratios}.

The effect is perhaps most pronounced when looking at \ce{HC3NH+}:\ce{HC3N}. Since \ce{HC3NH+} is a molecular ion that is produced in larger abundances with a very high $\zeta$, it is a good probe to investigate the effects of $\zeta$ on chemistry. In this case, models with a constant and elevated $\zeta$ throughout the source appear to produce more \ce{HC3NH+} than those with an attenuated $\zeta$. This can be seen when comparing the column density ratios of Model 5 and Model 7 in Table \ref{t:cd_ratios}. Both values are higher than the observational upper limit, but the ratio for Model 7 is significantly higher (\num{1.4e-2} vs. \num{3.8e-3}). This is caused by the lack of attenuation throughout the model, thus producing a higher $\zeta$ in the interior of the source. In fact, significant differences are seen for most column density ratios when comparing Model 5 to Model 7. Another example is the abundance of \ce{C2H5CN}, which is much lower in Model 7 due to the lack of attenuation. This highlights the need to effectively model the attenuation of $\zeta$ throughout astrochemical models, since it has such a significant impact on the chemistry. In fact, recent work from \citet{gaches19} has also shown that extinction-dependent $\zeta$ models are essential. 

We have shown that the behavior of molecules as $\zeta$ is altered is complex and often nonmonotonic. Since the focus of this paper is on developing the chemical network to include more complex isocyanides, we do not perform an in-depth analysis of the effect of changing $\zeta$ on cyanides and isocyanides. This will be left for future work. Other work is currently being done to investigate the effect of $\zeta$ on other complex organics as well \citep{bargergarrod}.

\subsection{Comparison of observations to models}

As noted in Sect. 6.3, the theoretical column densities obtained from our chemical models using rotational diagrams differ significantly from the observational values, and are in many cases orders of magnitude lower. These discrepancies could be due to a number of factors. It is possible that the physical profiles (density and temperature) that are being used in the spectroscopic model are not physically representative of the actual source. On very small scales, the hot core most likely exhibits nonuniformities in its physical structure that are not taken into account here. This sort of structure has been evidenced by the aforementioned recent detection of multiple new hot cores in Sgr B2(N) \citep{bonfand17}, as well as recent studies of Orion KL, to name one other hot-core source \citep{wright17}. In addition to this, the structure of the cloud is more complicated than we assume here, with evidence of filaments that converge toward the main hot core, as suggested recently by \citet{schworer19}.

It is worth noting that when calculating the \ce{H2} column density using these physical profiles based on a simple pencil-beam calculation, a value of $\sim$\num{2.5e24} cm$^{-2}$ is obtained, which is within a factor of two of the observational value of \num{1.4e24} cm$^{-2}$ \citep{bonfand19}. This seems to indicate that the lower column densities for these complex molecules are a result of either the chemistry or the radiative transfer calculation itself. Another possibility is that the assumptions of spherical symmetry and a power-law density profile used to construct the physical profile, which are simplifications, could lead to large discrepancies in the observed and calculated column densities.

It is also possible that chemical factors lead to these differences, though it is unlikely that a molecule as well-studied as \ce{CH3CN} would be underproduced by two orders of magnitude by purely chemical inaccuracies. Investigating fractional abundance and column density {ratios} provides a way to remove some of the inaccuracies from the physical model.

It is interesting to note the change that occurs in the R-NC:R-CN ratio when going from the fractional abundance ratios to the column density ratios derived from the radiative transfer model. For example, the \ce{CH3NC}:\ce{CH3CN} and \ce{HCCNC}:\ce{HC3N} ratios both increase significantly in the rotational diagrams. This is because the rotational diagram method includes a range of the abundance profiles of the species in the calculation. As mentioned previously, we integrate out to a radius of $\sim$\num{8e4} au, or a temperature of $\sim$50~K, in our chemical model. In both of these cases, at lower temperatures, the ratios between the NC and CN molecules are higher, and thus the convolved ratios are higher than the values at a specific temperature. The temperature dependence of these ratios is something that could potentially be very useful as an observational probe in the future. 

In general, it is useful to use both methods of observational comparison. Since ALMA is not sensitive to the large-scale emission included in our radiative transfer models, the column densities we obtain from our models are not directly comparable to those obtained with ALMA, though we have attempted to minimize this issue by only integrating out to ALMA's maximum recoverable scale in our calculations. However, simply comparing a fractional abundance ratio at a specific temperature is not robust on its own either, particularly for species that experience large abundance variations with temperature, such as \ce{CH3NC}. Therefore, using both methods allows us to a get a better idea for where our models agree with observations and where improvement is needed. 

Since the chemistry of \ce{CH3CN} is the best-understood of the complex molecules studied here, it is useful also to look at the column densities of species with respect to \ce{CH3CN}, as this can provide a clue as to which are least-well reproduced in our models. This of course assumes that the chemistry that we have for \ce{CH3CN} is correct in our network. Table \ref{t:rad_transfer} shows observational and theoretical column densities with respect to \ce{CH3CN} for all models.

\begin{table*}
 \begin{center}
 \caption{Observational and theoretical column densities with respect to \ce{CH3CN}.}
 \label{t:rad_transfer}
 \vspace*{-2.0ex}
 \begin{tabular}{lllllllll}
 \hline\hline\\[-1.0em]
 \multicolumn{1}{l}{Molecule} & \multicolumn{1}{l}{Observations} & \multicolumn{1}{l}{Mod. 1} & \multicolumn{1}{l}{Mod. 2} & \multicolumn{1}{l}{Mod. 3} & \multicolumn{1}{l}{Mod. 4} & \multicolumn{1}{l}{Mod. 5} & \multicolumn{1}{l}{Mod. 6} & \multicolumn{1}{l}{Mod. 7}
 \\ \hline
    \ce{CH3CN} & 1.0 & 1.0 & 1.0 & 1.0 & 1.0 & 1.0 & 1.0 & 1.0 \\
    \ce{CH3NC} & \num{4.7e-3} & \num{5.2e-2} & \num{5.5e-2} & \num{4.2e-3} & \num{4.5e-3} & \num{1.5e-2} & \num{5.0e-3} & \num{6.8e-3}\\ 
    \ce{C2H5CN} & 2.8 & 2.5 & 2.5 & \num{5.7e-1} & \num{2.0e-1} & 3.3 & \num{2.7e-1} & \num{1.7e-4}\\
    \ce{C2H5NC} & $<$\num{6.8e-4} & \num{4.3e-4} & \num{4.3e-4} & \num{1.2e-4} & \num{1.8e-5} & \num{2.6e-6} & \num{1.0e-5} & \num{4.2e-11}\\
    \ce{C2H3CN} & \num{1.9e-1} & \num{1.1e-2} & \num{1.1e-2} & \num{6.2e-4} & \num{1.1e-3} & \num{1.1e-2} & \num{1.0e-3} & \num{2.2e-5}\\
    \ce{C2H3NC} & $<$\num{1.4e-3} & \num{8.0e-7} & \num{8.0e-7} & \num{8.2e-8} & \num{5.1e-8} & \num{2.4e-8} & \num{2.7e-8} & \num{3.3e-11}\\
    \ce{HC3N} & \num{1.6e-1} & 1.1 & 1.1 & \num{1.0e-1} & \num{1.0e-1} & \num{1.5e-1} & \num{3.0e-2} & \num{3.2e-1}\\
    \ce{HCCNC} & \num{2.3e-4} & \num{3.9e-2} & \num{3.9e-2} & \num{1.2e-2} & \num{7.3e-3} & \num{1.3e-2} & \num{8.6e-3} & \num{2.6e-2}\\
    \ce{HNC3} & $<$\num{3.0e-5} & \num{7.5e-5} & \num{7.5e-5} & \num{6.0e-6} & \num{2.0e-6} & \num{4.5e-6} & \num{4.4e-6} & \num{4.9e-6}\\ 
    \ce{HC3NH+} & $<$\num{2.6e-4} & \num{1.4e-4} & \num{1.4e-4} & \num{1.8e-5} & \num{4.4e-5} & \num{6.0e-4} & \num{2.0e-5} & \num{4.5e-3}\\ \hline
 \end{tabular}
 \end{center}
 \vspace*{-2.5ex}
 \end{table*}

From looking at Table \ref{t:rad_transfer}, it can be seen that we produce column densities with respect to \ce{CH3CN} that are consistent with the upper limits for most species that were not detected. Exceptions are seen for \ce{HNC3} in Models 1 and 2, and once again for \ce{HC3NH+}, which is overproduced in models with very high $\zeta$. Although it is good that we are consistent with these upper limits, no significant further information on the chemistry can be determined from these ratios; detections of these species would allow us to make further constraints.

Regarding those species that are firmly detected, Models 1, 2, and 5 do a reasonably good job of reproducing the ratio of \ce{C2H5CN}:\ce{CH3CN}, and Models 3-5 do the best job at reproducing \ce{HC3N}. The ratio of \ce{HCCNC} to \ce{CH3CN} is consistently higher in our models, once again indicating that we are systematically overproducing \ce{HCCNC}. This is something that will be investigated in future studies. The opposite problem exists for \ce{C2H3CN}, which appears to be systematically underproduced relative to \ce{CH3CN} in our models. Overall, there appears to be a significant amount of work remaining on constraining the chemistry of the cyanides and isocyanides in hot-core models. 

\begin{figure*}[htb]
    \centering
    \includegraphics[width=.94\textwidth]{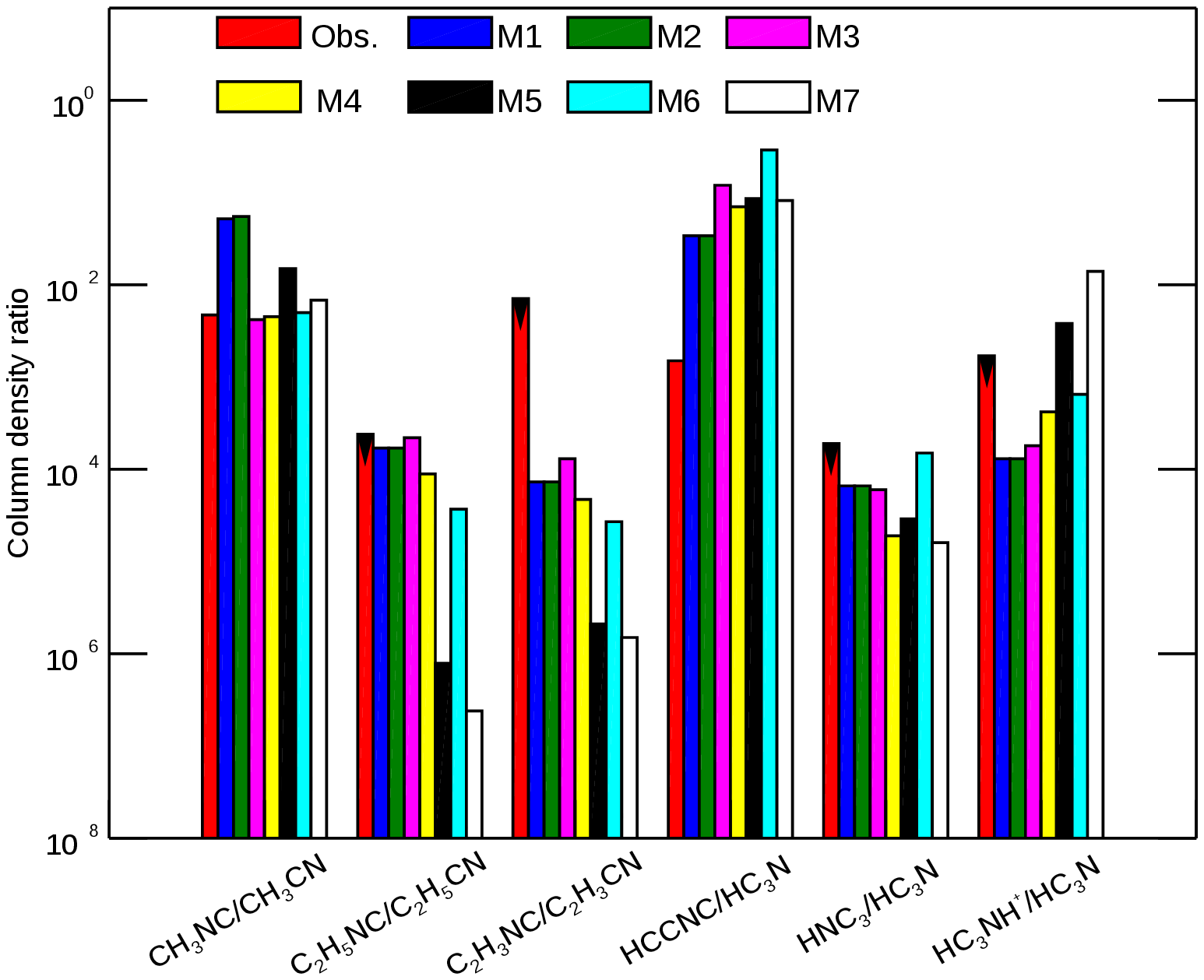}
    \caption{Comparison of column density ratios between observations and models.} 
    \label{fig:cd_comparison}
\end{figure*}

\begin{figure*}[htb]
    \centering
    \includegraphics[width=.94\textwidth]{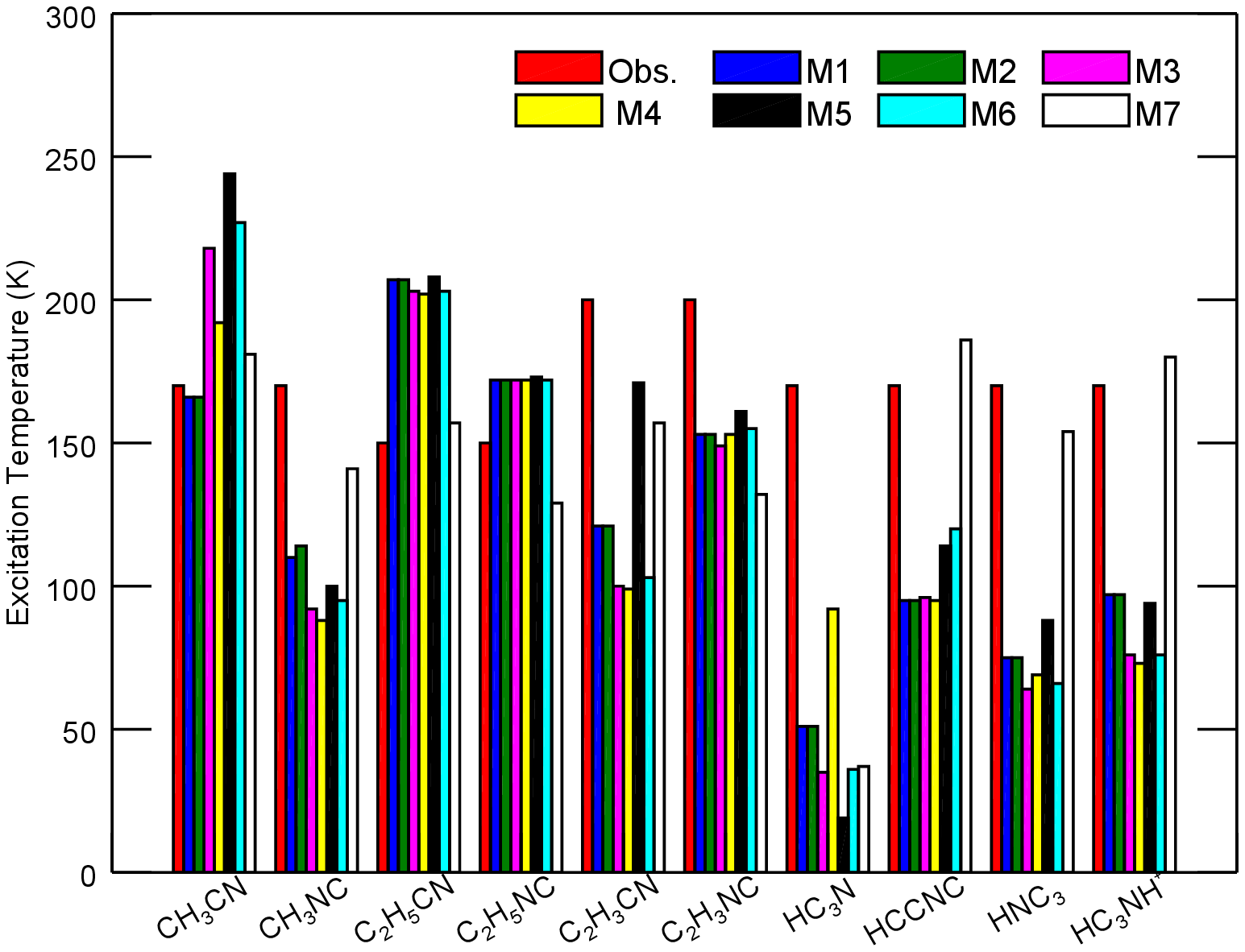}
    \caption{Comparison of excitation temperatures between observations and models.} 
    \label{fig:tex_comparison}
\end{figure*}

Based on all of the results presented here, it appears that an enhanced cosmic-ray ionization rate does the best job at reproducing the observational results towards Sgr B2(N2), particularly when investigating the column density ratios obtained from our chemical models. Model 4 is a very good match to the observational column density ratio of \ce{CH3NC}:\ce{CH3CN}, and is also consistent with all upper limits for ratios we do not have definitive values for. However, it does a poor job of reproducing the \ce{HCCNC}:\ce{HC3N} ratio. Nevertheless, this can be said of all the models, and is likely a result of some systematic inaccuracy in the chemical network for these species, most likely for \ce{HCCNC}. Model 4 has an $A_V$-dependent $\zeta$ which varies from $\sim$\num{2.0e-15} s$^{-1}$ to $\sim$\num{1.0e-16} s$^{-1}$. This result qualitatively agrees with the result of \citet{bonfand19}, which showed that chemical models with enhanced $\zeta$ more accurately reproduced the observations for several of the hot cores in Sgr B2(N). Therefore, we believe that enhanced, extinction-dependent cosmic-ray ionization rates should be considered in all models of the chemistry of Sgr B2(N) in the future. 

Figures \ref{fig:cd_comparison} and \ref{fig:tex_comparison} summarize the results of the modeling efforts presented here, showing a comparison between the observational and theoretical column density ratios and excitation temperatures. Regarding the excitation temperatures, many chemical species are reproduced within a factor of approximately two or better for most of the molecules, particularly those molecules traditionally associated with hot cores. However, many of the smaller molecules (\ce{HC3N, HCCNC, and HC3NH+}) have much poorer agreement, with theoretical excitation temperatures much lower or higher than the observational values. These discrepancies in $T_{Ex}$ could be related to an inaccurate physical profile, similarly to the inaccuracies in column densities. However, there are also potential contributions from inaccurate binding energies for many of these molecules, particularly the isocyanides, as there are not many data in the literature for these species. Yet another potential explanation for this is the assumption of LTE in our radiative transfer model. The density at the radius for which we truncate our radiative transfer calculations ($\sim$\num{8e4} au) is $\sim$\num{3e5} cm$^{-3}$, which may not be high enough to thermalize some transitions we are simulating. This would preferentially weigh low-temperature material in the model integration, leading to lower $T_{Ex}$ values. However, this would also tend to overestimate our calculated column densities, which are already too low. The discrepancies between theoretical and observational values also appear to be related to having too few lines available for some species in the wavelength range of ALMA Band 3. Therefore, it is difficult to tell at this point what is causing the disagreement between observations and our models, but it is likely a combined effect.
 
 \subsection{Comparison of Sgr B2(N2) to other sources}
\label{ss:literature}

\begin{sidewaystable}
 \begin{center}
 \caption{Column density ratios in different sources.}
 \label{t:ratios}
 \vspace*{-2.0ex}
 \begin{adjustbox}{width=\textwidth}
 \begin{tabular}{lllllllllllll}
 \hline\hline\\[-1.0em]
 \multicolumn{1}{c}{Column density} & \multicolumn{1}{c}{Sgr~B2 (N2)\tablefootmark{a}} &
 \multicolumn{1}{c}{Sgr~B2\tablefootmark{b}} & \multicolumn{1}{c}{\hspace*{-2ex} Orion~KL\tablefootmark{c}} & \multicolumn{1}{c}{\hspace*{-2ex} Orion~KL\tablefootmark{c}} & \multicolumn{1}{c}{\hspace*{-2ex} Orion~KL\tablefootmark{c}} & \multicolumn{1}{c}{\hspace*{-2ex} Orion~KL\tablefootmark{d}} & \multicolumn{1}{c}{IRAS16293A\tablefootmark{e}} & \multicolumn{1}{c}{IRAS16293B\tablefootmark{e}} & \multicolumn{1}{c}{TMC-1} & \multicolumn{1}{c}{L1544} & \multicolumn{1}{c}{L483\tablefootmark{f}}& \multicolumn{1}{c}{Horsehead} \\ 
 \multicolumn{1}{c}{ratio} & \multicolumn{1}{c}{} & \multicolumn{1}{c}{} & \multicolumn{1}{c}{\hspace*{-2ex} EtCN3} & \multicolumn{1}{c}{\hspace*{-2ex} EtCN1} & \multicolumn{1}{c}{\hspace*{-2ex} EtCN2} & \multicolumn{1}{c}{\hspace*{-2ex} (IRAM 30\,m)} & \multicolumn{1}{c}{} & & & \multicolumn{1}{c}{} & \multicolumn{1}{c}{} & \multicolumn{1}{c}{PDR} \\
 \hline\\[-1.0em]
 CH$_3$NC/CH$_3$CN & \num{4.7e-3} & \hspace*{-2ex} 0.02 & \hspace*{-2ex} \num{3e-3}$^\star$ & \hspace*{-2ex} $< \num{1.4e-3}$ & \hspace*{-2ex} \num{6e-4}$^\star$ & \hspace*{-2ex} \num{2(1)e-3}$^\star$ & $<$ \num{1.8e-4} & \num{5e-3} & $<$ 0.09\tablefootmark{g} & -- & \hspace*{-2ex} 0.09 & 0.15(2)\tablefootmark{n} \\[0.2ex]
 C$_2$H$_5$NC/C$_2$H$_5$CN & $< \num{2.4e-4}$ & -- & \hspace*{-2ex} $< \num{2.5e-3}$ & \hspace*{-2ex} $< \num{7e-3}$ & \hspace*{-2ex} $< \num{7e-3}$ & \hspace*{-2ex} $<$ \num{3(2)e-3} & -- & -- & -- &  -- & -- & -- \\[0.2ex]
 C$_2$H$_3$NC/C$_2$H$_3$CN & $< \num{7.1e-3}$  & -- & \hspace*{-2ex} $< 0.025$ & \hspace*{-2ex} $< 0.02$ & \hspace*{-2ex} $< 0.05$ & \hspace*{-2ex} $<$ 0.10(5) & -- & -- & -- & -- & -- & -- \\[0.2ex]
 HCCNC/HC$_3$N & \num{1.5e-3} & -- & -- & \hspace*{-2ex} -- & \hspace*{-2ex} -- & \hspace*{-2ex} $<$ \num{8(4)e-4} & -- & -- & 0.018(2)\tablefootmark{h,k} & 0.05--0.09\tablefootmark{l,m} & \hspace*{-2ex} 0.014 & $<$ 0.1\tablefootmark{n} \\[0.2ex]
 \hline\\[-1.0em]
 HNC$_3$/HC$_3$N & $< \num{1.9e-4}$ & -- & --  & \hspace*{-2ex} -- & \hspace*{-2ex} -- & \hspace*{-2ex} $<$ \num{8(4)e-4} & -- & -- & \num{2.4(4)e-3}\tablefootmark{i,k} & $5-8 \times 10^{-3}$\tablefootmark{l,m} & \hspace*{-2ex} \num{1.2e-3} & -- \\[0.2ex]
 HC$_3$NH$^+$/HC$_3$N & $< \num{1.7e-3}$  & -- & --  & \hspace*{-2ex} -- & \hspace*{-2ex} -- & \hspace*{-2ex} -- & -- & -- & \num{6(1)e-3}\tablefootmark{j,k} & \num{8.4(42)e-3}\tablefootmark{m} & \hspace*{-2ex} \num{5.5e-3} & -- \\
 \hline
 \end{tabular}
 \end{adjustbox}
 \end{center}
 \vspace*{-2.5ex}
 \tablefoot{Numbers in parentheses represent uncertainties in unit of the last 
 digit. Asterisks mark detections reported in the literature that we consider 
 as only tentative after inspection of the published spectra and fits. 
 References:
 \tablefoottext{a}{this work;}
 \tablefoottext{b}{\citet{remijan05};}
 \tablefoottext{c}{\citet{Margules18};}
 \tablefoottext{d}{\citet{Lopez14};}
 \tablefoottext{e}{\citet{calcutt18};}
 \tablefoottext{f}{\citet{agundez19};}
 \tablefoottext{g}{\citet{Irvine84};}
 \tablefoottext{h}{\citet{Kawaguchi92a};}
 \tablefoottext{i}{\citet{Kawaguchi92b};}
 \tablefoottext{j}{\citet{Kawaguchi94};}
 \tablefoottext{k}{\citet{Takano98};}
 \tablefoottext{l}{\citet{Vastel18};}
 \tablefoottext{m}{\citet{Quenard17};}
 \tablefoottext{n}{\citet{Gratier13}.}
 }
 \end{sidewaystable}

The observational column density ratios obtained for the pairs of cyanides/isocyanides or 
their upper limits toward Sgr~B2(N2) are compared to the ratios reported in the literature for a separate study of Sgr~B2(N) sensitive to larger scales, as well as for the hot core Orion~KL, the low-mass protostellar binary IRAS 16293-2422(A and B), the low-mass protostar L483, the dark cloud TMC-1, the prestellar core L1544, and the Horsehead photodissociation region (PDR) in Table~\ref{t:ratios}. The column density ratios reported for Sgr B2(N2) in this table account for the contribution of vibrationally excited states to the total partition function of the molecules, as in Table~\ref{t:coldens}. This is also the case for the data for IRAS 16293A and IRAS16293B, but may not be true for the other values reported in the literature.

It is interesting to note that the observed ratio of \ce{CH3NC}:\ce{CH3CN} varies significantly from source to source. Table \ref{t:ratios} shows a significant range of values, from 0.15 for the Horsehead PDR, to an upper limit of \num{1.8e-4} for IRAS16293A. There are many factors that could contribute to this difference. For example, these differences may be due to differing kinetic temperatures, differing UV fields, or differing cosmic-ray ionization rates. We note that the chemical models presented here (with the exception of Model 2) also predict a significant change in \ce{CH3NC}:\ce{CH3CN} ratio with temperature. It is likely that it is a combination of all parameters. The observations of \citet{remijan05} toward Sgr~B2 is a factor of approximately four higher than what we determine for Sgr~B2(N2). This is qualitatively consistent with the models we present here. \citet{remijan05} did not detect compact emission of \ce{CH3NC} from the hot core with BIMA. They instead detected extended emission in the cloud with the GBT. Our chemical models show a higher \ce{CH3NC}:\ce{CH3CN} ratio in the low-temperature regions of the cloud, which is what their GBT observations probe.

\citet{Gratier13} determined a high abundance ratio in the Horsehead PDR, but the excitation temperatures derived for \ce{CH3CN} in their work were $\sim$30-40~K, which is much lower than what we have derived here. In addition, the Horsehead PDR also has a significantly greater UV flux. So it is likely that both effects combine in this case. In the case of UV, these results appear to highlight the fact that UV and cosmic-ray chemistry behave differently, as it is shown that high UV flux appears to increase the \ce{CH3NC}:\ce{CH3CN} ratio, whereas higher $\zeta$ appears to decrease it. Differences in UV and cosmic-ray chemistry have been noted before. An example of this is the case of \ce{ArH+}, which has been shown to be formed from cosmic-ray-induced processes, but not from UV processes, albeit in diffuse atomic hydrogen environments that are quite different from the denser regions discussed here \citep{schilke14}.

It is also instructive to compare the \ce{HCCNC}:\ce{HC3N} and \ce{HNC3}:\ce{HC3N} ratios between sources. Observations of the cold regions TMC-1, L1544, and L483 reveal a \ce{HCCNC}:\ce{HC3N} ratio that is about an order of magnitude higher than Sgr B2(N2). This is in better agreement with our chemical model predictions for these species, albeit in a much colder environment than what we are modeling. This could indicate that we are missing some important temperature-sensitive reactions for the formation and destruction of these species. The same can be shown when looking at the \ce{HNC3}:\ce{HC3N} ratio, which is also about an order of magnitude higher in TMC-1, L1544, and L483 than in Sgr B2(N2). However, this higher ratio is in worse agreement with our models, contrary to the behavior shown in \ce{HCCNC}:\ce{HC3N}. It is clear that there are key ingredients missing from the chemical network in regards to these smaller cyanides and isocyanides.

 \section{Conclusion}
Here, we present a joint observational and modeling effort aimed at studying the chemistry of complex isocyanides in Sgr B2(N2). This is, to our knowledge, the most comprehensive effort aimed at understanding the chemistry of these species in the literature to date. We introduce a new, single-stage chemical model that combines the traditional two stages of a hot-core chemical model (collapse and warm-up) into a single concerted phase. We also introduce a visual extinction-dependent cosmic-ray ionization rate into hot-core chemical models.
Several new species and reactions were added to our chemical network, including \ce{C2H5NC}, \ce{C2H3NC}, and related radicals, and our models were compared with observations from the EMoCA survey. Our main conclusions are summarized below.

\begin{enumerate}
    \item We report tentative detections of \ce{CH3NC} and \ce{HCCNC} toward Sgr B2(N2) for the first time, with abundance ratios of \ce{CH3NC}:\ce{CH3CN} $\sim$ \num{5e-3} and \ce{HCCNC}:\ce{HC3N} $\sim$ \num{1.5e-3}. In addition, we calculate upper limits for \ce{C2H5NC}, \ce{C2H3NC}, \ce{HNC3}, and \ce{HC3NH+}.
    \item Using a variable and higher cosmic-ray ionization rate has a complex effect on the chemistry of the cyanides and isocyanides. Incorporating this into our chemical network increases agreement for some molecular ratios (\ce{CH3NC}:\ce{CH3CN}) to a point, but models with very high $\zeta$ do not show as good an agreement. The impact of changing $\zeta$ must be studied in greater detail.
    \item The best agreement with observations is reached using an enhanced, extinction-dependent cosmic-ray ionization rate, which is in line with other observational and modeling studies of Sgr B2(N2). Model 4 reproduces the ratio of \ce{CH3NC}:\ce{CH3CN} almost exactly when considering the theoretical column density ratios, and is also consistent with all upper limits. Models with high, constant $\zeta$ do not  reproduce observations particularly well. This highlights the need for extinction-dependent $\zeta$ profiles in chemical models.
    \item The \ce{HCCNC}:\ce{HC3N} ratio is too high across all models presented here. This appears to be due to a systematic overproduction of \ce{HCCNC}, as the ratio of \ce{HCCNC}:\ce{CH3CN} is also too high. This overproduction will be a topic of further study. 
    \item Molecular radiative transfer calculations that take account of source structure show that the column densities produced are multiple orders of magnitude too low in some cases. This implies that we are not producing enough complex molecules in our models, or that the physical profile we are using is inaccurate. Excitation temperatures are reproduced well for some species, especially the classic ``hot core'' molecules. Smaller molecules with few lines at Band 3 wavelengths have the worst agreement with observational $T_{Ex}$ values.
\end{enumerate}
    
From a chemical perspective, there is still much work to be done on the chemistry of the isocyanides. Although the model results are in agreement with the upper limits for the species that have not been firmly detected, this is not very constraining for the models. Regarding those species that have been tentatively detected, the abundance of \ce{CH3NC} is very dependent on the barrier used in the \ce{H + CH3NC} reaction, which is better constrained now due to the calculations of \citet{Nguyen19}, but still has uncertainty associated with it. Further observations are needed in different sources in order to constrain the abundance ratios of the isocyanides and cyanides, and experimental efforts and quantum chemical calculations will be invaluable in investigating the chemistry of these molecules across different physical environments. 

\begin{acknowledgements}
This paper makes use of the following ALMA data: 
ADS/JAO.ALMA\#2011.0.00017.S, ADS/JAO.ALMA\#2012.1.00012.S. 
ALMA is a partnership of ESO (representing its member states), NSF (USA) and 
NINS (Japan), together with NRC (Canada), NSC and ASIAA (Taiwan), and KASI 
(Republic of Korea), in cooperation with the Republic of Chile. The Joint ALMA 
Observatory is operated by ESO, AUI/NRAO and NAOJ. The interferometric data 
are available in the ALMA archive at https://almascience.eso.org/aq/. 
This work has been in part supported by the Collaborative Research Centre 956, sub-project B3, funded by the Deutsche Forschungsgemeinschaft (DFG) - project ID 184018867.
\end{acknowledgements}

\onecolumn
\begin{appendix}
\label{Appendix}
\section{Complementary observational figures and tables}

\begin{figure*}[!h]
\centerline{\resizebox{\hsize}{!}{\includegraphics[angle=0]{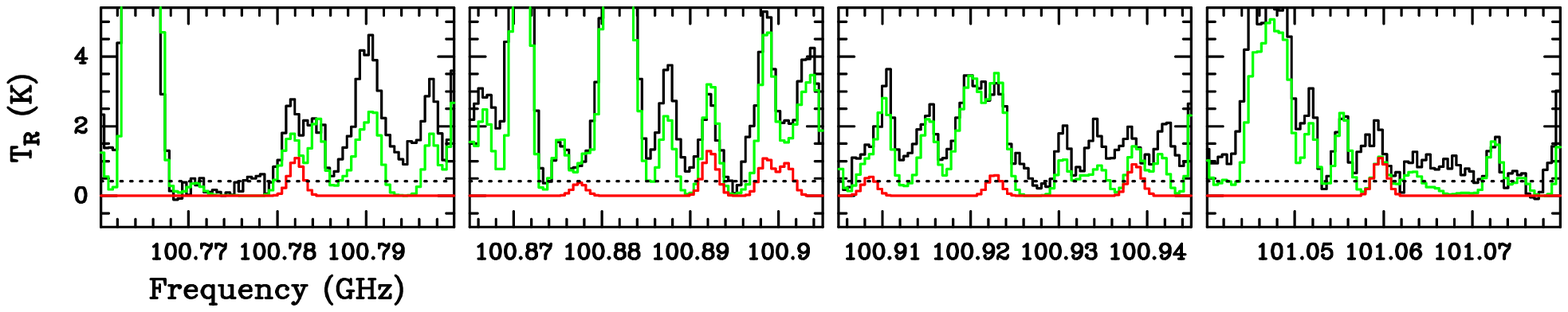}}}
\caption{Same as Fig.~\ref{f:spec_ch3nc_ve0} for CH$_3$NC, $\varv_8 = 1$.
}
\label{f:spec_ch3nc_ve1}
\end{figure*}

\begin{figure*}[!h]
\centerline{\resizebox{\hsize}{!}{\includegraphics[angle=0]{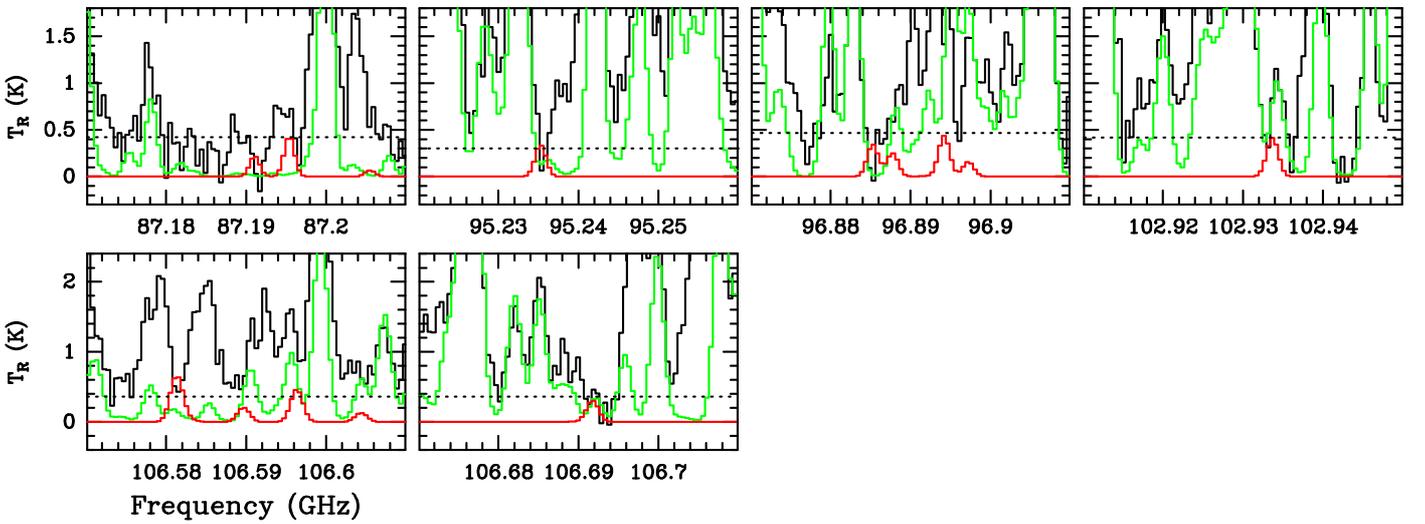}}}
\caption{Selection of transitions of C$_2$H$_5$NC, $\varv = 0$ covered by our 
ALMA survey. The LTE synthetic spectrum of C$_2$H$_5$NC, $\varv = 0$ used to
derive the upper limit on its column density is displayed in red and overlaid 
on the observed spectrum of Sgr~B2(N2) shown in black. The green synthetic 
spectrum contains the contributions of all molecules identified in our survey 
so far, but does not include the species shown in red. 
The central frequency and width are indicated in MHz below each panel. The y-axis is labeled in effective radiation temperature scale. The dotted line indicates the $3\sigma$ 
noise level.
}
\label{f:spec_c2h5nc_ve0}
\end{figure*}

\begin{figure*}[!h]
\centerline{\resizebox{\hsize}{!}{\includegraphics[angle=0]{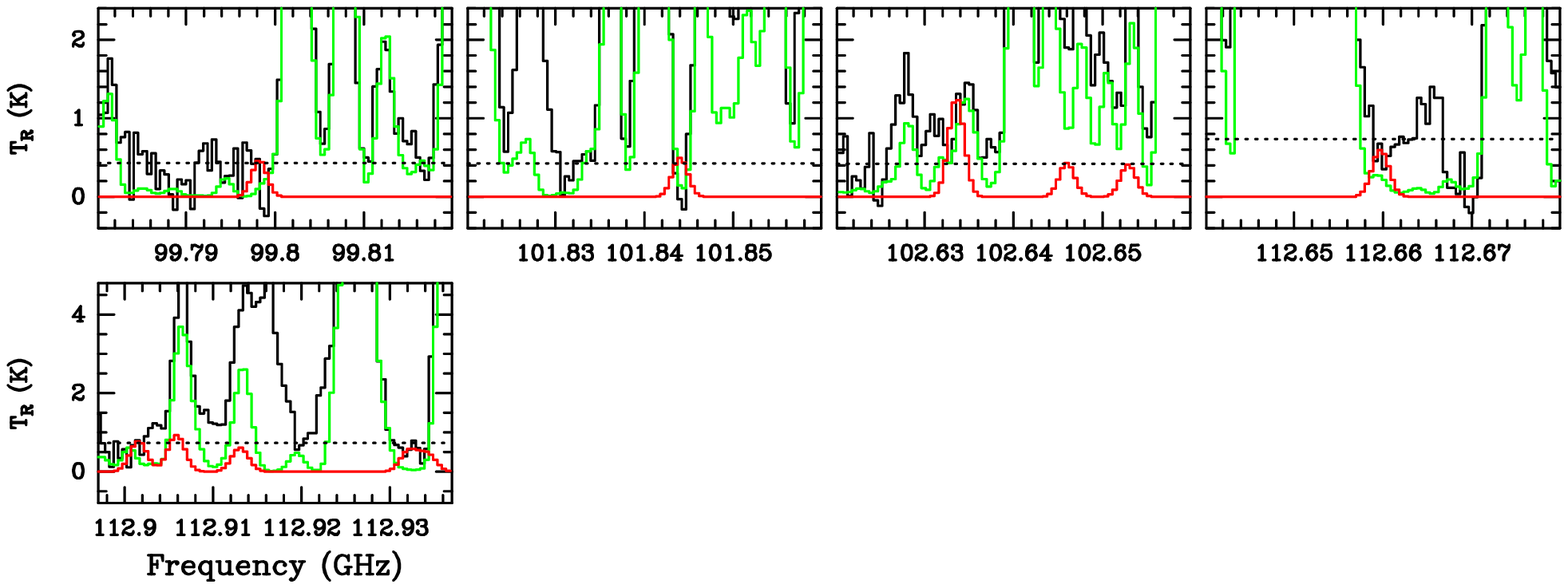}}}
\caption{Same as Fig.~\ref{f:spec_c2h5nc_ve0} but for C$_2$H$_3$NC, $\varv = 0$.
}
\label{f:spec_c2h3nc_ve0}
\end{figure*}

\begin{figure*}
\centerline{\resizebox{0.75\hsize}{!}{\includegraphics[angle=0]{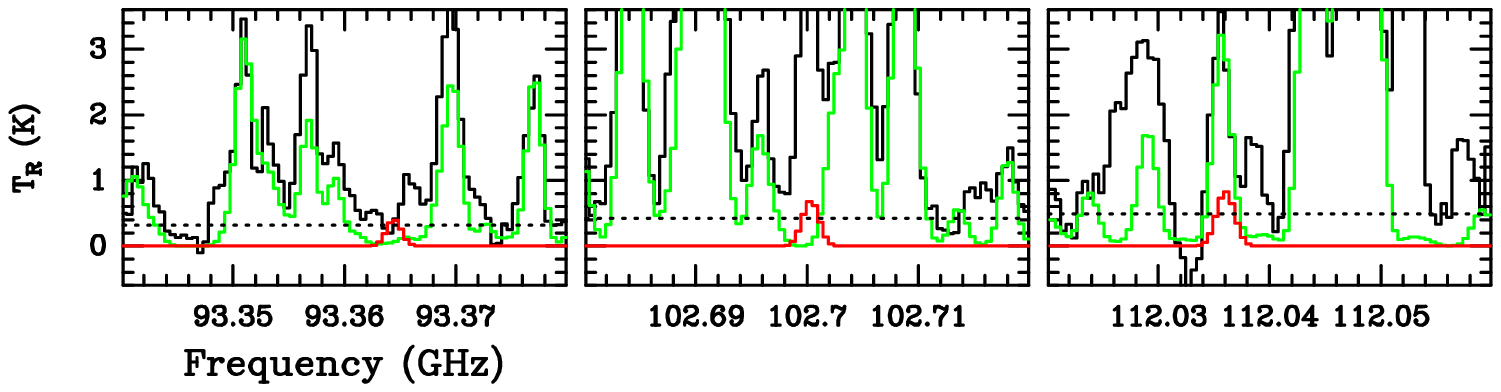}}}
\caption{Same as Fig.~\ref{f:spec_c2h5nc_ve0} but for HNC$_3$, $\varv = 0$.
}
\label{f:spec_hnc3_ve0}
\end{figure*}

\begin{figure*}[!h]
\centerline{\resizebox{\hsize}{!}{\includegraphics[angle=0]{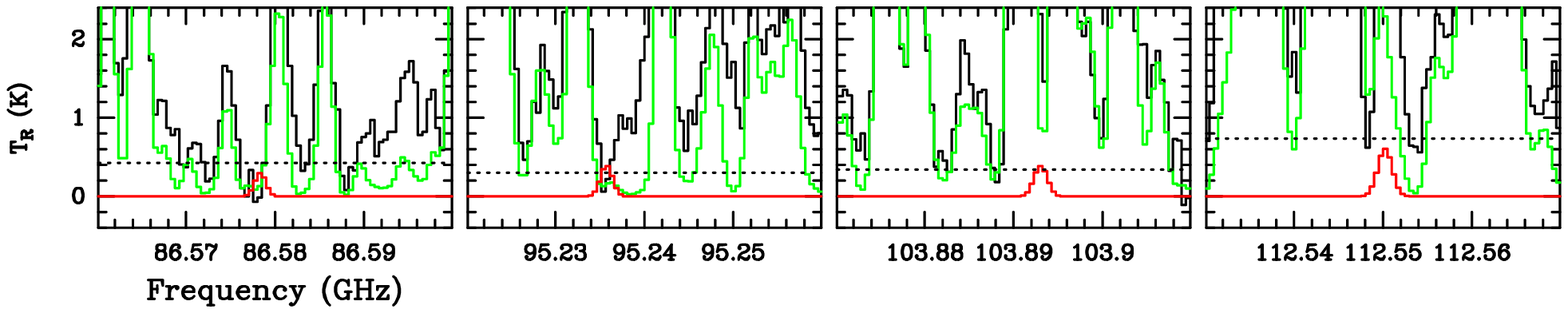}}}
\caption{Same as Fig.~\ref{f:spec_c2h5nc_ve0} but for HC$_3$NH$^+$, $\varv = 0$.
}
\label{f:spec_hc3nhp_ve0}
\end{figure*}

Figure~\ref{f:spec_ch3nc_ve1} shows all the transitions of CH$_3$NC, 
$\varv_8=1$ that are covered by the EMoCA survey toward 
Sgr~B2(N2). Figures~\ref{f:spec_c2h5nc_ve0}--\ref{f:spec_hc3nhp_ve0} show
a selection of transitions of C$_2$H$_5$NC, C$_2$H$_3$NC, HNC$_3$, and 
HC$_3$NH$^+$ that are covered by the survey and were used to derive an upper 
limit to their column density by comparing synthetic LTE spectra to
the EMoCA spectra. Table~\ref{t:list_r-nc} lists the lines of CH$_3$NC and HCCNC that we count as detected toward Sgr~B2(N2).

\begin{table*}
 {\centering
 \caption{
 Lines of CH$_3$NC and HCCNC detected in the EMoCA spectrum of Sgr B2(N2).
}
 \label{t:list_r-nc}
 \vspace*{0.0ex}
 \begin{tabular}{lrcccrccrrcrr}
 \hline\hline
 \multicolumn{1}{c}{Transition\tablefootmark{a}} & \multicolumn{1}{c}{Frequency} & \multicolumn{1}{c}{Unc.\tablefootmark{b}} & \multicolumn{1}{c}{$E_{\rm up}$\tablefootmark{c}} & \multicolumn{1}{c}{$g_{\rm up}$\tablefootmark{d}} & \multicolumn{1}{c}{$A_{\rm ul}$\tablefootmark{e}} & \multicolumn{1}{c}{$\sigma$\tablefootmark{f}} & \multicolumn{1}{c}{$\tau_{\rm peak}$\tablefootmark{g}} & \multicolumn{2}{c}{Frequency range\tablefootmark{h}} & \multicolumn{1}{c}{$I_{\rm obs}$\tablefootmark{i}} & \multicolumn{1}{c}{$I_{\rm mod}$\tablefootmark{j}} & \multicolumn{1}{c}{$I_{\rm all}$\tablefootmark{k}} \\ 
  & \multicolumn{1}{c}{\scriptsize (MHz)} & \multicolumn{1}{c}{\scriptsize (kHz)} &  \multicolumn{1}{c}{\scriptsize (K)} & & \multicolumn{1}{c}{\scriptsize (s$^{-1}$)} & \multicolumn{1}{c}{\scriptsize (mK)} & & \multicolumn{1}{c}{\scriptsize (MHz)} & \multicolumn{1}{c}{\scriptsize (MHz)} & \multicolumn{1}{c}{\scriptsize (K km s$^{-1}$)} & \multicolumn{2}{c}{\scriptsize (K km s$^{-1}$)} \\ 
 \hline
\multicolumn{13}{l}{CH$_3$NC} \\ 
5$_{2}$ -- 4$_{2}$ &  100517.433 &  90 &   43 & 22 &  6.8($-5$) &  141 &  0.109 & 100515.3 & 100519.2 &  48.5(6) &  42.5 &  49.3 \\ 
5$_{1}$ -- 4$_{1}$ &  100524.249 &  90 &   22 & 22 &  7.8($-5$) &  141 &  0.144 & 100523.1 & 100525.6 &  66.9(5) &  52.7 &  57.2 \\ 
\hline 
\multicolumn{13}{l}{HCCNC} \\ 
10$_{9}$ -- 9$_{8}$ &   99354.250 &  15 &   26 & 19 &  4.6($-5$) &  162 &  0.022 &  99352.7 &  99356.1 &   7.1(6) &   7.6 &   7.8 \\ 
10$_{9}$ -- 9$_{9}$ &   99354.250 &  15 &   26 & 19 &  5.2($-7$) & -- & -- & -- & -- & -- & -- & -- \\ 
10$_{9}$ -- 9$_{10}$ &   99354.250 &  15 &   26 & 19 &  1.3($-9$) & -- & -- & -- & -- & -- & -- & -- \\ 
10$_{10}$ -- 9$_{9}$ &   99354.250 &  15 &   26 & 21 &  4.6($-5$) & -- & -- & -- & -- & -- & -- & -- \\ 
10$_{10}$ -- 9$_{10}$ &   99354.250 &  15 &   26 & 21 &  4.7($-7$) & -- & -- & -- & -- & -- & -- & -- \\ 
10$_{11}$ -- 9$_{10}$ &   99354.250 &  15 &   26 & 23 &  4.7($-5$) & -- & -- & -- & -- & -- & -- & -- \\ 
11$_{10}$ -- 1$_{0}$ &  109289.095 &  15 &   32 & 21 &  6.2($-5$) &  123 &  0.026 & 109287.6 & 109291.0 &  14.7(4) &   9.7 &  12.6 \\ 
11$_{10}$ -- 1$_{1}$ &  109289.095 &  15 &   32 & 21 &  5.7($-7$) & -- & -- & -- & -- & -- & -- & -- \\ 
11$_{10}$ -- 1$_{1}$ &  109289.095 &  15 &   32 & 21 &  1.2($-9$) & -- & -- & -- & -- & -- & -- & -- \\ 
11$_{11}$ -- 1$_{1}$ &  109289.095 &  15 &   32 & 23 &  6.2($-5$) & -- & -- & -- & -- & -- & -- & -- \\ 
11$_{11}$ -- 1$_{1}$ &  109289.095 &  15 &   32 & 23 &  5.2($-7$) & -- & -- & -- & -- & -- & -- & -- \\ 
11$_{12}$ -- 1$_{1}$ &  109289.095 &  15 &   32 & 25 &  6.2($-5$) & -- & -- & -- & -- & -- & -- & -- \\ 
 \hline
 \end{tabular}
 }\\[1ex] 
 \tablefoot{
 \tablefoottext{a}{Quantum numbers of the upper and lower levels.}
 \tablefoottext{b}{Frequency uncertainty.}
 \tablefoottext{c}{Upper level energy.}
 \tablefoottext{d}{Upper level degeneracy.}
 \tablefoottext{e}{Einstein coefficient for spontaneous emission. $X$($Y$) means $X \times 10^Y$.}
 \tablefoottext{f}{Measured rms noise level.}
 \tablefoottext{g}{Peak opacity of the synthetic line.}
 \tablefoottext{h}{Frequency range over which the emission was integrated.}
 \tablefoottext{i}{Integrated intensity of the observed spectrum in brightness temperature scale. The statistical standard deviation is given in parentheses in unit of the last digit.}
 \tablefoottext{j}{Integrated intensity of the synthetic spectrum of the selected state.}
 \tablefoottext{k}{Integrated intensity of the model that contains the contribution of all identified molecules, including CH$_3$NC and HCCNC.}
 }
 \end{table*}

\clearpage\section{Additional tables}
\begin{table*}[!h]
 \begin{center}
 \caption{Physical quantities of new and related chemical species.}
 \label{t:physical_quantities}
 \vspace*{-2.0ex}
 \begin{tabular}{llll}
 \hline\hline\\[-1.0em]
 \multicolumn{1}{l}{Species} & \multicolumn{1}{l}{Binding Energy (K)} & \multicolumn{1}{l}{\hspace*{-2ex} \thead[l]{Enthalpy of Formation, $\Delta H_f$(298K) \\ (kcal mol$^{-1}$)}} &
 \multicolumn{1}{c}{\hspace*{-2ex} Notes} \\
 \hline\\[-1.0em]
 \ce{HCN} & 2050 & +32.30 & \\
 \ce{HNC} & 2050 & +32.30 & \\
 \ce{CH3CN} & 6150 & +17.70 & Binding energy from \citet{bertin17} \\
 \ce{CH3NC} & 5686 & +17.70 & Binding energy from \citet{bertin17} \\
 \ce{CH2CN} & 4230 & +59.00 & Based on \ce{CH3CN}-\ce{H} \\
 \ce{CH2NC} & 4230 & +74.00 & Based on \ce{CH2CN} \\
 \ce{C2H2CN} & 4187 & +105.84 & Based on \ce{C2H2}+\ce{CN} \\
 \ce{C2H2NC} & 4187 & +105.84 & Based on \ce{C2H2CN} \\
 \ce{C2H3CN} & 4637 & +42.95 & Based on \ce{C2H2CN}+\ce{H} \\
 \ce{C2H3NC} & 4637 & +63.20 & Based on \ce{C2H3CN} \\
 \ce{C2H4CN} & 5087 & +55.13 & Based on \ce{C2H3CN}+\ce{H} \\
 \ce{C2H4NC} & 5087 & +55.13 & Based on \ce{C2H4CN} \\
 \ce{CH3CHCN} & 5087 & +53.23 & Based on \ce{C2H4CN} \\
 \ce{CH3CHNC} & 5087 & +53.23 & Based on \ce{C2H4CN} \\
 \ce{C2H5CN} & 5537 & +12.71 & Based on \ce{C2H4CN}+\ce{H} \\
 \ce{C2H5NC} & 5537 & +31.69 & Based on \ce{C2H5CN} \\
 \ce{HC3N} & 4580 & +84.63 & Based on \ce{HCCN}+\ce{C} \\
 \ce{HCNC2} & 4580 & +84.60 & Based on \ce{HC3N} \\
 \ce{HCCNC} & 4580 & +84.60 & Based on \ce{HC3N} \\
 \ce{HNC3} & 4580 & +84.60 & Based on \ce{HC3N} \\
 \hline
 \end{tabular}
 \end{center}
 \vspace*{-2.5ex}
 \tablefoot{Binding energies are representative of physisorption on amorphous water ice. Enthalpies of formation are taken from the NIST WebBook database where available; otherwise estimates are made as described in the Notes column.}
 \end{table*}
\begin{table*}[!h]
 \begin{center}
 \caption{New grain-surface/ice-mantle reactions involved in formation and destruction of new and related species.}
 \label{t:grain_table}
 \vspace*{-2.0ex}
 \begin{tabular}{llll}
 \hline\hline\\[-1.0em]
 \multicolumn{1}{l}{\#} & \multicolumn{1}{l}{Reaction} & \multicolumn{1}{l}{\hspace*{-2ex} $E_A$ (K)} &
 \multicolumn{1}{c}{\hspace*{-2ex} Ref.} \\
 \hline\\[-1.0em]
 1 & \ce{CH2 + CH2NC -> C2H4NC} & 0 & From analogous -CN reaction \\
 2 & \ce{CH2OH + C2H4NC -> C2H5NC + H2CO} & 0 & From analogous -CN reaction \\
 3 & \ce{CH2OH + C2H5NC -> CH3OH + C2H4NC} & 6490 & From analogous -CN reaction \\
 4 & \ce{CH2OH + C2H5NC -> CH3OH + CH3CHNC} & 5990 & From analogous -CN reaction \\
 5 & \ce{CH2OH + CH2NC -> CH3NC + H2CO} & 0 & From analogous -CN reaction \\
 6 & \ce{CH2OH + CH3NC -> CH3OH + CH2NC} & 6200 & From analogous -CN reaction \\
 7 & \ce{CH2OH + CH3CHNC -> C2H5NC + H2CO} & 0 & From analogous -CN reaction \\
 8 & \ce{CH3 + CH2NC -> C2H5NC} & 0 & From analogous -CN reaction \\
 9 & \ce{CH3O + C2H4NC -> C2H5NC + H2CO} & 0 & From analogous -CN reaction \\
 10 & \ce{CH3O + C2H5NC -> CH3OH + C2H4NC} & 2340 & From analogous -CN reaction \\
 11 & \ce{CH3O + C2H5NC -> CH3OH + CH3CHNC} & 1950 & From analogous -CN reaction \\
 12 & \ce{CH3O + CH2NC -> CH3NC + H2CO} & 0 & From analogous -CN reaction \\
 13 & \ce{CH3O + CH3NC -> CH3OH + CH2NC} & 2070 & From analogous -CN reaction \\
 14 & \ce{CH3O + CH3CHNC -> C2H5NC + H2CO} & 0 & From analogous -CN reaction \\
 15 & \ce{COOH + C2H4NC -> C2H5NC + CO2} & 0 & From analogous -CN reaction \\
 16 & \ce{COOH + CH2NC -> CH3NC + CO2} & 0 & From analogous -CN reaction \\
 17 & \ce{COOH + CH3CHNC -> C2H5NC + CO2} & 0 & From analogous -CN reaction \\
 18 & \ce{HCO + C2H4NC -> C2H5NC + CO} & 0 & From analogous -CN reaction \\
 19 & \ce{HCO + CH2NC -> CH3NC + CO} & 0 & From analogous -CN reaction \\
 20 & \ce{HCO + CH3CHNC -> C2H5NC + CO} & 0 & From analogous -CN reaction \\
 21 & \ce{H + C2H2NC -> C2H3NC} & 0 & From analogous -CN reaction \\
 22 & \ce{H + C2H3NC -> C2H4NC} & 1320 & From analogous -CN reaction \\
 23 & \ce{H + C2H3NC -> CH3CHNC} & 619 & From analogous -CN reaction \\
 24 & \ce{H + C2H4NC -> C2H5NC} & 0 & From analogous -CN reaction \\
 25 & \ce{H + CH3CHNC -> C2H5NC} & 0 & From analogous -CN reaction \\
 26 & \ce{H + CH2NC -> CH3NC} & 0 & From analogous -CN reaction \\
 27 & \ce{H + CH3NC -> HCN + CH3} & 1200 & Based on \citet{graninger14} \\
 28 & \ce{H + HCCNC -> C2H2NC} & 1710 & From analogous -CN reaction \\
 29 & \ce{H + HNC -> H + HCN} & 1200 & \citet{graninger14} \\
 30 & \ce{NH + C2H5NC -> NH2 + C2H4NC} & 7200 & From analogous -CN reaction \\
 31 & \ce{NH + C2H5NC -> NH2 + CH3CHNC} & 7000 & From analogous -CN reaction \\
 32 & \ce{NH + CH3NC -> NH2 + CH2NC} & 7000 & From analogous -CN reaction \\
 33 & \ce{NH2 + C2H5NC -> NH3 + C2H4NC} & 3280 & From analogous -CN reaction \\
 34 & \ce{NH2 + C2H5NC -> NH3 + CH3CHNC} & 2480 & From analogous -CN reaction \\
 35 & \ce{NH2 + CH3NC -> NH3 + CH2NC} & 2680 & From analogous -CN reaction \\
 36 & \ce{O + HNC -> CO + NH} & 1100 & \citet{graninger14} \\
 37 & \ce{OH + C2H5NC -> H2O + C2H4NC} & 1200 & From analogous -CN reaction \\
 38 & \ce{OH + C2H5NC -> H2O + CH3CHNC} & 1000 & From analogous -CN reaction \\
 39 & \ce{OH + C2H3NC -> H2O + C2H2NC} & 4000 & From analogous -CN reaction \\
 40 & \ce{OH + CH3NC -> H2O + CH2NC} & 500 & From analogous -CN reaction \\
 \hline
 \end{tabular}
 \end{center}
 \vspace*{-2.5ex}
 \end{table*}
\begin{table*}[!h]
 \begin{center}
 \caption{New gas-phase reactions involved in formation and destruction of new and related species.}
 \label{t:reaction_table}
 \vspace*{-2.0ex}
 \begin{tabular}{llllll}
 \hline\hline\\[-1.0em]
 \multicolumn{1}{l}{\#} & \multicolumn{1}{l}{Reaction} & \multicolumn{1}{l}{\hspace*{-2ex} $\alpha$} &
 \multicolumn{1}{l}{\hspace*{-2ex} $\beta$} &
 \multicolumn{1}{l}{\hspace*{-2ex} $\gamma$} &
 \multicolumn{1}{l}{\hspace*{-2ex} Ref.}\\
 \hline\\[-1.0em]
 1 & \ce{CNC+ + CH4 -> HC2NCH+ + H2} & 2.10E-10 & 0 & 0 & \citet{osamura99} \\
 2 & \ce{CH3+ + HCN -> CH3CNH+} & 7.65E-9 & -0.5 & 0 & \citet{defrees85} \\
 3 & \ce{CH3+ + HCN -> CH3NCH+} & 1.35E-9 & -0.5 & 0 & \citet{defrees85} \\
 4 & \ce{CH3CN+ + CO -> CH2NC + HCO+} & 3.00E-10 & -0.5 & 0 & From analogous -CN reaction \\
 5 & \ce{CH2OH + C2H4NC -> C2H5NC + H2CO} & 1.00E-11 & 0 & 0 & From analogous -CN reaction \\
 6 & \ce{CH2OH + CH3CHNC -> C2H5NC + H2CO} & 1.00E-11 & 0 & 0 & From analogous -CN reaction \\
 7 & \ce{CH3O + C2H4NC -> C2H5NC + H2CO} & 1.00E-11 & 0 & 0 & From analogous -CN reaction \\
 8 & \ce{CH3O + CH3CHNC -> C2H5NC + H2CO} & 1.00E-11 & 0 & 0 & From analogous -CN reaction \\
 9 & \ce{C + HNC -> C2N + H} & 4.00E-11 & 0 & 0 & \citet{graninger14} \\
 10 & \ce{C + HNC -> HCN + C} & 1.60E-10 & 0 & 0 & \citet{graninger14} \\
 11 & \ce{C + CH2NH -> CH2CN + H} & 1.00E-10 & 0 & 0 & \citet{loison14} \\
 12 & \ce{C + CH2NH -> HCN + CH2} & 1.00E-10 & 0 & 0 & \citet{loison14} \\
 13 & \ce{COOH + C2H4NC -> C2H5NC + CO2} & 1.00E-11 & 0 & 0 & From analogous -CN reaction \\
 14 & \ce{COOH + CH3CHNC -> C2H5NC + CO2} & 1.00E-11 & 0 & 0 & From analogous -CN reaction \\
 15 & \ce{H + CH3NC -> HCN + CH3} & 1.00E-10 & 0 & 0 & \citet{graninger14} \\
 16 & \ce{HCO + C2H4NC -> C2H5NC + CO} & 1.00E-11 & 0 & 0 & From analogous -CN reaction \\
 17 & \ce{HCO + CH3CHNC -> C2H5NC + CO} & 1.00E-11 & 0 & 0 & From analogous -CN reaction \\
 18 & \ce{H + HNC -> H + HCN} & 1.00E-10 & 0 & 1200 & \citet{graninger14} \\
 19 & \ce{H + C2N -> C + HCN} & 2.00E-10 & 0 & 0 & \citet{loison14} \\
 20 & \ce{H + H2CN -> H2 + HNC} & 1.20E-11 & 0 & 0 & \citet{loison14} \\
 21 & \ce{O + HNC -> CO + NH} & 7.64E-10 & 0 & 1120 & \citet{graninger14} \\
 \hline
 \end{tabular}
 \end{center}
 \vspace*{-2.5ex}
 \tablefoot{Dissociative recombination reactions are included for all new species, but are omitted from these tables for brevity.}
 \end{table*}

\end{appendix}

\end{document}